\documentclass[12pt,reqno]{amsart}

\usepackage{latexsym}
\usepackage{amssymb}

\renewcommand{\Re}{{\operatorname{Re}\,}}
\renewcommand{\Im}{{\operatorname{Im}\,}}

\newcommand{\Diff}{{\operatorname{Diff}}}

\newcommand{\sym}{{\operatorname{Sym}}}

\newcommand{\crit}{{\operatorname {crit}}}

\newcommand{\sm}{\smallsetminus}
\newcommand{\szego}{Szeg\"o }
\newcommand{\nhat}{\raisebox{2pt}{$\wh{\ }$}}

\newcommand{\inv}{^{-1}}
\newcommand{\kahler}{K\"ahler }

\newcommand{\wt}{\widetilde}
\newcommand{\wh}{\widehat}
\newcommand{\PP}{{\mathbb P}}
\newcommand{\R}{{\mathbb R}}
\newcommand{\C}{{\mathbb C}}

\newcommand{\Z}{{\mathbb Z}}

\newcommand{\CP}{\C\PP}
\renewcommand{\d}{\partial}
\newcommand{\dbar}{\bar\partial}
\newcommand{\ddbar}{\partial\dbar}

\newcommand{\E}{{\mathbf E}\,}

\newcommand{\half}{{\frac{1}{2}}}
\newcommand{\vol}{{\operatorname{Vol}}}

\newcommand{\supp}{{\operatorname{Supp\,}}}
\renewcommand{\phi}{\varphi}
\renewcommand{\epsilon}{\varepsilon}

\newcommand{\tcal}{\mathcal{T}}
\newcommand{\ccal}{\mathcal{C}}
\newcommand{\dcal}{\mathcal{D}}
\newcommand{\ecal}{\mathcal{E}}
\newcommand{\fcal}{\mathcal{F}}

\newcommand{\hcal}{\mathcal{H}}
\newcommand{\ical}{\mathcal{I}}
\newcommand{\lcal}{\mathcal{L}}
\newcommand{\mcal}{\mathcal{M}}
\newcommand{\kcal}{\mathcal{K}}
\newcommand{\ncal}{\mathcal{N}}

\newcommand{\rcal}{\mathcal{R}}
\newcommand{\ocal}{\mathcal{O}}
\newcommand{\scal}{\mathcal{S}}

\newcommand{\wcal}{\mathcal{W}}

\newcommand{\al}{\alpha}
\newcommand{\be}{\beta}
\newcommand{\ga}{\gamma}

\newcommand{\La}{\Lambda}
\newcommand{\la}{\lambda}
\newcommand{\ep}{\varepsilon}
\newcommand{\de}{\delta}

\newcommand{\om}{\omega}
\newcommand{\Om}{\Omega}

\newtheorem{theo}{{\sc Theorem}}[section]
\newtheorem{cor}[theo]{{\sc Corollary}}

\newtheorem{prob}[theo]{{\sc Problem}}
\newtheorem{lem}[theo]{{\sc Lemma}}
\newtheorem{prop}[theo]{{\sc Proposition}}

\newenvironment{rem}{\medskip\noindent{\it Remark:\/} }{\medskip}

\title[Critical points and supersymmetric vacua, III ]
{Critical points and supersymmetric vacua, III:\\ String/M models}

\author{Michael R. Douglas}
\address{NHETC and Department of Physics and Astronomy,
Rutgers University,
Piscataway, NJ 08855--0849, USA;
and I.H.E.S., Bures-sur-Yvette, France} \email{mrd@physics.rutgers.edu}

\author{Bernard Shiffman}
\address{Department of Mathematics, Johns Hopkins University, Baltimore,
MD 21218, USA} \email{shiffman@math.jhu.edu}

\author{Steve Zelditch }
\address{Department of Mathematics, Johns Hopkins University, Baltimore,
MD 21218, USA} \email{zelditch@math.jhu.edu}

\thanks{Research partially supported by
DOE grant DE-FG02-96ER40959  (first author) and NSF grants
DMS-0100474 (second author) and DMS-0302518 (third author).}

\begin{document}

\begin{abstract} A fundamental problem in contemporary string/M
theory is to count the number of inequivalent vacua satisfying
constraints in a string theory model. This article contains the
first rigorous results on the number and distribution of
supersymmetric vacua of type IIb string theories compactified on a
Calabi-Yau $3$-fold $X$ with flux.  In particular,  complete
proofs of the counting formulas in Ashok-Douglas \cite{AD} and
Denef-Douglas \cite{DD} are given, together with van der Corput
style remainder estimates.

Supersymmetric vacua are critical points of certain holomorphic
sections (flux superpotentials) of a line bundle $\lcal \to \ccal$
over the moduli space  of complex structures on $X \times T^2$
with respect to the Weil-Petersson connection. Flux
superpotentials form a lattice  of full rank in a
 $2 b_3(X)$-dimensional real subspace $\scal \subset H^0(\ccal,
 \lcal)$. We show that the density of  critical points in $\ccal$ for this
  lattice of sections is well approximated by
  Gaussian measures of the kind studied in \cite{DSZ1, DSZ2, AD, DD}.

\end{abstract}

\maketitle

\tableofcontents

\section{Introduction}

This is the third in a series of
articles  \cite{DSZ1, DSZ2} (see also \cite{Z2}) by the authors on
statistics of critical points of
random holomorphic sections  and their applications to
the  vacuum selection problem in string/M theory. We recall that,
in these articles, a `vacuum' in string theory
is a Calabi-Yau manifold of complex dimension $d = 3$  which  forms the $6$ `small
dimensions' of the $10$-dimensional  universe, together with a choice of
orientifolding and flux.  Mathematically,
vacua are critical points of a {\it superpotential} $W$, a
holomorphic section of a line bundle ${\mathcal L} \to \ccal$ over
the configuration space $\ccal$ which will be recalled in \S
\ref{BGR}. The `vacuum selection problem' is that there exists no
principle at present which selects a unique superpotential, nor a
unique critical point of a given superpotential, out of a large
ensemble of possible vacua. This motivates the  program of
studying statistics
 of vacua, whose  basic problems are to
count the number of vacua satisfying physically natural
constraints and to determine how they are distributed in $\ccal$
(see
 \cite{Doug, DD, AD, DGKT,  KL, Sil}). In this article, we present
 the first rigorous results on counting vacua with remainder
 estimates. In particular, we  justify and improve on the approximations made in
 \cite{DD}.

Our previous articles  \cite{DSZ1, DSZ2} were devoted to the
statistics  of critical points of Gaussian random holomorphic
sections of line bundles over complex manifolds. The principal
issue we face in this article is that the physically relevant
ensembles of superpotentials are not Gaussian but rather are
discrete ensembles of `quantized flux' superpotentials which form
a set of lattice points in a hyperbolic shell in $H^3(X, \C)$.
This hyperbolic shell is defined by  the inequality (known as the
{\it tadpole constraint})
\begin{equation} \label{TC}  0 \leq Q[\phi] \leq L,\;\;\;
\end{equation} where
\begin{equation} \label{HRFORM} Q[\phi]= Q(\phi,\bar\phi) = -\sqrt{-1}\, \int_X \phi
\wedge
\bar{\phi} \end{equation} is  the  Hodge-Riemann bilinear form.
As will be recalled in  \ \S \ref{IP}, $Q$ is an indefinite
quadratic form, whose `null cone' $\{G: Q[G] = 0 \}$ is a real
quadric hypersurface which separates $H^3(X, \C)$ into the
interior $\{W : Q[G]
> 0\} $ and the exterior where $Q[G] < 0$. As will be seen below (Propositions
 \ref{POS} and \ref{HRPLUS})
 , only
flux superpotentials corresponding to lattice points in $\{G :
Q[G]
> 0\} $
contribute vacua, and that is why we consider the shell
(\ref{TC}).

Our main results show that as $L \to \infty$, the statistics of
critical points relative to the discrete lattice ensemble is well
approximated by the statistics of critical points relative to the
continuum ensemble in the shell, which is dual to the Gaussian
ensembles of \cite{DSZ1, DSZ2} and is therefore well understood.
Thus, the vacuum statistics problem in string/M theory is a
mixture of two kinds of equidistribution problems:
\begin{enumerate}

\item The distribution of radial projections of lattice points
onto a quadric hypersurface;

\item The distribution of critical points of a continuous ensemble
of random holomorphic sections (related to a Gaussian ensemble) of
a negative line bundle, and their interpretation in the special
geometry of Calabi-Yau moduli spaces.

\end{enumerate}
The   equidistribution problem in (2) is analyzed in detail in
\cite{DSZ1, DD}, so the main purpose of this paper is to analyze
(1) and to combine it with the previous analysis of (2).

At the end of this article in \S \ref{FCFP} and in \cite{Z2}, we
compare the mathematical results of this article to discussions of
vacua in the string theory literature.

\subsection{\label{BGR}Background to the results}

To state our results, we will need some notation  (see \S \ref{CY}
for more details). The models we consider in this article are
called type IIb flux compactifications \cite{GVW, GKP}. We fix a
complex $3$-dimensional Calabi-Yau manifold $X$, i.e. a complex
manifold with trivial canonical bundle $K_X \simeq \ocal$ and with
first Betti number $b_1(X)=0$. In some of the physics literature,
it is also assumed that $H^{2,0}(X)=0$, but our results hold
without this assumption. For each complex structure $z$ on $X$,
there is a corresponding Hodge decomposition
\begin{equation} \label{HD} H^3(X, \C) = H^{3,0}_z(X) \oplus H^{2,1}_z(X)\oplus
H^{1,2}_z(X)\oplus H^{0, 3}_z(X). \end{equation} The space
$H^{3,0}_z(X)$ of $(3, 0)$-forms relative to $z$ is
one-dimensional and is spanned by a nowhere vanishing holomorphic
volume form $\Omega_z.$ We also put $b_3 =b_3(X)= \dim H^3(X, \R)$,
$h^{p,q} =h^{p,q}(X) = \dim_{\C} H^{p,q}(X)$. Thus, $b_3 = 2(h^{2,1} + 1)$.

When we speak of vacua of string theory compactified on the
Calabi-Yau space $X$, we refer to classical vacua of the effective
supergravity theory it determines. As discussed in \cite{St2}, the
effective supergravity Lagrangian  is derived by `integrating out'
or neglecting the massive modes (positive
 eigenvalues) of various operators. The data of effective supergravity   consists
of $(\ccal, \lcal, W)$ where:
\begin{enumerate}

\item $\ccal$ is the configuration space;

\item $\lcal \to \ccal$ is a holomorphic line bundle.

\item the  superpotential $W$ is a holomorphic section of $\lcal$.
\end{enumerate}
In  type IIb flux compactifications the configuration space is the
moduli
  space of Calabi-Yau (Ricci flat \kahler) product metrics on $X \times T^2$.  At this time of writing,
  the study of  vacua  in string theory is  simplified by
  replacing the moduli space of Calabi-Yau metrics by the moduli space  of complex
  structures on $X$ (see e.g. \cite{Doug, AD}). In the case where
$h^{2,0}(X)=0$, this is equivalent to fixing the   \kahler class
$[\om]\in H^2(X,\R)$ of the Calabi-Yau metrics. Hence we define
the configuration space to be
 \begin{equation} \label{CCAL} \ccal = \mcal \times \ecal, \end{equation}
where $\mcal$ is the moduli space of complex structures on $X$ and
where $\ecal = \hcal/ SL(2, \Z)$ is the moduli space of elliptic
curves. Throughout this paper we identify $\ccal=\mcal\times
\ecal$ with a fundamental domain $\dcal$ for the modular group
$\Gamma$ in  the Teichm\"uller space $\tcal\! eich(X)\times\hcal$
 of complex structures (see \S\ref{GCY}). For simplicity of
 exposition,
  we refer to   restrictions  to $\dcal$ of holomorphic
objects on $\tcal\! eich(X)\times\hcal$ as holomorphic objects
over $\ccal$.

The line bundle $\lcal$ is defined to be the dual line bundle to
the Hodge bundle $H^{3,0}(X) \otimes H^{1,0}(T^2) \to \ccal$,
where $T^2 = \R^2/\Z^2$.  We give $\ccal$ the {\it Weil-Petersson
\kahler form\/} $\om_{WP}$ induced from the Weil-Petersson metric
on $\lcal$ (see \S \ref{SUSYCRIT}). To be precise, $\lcal$ is a
holomorphic  line bundle over $\tcal\! eich(X)\times\hcal$, and
$W$ is a holomorphic section of $\tcal\! eich(X)\times\hcal$. But
as mentioned above, by holomorphic sections  $W\in
H^0(\ccal,\lcal)$ we mean  restrictions to $\dcal$ of holomorphic
sections of $H^0(\tcal\! eich(X)\times\hcal, \lcal).$

 Type IIb flux compactifications
contain two non-zero harmonic $3$-forms $F, H \in H^3(X, \Z)$
which are known respectively as the RR (Ramond-Ramond) and NS
(Neveu-Schwartz) $3$-form field strengths. We combine them into a
complex flux $G = F + i H \in H^3(X, \Z \oplus i \Z)$.  The
parameter  $\tau \in \ecal$ is known as the dilaton-axion and may
be viewed as the period of  $\omega_{\tau} = dx + \tau dy$ over
the one-cycle dual to $dy$ in $T^2$.  Given $G \in H^3(X, \Z
\oplus \sqrt{-1} \Z),$ physicists define the corresponding flux
superpotential $W_G$ by:
\begin{equation} \label{WSUBG} W_{G}(z, \tau) = \int_X ( F + \tau H)  \wedge
\Omega_{z}, \end{equation} where $\Omega_z \in H^{3, 0}(X)$.  This
is not well-defined as a function on $\ccal$ since $\Omega_z$ and
$\tau$ depend on a choice of frame. To be more precise,  $G \in
H^3(X, \C)$ determines a section $W_{G}$ of the line bundle
$$\lcal=(H^{3,0}(X) \otimes H^{1,0}(T^2))^* \to \tcal\! eich(X)\times\hcal$$
by making $G$
into the following linear functional on $H^{3,0}_z (X) \otimes
H^{1,0}_{\tau}(T^2):$
\begin{equation} \label{WGG} \langle W_{G} (z, \tau), \Omega_{z} \otimes \omega_{\tau} \rangle
= \int_{X \times T^2 }  (F \wedge dy -  H \wedge dx) \wedge (
\Omega_{z} \wedge \omega_{\tau}).
\end{equation}

 The map $G\to W_{G}$ defines an injective real (but not complex) linear map
which embeds complex integral fluxes
\begin{equation} H^3(X, \Z \oplus \sqrt{-1} \Z) \to H^0(\ccal,
\lcal) \end{equation} as a lattice  of rank $2 b_3$ in $H^0(\mcal
\times \ecal, \lcal)$  which we call the lattice
 $\scal^{\Z}$ of {\it integral flux
superpotentials}. The real  span \begin{equation} \scal = \R
\scal^{\Z} \subset H^0(\mcal, \lcal) \end{equation}  of
$\scal^{\Z}$ is also important, and will be referred as the space
of {\it flux superpotentials}.  We emphasize here that  $\scal$ is
not a complex vector space, nor are any of the associated spaces
discussed below.  We also use the (real-linear) map $G\mapsto W_G$ to
regard
$Q$ as a quadratic form on
$\scal$, writing
\begin{equation} \label{QW} Q[W_G]:=Q[G]= -\sqrt{-1}\int_X G\wedge\overline G =
2\int_X F\wedge H\;,\qquad G=F+iH\in H^3(X,\C)\;.\end{equation}

The bundles $H_{z}^{3,0} \to \mcal$ and $H_{\tau}^{1, 0} \to
\ecal$  carry   Weil-Petersson Hermitian metrics $h_{WP}$ defined
by
\begin{equation}\label{HWP}  h_{WP}(\Omega_{z}, \Omega_{z}) = e^{-K(z, \bar{z})} =
i\int_X
\Omega_{z} \wedge \overline{\Omega}_{z},\end{equation} and their
associated Chern connections $\nabla_{WP}$. They induce dual
metrics and connections on $\lcal$. We denote the connection
simply by $\nabla$.

\subsection{Statement of the problem}

Given a flux superpotential $W$, there is an associated potential
energy on $\ccal$ defined by
\begin{equation} \label{V} V_W(Z) = |\nabla W(Z)|^2 - 3 |W(Z)|^2.
 \end{equation}
 (See \cite{WB} for background on $V$).
  By a vacuum we mean a critical point of $V(Z)$ on $\ccal$.
In this paper, we only study supersymmetric vacua,
namely  $Z\in \ccal$ which are connection critical points in the
sense that $\nabla_{WP} W(Z) = 0.$ We denote the set of
supersymmetric  vacua of $W$ by
\begin{equation}\label{CRITSET} Crit(W) = \{Z \in \ccal: \nabla_{WP} W(Z) = 0\}.
\end{equation}

Our goal is thus to count and find the distribution law of the
supersymmetric vacua
\begin{equation} \label{VACUAL} \{\mbox{SUSY vacua}\} =
\bigcup_{\textstyle G\in\scal^\Z: Q[G]
\leq L} Crit (W_G)
\end{equation}  as $W_{G}$ varies over the lattice $\scal^{\Z}$
within the hyperbolic shell (\ref{TC}). To define the distribution
law, we  introduce the {\it incidence relation}
\begin{equation} \label{ICAL} \ical = \{(W_G, Z) \in \scal \times \ccal: \nabla W_G (Z) = 0 \}.
\end{equation}
We shall view $\ccal$ as a fundamental domain for the modular
group $\Gamma$ in Teichm\"uller space (cf. \S \ref{CY}). The
incidence variety $\ical$ is then a real $2m$-dimensional
 subvariety of $\ccal\times \scal$ with the following diagram
of projections:
\begin{equation}\label{DIAGRAM} \begin{array}{ccccc} &\hspace{-.4in}
\ical& \hspace{-.3in}
\subset \ccal \times \scal &   \\
 \rho \swarrow & \searrow \pi & \\ \ccal &  \scal
\end{array} \end{equation}  The fiber $\pi\inv(W)$ is the set
$Crit(W)$ of critical points of $W$ in $\ccal$. Since $\ccal$ is
regarded as a fundamental domain in Teichm\"uller space, the map
$\pi$ is not surjective: there exist $W$ with no critical points
in $\ccal$; hence $\pi(\ccal)$ is a domain with boundary in
$\scal$ (see \S \ref{EXZERO}). Critical points can move out of
$\ccal$ as $W$ varies in $\scal$. (There is a similar but more
complicated theory of non-supersymmetric vacua \cite{DD2}.)

The fibers of $\rho$ are the
subspaces
\begin{equation} \label{QZ} \scal_Z : = \{W \in \scal: \nabla_{WP} W(Z) = 0
\},
\end{equation}
which play a crucial role in this article.   They  have the
remarkable Hodge theoretic identifications,
\begin{equation}\label{FZHODGE}  \scal_{z,\tau} \equiv  H^{2,1}_z(X)
 \oplus H^{0, 3}_z(X)\quad (\mbox{Proposition}\;\; \ref{POS}).
\end{equation}
 It then follows (see Proposition \ref{H3XR2}) that
$\ical\buildrel{\rho} \over \to \ccal$ is a vector bundle (with
fiber $\approx\C^{b_3/2}$) over a manifold with boundary. Another
key point is that
 the  restrictions of $Q$ to the fibers are  always positive
 definite:
\begin{equation}\label{QZPOS}  Q |_{ H^{2,1}_z(X) \oplus H^{0, 3}_z(X)}
\gg  0 \quad (\mbox{Proposition}\;\; \ref{HRPLUS}), \end{equation}
i.e. $\scal_Z$ lies in the positive cone
$\{Q(\phi,\overline{\phi}) > 0\}$ of the indefinite quadratic
(Hodge-Riemann) form
 (\ref{HRFORM}) (cf.  \S \ref{IP}).

We  now define the {\it discriminant locus}  $$\wt\dcal=\{(Z,W)\in
\ical:\det H^c W(Z)=0\} $$ of  points $(Z,W)\in\ical$ such that
$Z$ is a degenerate critical point of $W$, where $H^c W(Z)$ is the
{\it complex Hessian\/}  of $W$ at the critical point $Z$ as
defined in (\ref{HmatriX})--(\ref{H''}). Equivalently, $\wt\dcal$
is the set of critical points of the second projection
$\ical\buildrel{\pi} \over \to \scal$ together with the singular
points of $\ical$. Its image $\dcal = \pi(\wt \dcal)$ under $\pi$
is the discriminant variety of superpotentials with degenerate
critical points.

For each $W \in \scal\sm\{0\}$, we
 define its distribution of (non-degenerate) critical points as the measure $C_W$ on
$\ical\sm\wt\dcal$ defined by
\begin{equation} \langle C_W, \psi \rangle = \sum_{Z \in Crit(W)} \psi(Z,
W),\end{equation} for $\psi \in \ccal(\ical)$ such that
$\rho(\supp\psi)$ is relatively compact in $\ccal$ and $\supp\psi$
is disjoint from $\wt\dcal$. A more general definition of $C_W$ is
\begin{equation}\label{CWintro}  C_W = |\det H^c W(Z)| \;\; \nabla W^* \delta_0
\end{equation}
which will be discussed in \S \ref{HDCP}. We make these
assumptions on $\psi$ so that the sum on the right side is a
finite and well-defined sum. Indeed,  the pull back is not
 well-defined (without further work) on $\wt \dcal$. We will say
 more about $\wt \dcal$ after the statement of  Theorem \ref{MAIN}.

 The basic  sums we study are :
\begin{eqnarray}
\ncal_{\psi}(L) & = & \sum \big\{\langle C_N, \psi
\rangle :N \in \scal^\Z ,\ Q[N] \leq L\} \nonumber \\
& = & \sum \big\{\psi(Z, N):{ (Z,N) \in \ical,\ N\in\scal^\Z,\ 0 \leq Q[N]
\leq L }\big\}\;.\label{Nsum}
\end{eqnarray} For instance, when $\psi \equiv \chi_{K}$ is the characteristic
function of a compact subset $K \subset \subset \ical \sm \wt
\dcal$, $N_{\psi}(L)$  counts the total number of non-degenerate
critical points lying over $\rho(K)$  coming from all integral
flux superpotentials with $Q[W] \leq L$.   Physicists are
naturally interested in counting the number of vacua with close to
the observed  values of the cosmological constant and other
physical quantities, and hence would study sums relevant to such
quantities. For instance, the {\it cosmological constant} of the
theory defined by a vacuum $Z$ is the value $V(Z)$ of the
potential there (see \cite{DD}, \S 3.3).  Thus, we may state the
main problem of this paper:

\begin{prob} \label{CRITPROB} Find the asymptotics and remainder for
$\ncal_{\psi}(L)$ as $L \to \infty. $ \end{prob}

As indicated above, this problem is very closely related to the
pure lattice point problem of measuring the rate of uniform
distribution of radial projections of lattice points onto the
surface of a quadric  hypersurface. More generally one could
consider any  smooth strictly convex set $Q\subset \R^n$ ($n\ge
2)$  with $0\in Q^\circ$. Associated to $Q$ is the norm $|X|_Q$ of
$X\in \R^n$ defined by
$$Q=\{X\in\R^n:|X|_Q<1 \}\,.$$ To measure the equidistribution
of radial  projections of lattice points to $\d Q$, one considers
the sums \begin{equation} \label{SF} S_f (t) = \sum_{k \in
\Z^n\cap tQ\sm\{0\}} f\left(\frac{k}{|k|_Q}\right), \quad
\mbox{with } \ f \in C^{\infty}(\d Q),\ t>0.  \end{equation}

The
parallel lattice point problem is then
\begin{prob} \label{LATTICEPROB} Find the asymptotics and remainder for
$S_f(t)$ as $t \to \infty. $ \end{prob}

\subsection{Statement of the results} In
Theorem \ref{localvdC}, we obtain a van der Corput type estimate for the
lattice point problem \ref{LATTICEPROB}.  For the critical point
problem, we first give an elementary formula which is based on a trivial
lattice counting estimate (which is useful since it is sometimes sharp),
namely where the remainder term is simply a count of the cubes of the
lattice which intersect the boundary.  We  denote by
$\chi_{Q_Z}$ the characteristic function of the shell $\{W\in\scal_Z:0 <
Q_Z[W] < 1\}
$.

\begin{prop} \label{LNMO} Suppose that $\psi = \chi_K$
where $K\subset\ical$  such that $(Z,W)\in K \Leftrightarrow
(Z,rW) \in K$ for $r\in\R^+$.  Assume further that $\rho(K)$ is
relatively compact in $\ccal$ and $\pi(\d K)$ is piecewise smooth.
Then
$$\ncal_{\psi}(L) = L^{b_3}\left[
 \int_{\ccal}
\int_{\scal_{Z}} \psi(Z,W)\,|\det H^c W(Z)|
  \chi_{Q_{Z}}(W) \,dW\,d\vol_{WP}(Z) + O\left(L^{-1/2} \right)\right].
$$
\end{prop}

Here and in Theorem \ref{MAIN} below, $dW$ means the multiple of
Lebesgue measure on $\scal_Z$ which gives the volume form for  the
positive-definite quadratic form $Q_Z=Q|_{\scal_Z}$.
 We note that the
integral converges, since  by (\ref{QZPOS}), $\{Q_Z \leq 1\}$ is
an ellipsoid of finite volume.

It would be interesting to know if the remainder estimate is sharp
for any domain $K \subset \ical$. In the pure  lattice point
Problem \ref{LATTICEPROB}, the corresponding `trivial estimate' is
sharp. For instance, consider the domain $K = S^{n-1}_+ \subset
S^{n-1}$ formed by the northern hemisphere and put $\psi =
\chi_K$. Then the remainder term
$$\sum_{k \in \Z^n, |k|\leq \sqrt{L}} \chi_K\left(\frac{k}{|k|}\right)
- L^{\frac{n}{2}} \int_K f dA
$$
reflects the concentration of projections of lattice points on the
boundary $\d S^{n-1}_+$, namely a great equatorial sphere. When
the equator is defined by $x_{n}= 0$, the lattice points
projecting over the equator are the lattice points in $\Z^{n-1}
\subset \R^{n-1}$ and the number with $|k| \leq \sqrt{L}$ is of
size $\sim L^{\frac{n-1}{2}}.$ Analogously one may ask if there
are domains $K \subset \ccal$ along which critical points
concentrate to the same maximal degree. Some evidence that the
answer is `no' will be presented in \S \ref{LNMOPROOF}.

Our  main result stated below is a much sharper  van der Corput
type asymptotic estimate of $\ncal_{\psi}(L)$ as $L \to \infty$
for homogeneous  test functions which vanish near the discriminant
locus. Here, we say that a function $\psi\in\ccal(\ical)$ is
homogeneous of order $\al$ if
$$\psi(Z,rW) = r^\al\psi(Z,W),\qquad (Z,W)\in \ical,\ r\in\R^+\;.$$
We consider homogeneous functions since they include  (smoothed)
characteristic functions as well as the  cosmological constant
(which is homogeneous of degree $2$).

\begin{theo}\label{MAIN} Let $\psi\in\ccal^\infty(\ical)$ be homogeneous of order
$\al\ge 0$ and suppose that $\rho(\supp\psi)$ is a compact subset of
$\ccal$ and $\supp\psi
\cap\wt\dcal=\emptyset$. Then
$$\ncal_{\psi}(L) = L^{b_3+\al/2}\left[\int_{\ccal}
\int_{\scal_{Z}}\psi(Z,W)\, |\det H^c W(Z)|
\,  \chi_{Q_{Z}}(W) \,dW\,d\vol_{WP}(Z) +
O\left(L^{-\frac{2b_3}{2b_3+1}}\right)\right].
$$

\end{theo}

It is reasonable to make the assumption  $\supp\psi
\cap\wt\dcal=\emptyset$, because  degenerate critical points
cannot be physically acceptable vacua in string/M theory. Indeed,
 the  Hessian of $W$ at a critical point defines the `fermionic
mass matrix' of the theory, and a degenerate critical point would
give rise to massless fermions which are not observed in physics.
(See \cite{WB} for definitions of the mass matrix.)

Let us note some key features of the geometry of $\wt \dcal$ which
play a role in  the assumptions (and proofs) of Proposition
\ref{LNMO}  and Theorem \ref{MAIN}.
 First,   as observed in \cite{DSZ1,DSZ2}, its defining
 equation
\begin{equation}\label{detHc}\det H^c W(Z) = \det(H^*H-|W|^2I) =
0\end{equation} is real valued; here,  $H$ is the holomorphic Hessian (see
\S
\ref{CRITHESS}). Hence, $\wt \dcal \subset \ical$ is a real
analytic
 hypersurface (with boundary). For test functions $\psi$ which do not vanish on $\wt \dcal$,
 the expression $\langle C_W, \psi \rangle$ (when well-defined) can jump
 as one passes from one component of  $\scal\sm \dcal$ to another
 or across the boundary of $\ccal$.
It follows from \eqref{detHc} that
$\wt \dcal \cap (\{Z\} \times \scal_Z)$ is a real conic hypersurface
for all
$Z\in\ccal$. Thus $\wt \dcal \to \ccal$ is a bundle of
conic hypersurfaces and  $\rho(\wt \dcal) = \ccal$; i.e., every point
of moduli space is a degenerate critical point of some superpotential.
We further note that $\scal
\sm \dcal$ consists of a finite number of connected components,
and that
  $\pi:\ical\sm\wt\dcal\to \pi(\scal)\sm \dcal$ is
a finite covering over each connected component of $\pi(\scal) \sm
\dcal$.

\subsection{\label{SGCPD}Special geometry and critical point density}
 In obtaining reliable order of magnitude results on numbers of
vacua in a given string/M
 model, it is important   to estimate the  size of the leading coefficient
 $$ \int_{\ccal} \psi(Z)
\int_{\scal_{Z}} |\det H^c W(Z)|
  \chi_{Q_{Z}}(W) \,dW\,d\vol_{WP}(Z) $$
  and of the remainder. Since little is known about the volume of
  $\ccal$ at present (cf. \cite{LuS1}), we concentrate on estimating the integrand
\begin{equation} \label{LEADDEN} \kcal^\crit(Z): = \int_{\scal_{Z}}
|\det H^c W(Z)|
  \chi_{Q_{Z}} dW  \end{equation}
 in the  $b_3$ aspect. It is also important to study the behavior
 of the $\kcal^\crit(Z)$ as $Z$ tends to `infinity' in  $\ccal$,
 or to a singular point such as a conifold point (when one
 exists).

A key feature of   $\kcal^\crit(Z)$ is that it is  the integral of
a  homogeneous function of order $b_3$ over a space of dimension
$\dim_\R\scal_Z= b_3  = 2 (h^{2,1} + 1) $. Among the known
Calabi-Yau $3$-folds it is common to have $300 < b_3 < 1000$,
hence the integral is often over a space of large dimension.   The
 $b_3$-dependence is  sensitive since (e.g.) the ratio of the
 $L^{\infty}$ norm to the $L^2$ norm of a homogeneous function of
 degree $b_3$ in $b_3$ variables can be of order $b_3^{b_3}.$
 It is useful to have  alternative
 formulas for the leading coefficient, and we now present a few.
 We will use them to suggest conjectures on the order of magnitude
 of $\kcal^\crit(Z)$ in the $b_3$ aspect in \S \ref{FCFP}.

First, using the homogeneity of the integrand, we may rewrite the
integral  in terms of a Gaussian density
\begin{eqnarray}\label{Kcritgauss}
\kcal^\crit(Z)  &=&  \frac1{b_3!}
 \int_{\scal_Z} |\det H^c  W(Z)| e^{-  \langle Q_Z W, W \rangle}
 dW\,. \end{eqnarray}
 This formula shows that
$\kcal^\crit$ is formally analogous to
 density of critical points of random holomorphic sections relative to a Gaussian
 measure
 studied in  \cite{DSZ1}. For this reason, we call
(\ref{LEADDEN})  the {\it critical point density}. However, the
measure $e^{- Q[W]}\chi_{\{0<Q<1\}}(W)dW$
 is of infinite volume, so the analogy should not be taken too literally.  The density $\kcal^\crit(Z)$ is
 well-defined despite the infinite volume of the underlying
 measure on $\scal$ because the fibers $Q_Z$ of $\rho|_Q$ are of
 finite volume.  Indeed, the conditional measures of $e^{- Q[W]}
dW$ are standard
 (un-normalized) Gaussian measures $e^{- Q_Z(W)} dW$.

Next, we  rewrite the integrals by the methods
 in \cite{DSZ1, DSZ2}.
The first method is  to change variables to the Hessian $H^c
W(Z)$, i.e. to
 `push-forward' the
$\scal_{Z}$ integral under the Hessian map \begin{equation}
\label{HSUBZ} H_Z: \scal_{Z} \to \sym(m, \C) \oplus \C,\;\;\;
H_Z(W) = H^c W(Z), \end{equation}  where $m = \dim
\ccal=h^{2,1}+1$. In \cite{DSZ1, DSZ2}, we used this change of
variables to simplify the formulas for the density of critical
points. There, however, the spaces of holomorphic sections of the
line bundles $L \to M$ were so large that the image of the Hessian
map was the entire space $\sym(m, \C) \oplus \C$ of complex
Hessians of rank equal to the dimension $m = \dim M$.  In the case
of type IIb flux compactifications, the dimension of the
configuration space $\ccal$ is as large as the dimension of the
space $\scal$ of sections, and the Hessian map is by no means
surjective. Indeed, in Lemma \ref{RANGE}, we prove that  the
Hessian map is an isomorphism to a real $b_3$-dimensional space
$\hcal_Z\oplus\C$, where  $\hcal_Z$  is spanned (over $\R$) by the
$2h^{2,1}$ Hermitian matrices
\begin{equation}\label{HCALZ}\xi^j:=  \left(
 \begin{array}{cc}
0&   e_j    \\
     e_j^t  & \fcal^j(z)
 \end{array}
 \right), \qquad \xi^{h^{2,1}+j}:= \left(
 \begin{array}{cc}
0&  i e_j    \\
     ie_j^t  &-i \fcal^j(z)
 \end{array}
 \right),\qquad {j = 1, \dots, h^{2,1}}\ .
\end{equation}
Here, $e_j$ is the $j$-th standard basis element of $\C^{h^{2,1}}$
and
 $\fcal^j(z) \in \sym(h^{2,1}, \C)$ is the matrix $ \left( \fcal^{\bar j}_{i
k} (z) \right)$ whose entries define the  `Yukawa couplings' on
$\mcal$ (see (\ref{CANDY}),  \S\ref{SGMS} or \cite{St1, Can1})
with respect to normal coordinates at the point $z\in\mcal$.

Since $\hcal_Z$ is not a complex subspace of $\sym(m, \C)$, we
regard $\sym(m, \C)$ as a real vector space with inner product
\begin{equation}\label{real}(A,B)_\R=\Re\langle A, B \rangle_{HS}
=\Re(\mbox{Trace}\, A B^*)\;.\end{equation}

To state our next result, we let $ \Lambda_Z$ be the operator
given by the distortion under the Hessian map (see \S
\ref{DISTORTION}):
\begin{equation}\label{defC}\big((\La_Z\oplus I_\C)^{-1} H_Z W,\, H_Z W \big)_\R
=Q[W]\qquad (W\in\scal_Z),
\end{equation} where $Q[W]$ is given by \eqref{QW}.  In terms of the basis
$\{\xi^a\}_{1\le a\le 2h^{2,1}}$,
$$\La_Z\xi^a=\sum_{b = 1}^{2 h^{2,1}}\La_{ab}\xi^b\;,
\quad \La_{ab}= (\xi^a,\xi^b)_\R \;.$$ The $\Lambda$ matrix has
the block form
\begin{equation}\label{LAZY}  (\La_{ab})  = \begin{pmatrix} \La' & \La'' \\\La'' &
\La'\end{pmatrix}, \qquad \La'_{jk}= 2\de_{jk} + \Re \;
\mbox{Tr}\; \fcal^j \fcal^{k*},\ \  \La''_{jk}= \Im \; \mbox{Tr}\;
\fcal^j \fcal^{k*}\;.\end{equation}
In Proposition \ref{LAMBDARICCI}, we show that the $(1,1)$ form
\begin{equation} \label{LAMBDAFORM} \omega_\Lambda:=\frac i2 \sum (\Lambda'_{jk}
+i\Lambda''_{jk}) dz^j \wedge d\bar z^k=\frac i2 \sum
\left[2\de_{jk} + \mbox{Tr}\; \fcal^j(z_0) \fcal^{k*}(z_0)\right]
dz^j \wedge d\bar z^k  \end{equation}
is the so-called  Hodge metric $(m + 3) \omega_{WP} + Ric (\omega_{WP})$ of
 the Weil-Petersson metric  \cite{Lu, W2}.

By the injectivity of the Hessian map (stated in Lemma
\ref{RANGE}), we can make the change of variables $W{\buildrel
{H_Z}\over \mapsto}(H,x)$ in \eqref{LEADDEN}--\eqref{Kcritgauss}
to obtain the following alternate formulas for $\kcal^\crit (Z)$:
\begin{eqnarray}\kcal^\crit (Z)& = &\frac 1 {\sqrt{\det
\La_Z}} \int_{\hcal_Z \oplus \C} \left|\det H^*H - |x|^2 I\right|
\chi_{\La_Z} (H, x) dH dx,\nonumber  \\ & = &\frac 1
{b_3!\sqrt{\det \La_Z}} \int_{\hcal_Z \oplus \C} \left|\det H^*H -
|x|^2 I\right|\;\; e^{-(\La\inv_Z H, H)_\R - |x|^2}\,dH\,dx
\label{PF}
\end{eqnarray}
where $\chi_{\La_Z}$ is the characteristic function of the
ellipsoid $\{(\La_Z\inv H, H)_\R + |x|^2 \leq 1\}.$ These formulas
are analogous to  Theorem 1 and Corollary 2 of \cite{DSZ1}, the
key difference being that here we integrate over a moving subspace
$\hcal_Z$ of symmetric matrices.

We similarly have the following alternative formulations of
Proposition \ref{LNMO} and Theorem \ref{MAIN}:

\begin{cor}\label{MAIN2}
Let  $\psi = \chi_{K}$, where $K \subset \ical $ is as in Proposition \ref{LNMO},
and let $\tilde{\psi}(Z, H_Z W) = \psi(Z, W)$. Then,
{\small\begin{eqnarray*}\ncal_{\psi}(L) = \frac{L^{b_3}}{b_3!} \Big[\int_{\ccal}
\frac{1}{\sqrt{\det \La_Z}} \int_{\hcal_Z
\oplus \C} \tilde{\psi}(Z; H, x) \left|\det H^*H - |x|^2
I\right|\;
e^{-(\La\inv_Z H, H)_\R - |x|^2}\,dH\,dx \,d\vol_{WP}(Z) \\ +O(L^{-1/2})\Big].
\end{eqnarray*}}\end{cor}

\begin{cor}\label{MAIN3} Let $\psi\in\ccal^\infty(\ical)$ be homogeneous of order
$\al\ge 0$ and suppose that $\rho(\supp\psi)$ is a compact subset of
$\ccal$ and $\supp\psi
\cap\wt\dcal=\emptyset$.
Let $\tilde{\psi}(Z, H_Z W) = \psi(Z, W)$. Then,
\begin{eqnarray*}\ncal_{\psi}(L) &=& \frac{L^{b_3+\al/2}}{\Gamma(b_3+\al/2+1)}
 \left[\int_{\ccal}
\frac{1}{\sqrt{\det \La_Z}} \int_{\hcal_Z \oplus \C}
\tilde{\psi}(Z; H, x)\right.\\&&\quad\left. \times \left|\det H^*H
- |x|^2 I\right|\; e^{-(\La\inv_Z H, H)_\R - |x|^2}\,dH\,dx
\,d\vol_{WP}(Z)  + O\left(L^{-\frac{2b_3}{2b_3+1}}\right)\right].
\end{eqnarray*}\end{cor}

It is not obvious how to estimate the dependence of the integral
for $\kcal^\crit(Z)$ on the subspace $\hcal_Z$. There are two
natural ways to parameterize this space. One (which is used in
\cite{DD}) is to use as a basis of $\hcal_Z$ the  Hessians of a
$Q_Z$-orthonormal basis of $\scal_Z$. A second method is to use
the orthonormal basis of eigenmatrices $\{H_j\}$ of $\La_Z$ with
respect to the inner product \eqref{real}. We thus put $\La_Z
H_j(Z) = \mu_j(Z) H_j(Z)$, and $H(y, Z) = \sum_j y_j H_j(Z)$. We
also let $D(\mu)$ denote the diagonal matrix with entries $\mu_j$.
Changing variables to $\mu_j^{1/2} y$ cancels $\frac 1 {\sqrt{\det
\La_Z}}$ and we obtain:

\begin{cor} \label{CORLEAD1} We have:
$$\kcal^\crit (Z) = \int_{|y|^2 + |x|^2 \leq 1} \left|\det H(D(\mu)y, Z) ^*H(D(\mu) y, Z) - |x|^2
I\right|dy dx. $$
\end{cor}

In \S \ref{FCFP} we will discuss some conjectural bounds on the
density of critical point based on the assumption  that the
subspaces $\hcal_Z$ are sufficiently random subspaces of
$\sym(h^{2,1}, \C)$.

\subsection{\label{INDEXDENSITY}Index density}

The absolute value  in the expressions for the distribution of
 critical points $C_W$ of a single section  (\ref{CWintro}) and the
 expected distribution of critical points of a random section
 (e.g., \eqref{PF}) make it very difficult to
estimate  the order of magnitude of the density of critical
points. A simplifying `approximation'
 is to drop the absolute value around the determinant. The
 resulting density is
  {\it index density} for critical points. It  was used
in \cite{AD} and    \cite{DD} to give  a lower bound for the
critical point density.

To be precise, we modify (\ref{CWintro}) by defining the signed
distribution of critical points of $W$ as the measure $C_W$ on
$\ical\sm\wt\dcal$ given  by
\begin{equation} \langle Ind_W, \psi \rangle = \sum_{Z \in Crit(W)} \left(\mbox{sign} \det D^2 W(Z)\right)  \psi(Z,
W),\end{equation} where sign$\,a=1,0,-1$ if $a$ is positive, 0, or
negative, respectively. We then study the sums
\begin{eqnarray}
\ical nd_{\psi}(L) & = & \sum \big\{\langle Ind_N, \psi \rangle :N
\in \scal^\Z ,\ Q[N] \leq L\}.\label{Isum}
\end{eqnarray} For instance, if $\psi(Z,W)= \chi_{K}(Z)$ is the characteristic
function of a compact set $K \subset \ccal$, then $\ical
nd_{\psi}(L)$ is the sum  $ \sum_{Z \in Crit(W) \cap K}
\left(\mbox{sign} \det D^2 W(Z)\right)$ over all non-degenerate
critical points lying over $K$  of  all integral flux
superpotentials with $Q[W] \leq L$.

Simultaneously with Proposition \ref{LNMO}, we
obtain formula (1.5) of  Ashok-Douglas \cite{AD} with an estimate for the error
produced by passing from the sum to the integral (cf.\ \S \ref{COUNTING}):

\begin{theo}\label{MAININD} Let  $K$ be a compact subset of $\ccal$ with piecewise
smooth boundary.    Then
$$\ical nd_{\chi_K}(L) = \frac{(\pi L)^{b_3}}{b_3!\,  2^{b_3/2}} \left[\int_K
c_m(T^{* (1,0)}(\ccal)\otimes \lcal,\om_{WP}\otimes h^*_{WP})  +
O\left(L^{-1/2}\right)\right],
$$
where $m=\dim\ccal=b_3/2$ and $c_m(T^{* (1,0)}(\ccal)\otimes
\lcal,\,\om_{WP}\otimes h^*_{WP}) =\frac 1{\pi^m}\det\left(-R-
\omega_{WP} \otimes I  \right)$ is the $m$-th Chern form  of $T^{*
(1,0)}(\ccal)\otimes \lcal$ with respect to the Weil-Petersson
metric $\om_{WP}\otimes h^*_{WP}$.
\end{theo}
Here, $R = \sum_{i j} R^{k}_{\ell i \bar{j} } dz^i \wedge
d\bar{z}^{\bar{j}}$ is the curvature $(1,1)$ form of $T^{*
(1,0)}(\ccal)$ regarded as an $m \times m$ Hermitian-matrix-valued
$2$-form (with $m = \dim \ccal$= $b_3/2$) and $\omega_{WP} \otimes
I$ is a scalar 2-form times the $m \times m$ identity matrix. The
determinant is defined as in  Chern-Weil theory.  The only
additional step in the proof is the evaluation (given in Lemma
\ref{INDCURV}) of the analogue of (\ref{Kcritgauss}) in terms of
the curvature form:
\begin{equation}\label{Indcritgauss}
\int_{\scal_Z} \det H^c  W(Z) e^{-  \langle Q_Z W, W \rangle} dW
=\left(\frac \pi 2\right)^m\;\frac{  \det\left(-R-  \omega_{WP}
\otimes I \right)}{d\vol_{WP}}\,.\end{equation}

Recall that  the Chern-Gauss-Bonnet theorem tells us
that if $W$ is a holomorphic  section of a complex line
bundle $L\to M_m$ over a compact complex manifold such that
$\nabla W$ has   only non-degenerate zeros, then
$$c_m( T^{* (1,0)} M \otimes L) = Ind\, \nabla W:= \sum_{p: \nabla W (p) = 0}
\mbox{sign}\; \det\;  H^c W(p). $$ However, the Chern-Gauss-Bonnet theorem does
not apply in our setting,  and indeed  $Ind\, \nabla W$ is not constant in $W$,
since
 $\ccal$ is an incomplete \kahler manifold and critical points can occur on the boundary
 or  disappear.  There exists a
 Chern-Gauss-Bonnet theorem for manifolds with boundary which
 expresses $Ind \nabla W$ as $c_n(E)$ plus a boundary
 correction depending on $W$, but the correction term involves
 integrating a differential form over the boundary and that
 becomes problematic when the boundary is highly irregular as in the case of $\ccal$.
Nevertheless, the theorem shows that asymptotically the average index
density equals the Chern-Gauss-Bonnet form.

\subsection{\label{RELATIONS}Relations to prior results in the physics and mathematics literature}

We now relate   our results to the  physics literature on the
number of vacua and the complexity of the string theory landscape
as well as to the mathematical literature on lattice points. A
more detailed discussion of the landscape aspects  is given in \S
\ref{FCFP}.

First, the string/M aspects.  Over the last five years or so, many
 physics articles have been devoted to estimating the number of
 candidate vacua $N_{vac}$ of string/M theory, in particular those
 which are consistent with the standard model.  The candidate vacua
 are often pictured as valleys in a `string theory landscape', which
 is the graph of the effective potential.  The number of vacua is
 often stated as being around $10^{500}$. In \cite{BP}
 Bousso-Polchinski related the number of vacua to the number of
 quantized fluxes $N$ satisfying a constraint $|N| \leq L$, which
 implies $N_{vac}(L) \sim \frac{L^{b_3}}{b_3!}$ (see also \cite{AD,
 Sil}). In the specific type IIb flux compactifications studied in
 this paper, the constraint is hyperbolic rather than elliptic (as
 imagined in \cite{BP}), and the more precise estimate $N_{vac}(L)
 \sim \frac{L^{b_3}}{b_3!} f(b_3)$ was given in \cite{AD, DD} where
 $f(b_3)$ is the moduli space integral of the Gaussian integral in
 \eqref{PF}; it will be discussed further in \S
 \ref{FCFP}. There we will also review the heuristics and the
 mathematics of the landscape in more detail.

What do our results imply about the number of vacua? Since
Proposition \ref{LNMO} and Theorem \ref{MAIN} are asymptotic
results as $L \to \infty$, they   are most useful when $L^{b_3}$
is very large. But it is difficult to quantify `very large' due to
the complexity of the leading coefficient (\ref{LEADDEN}),  of the
remainder and of the volume of $\ccal$. Hence,  we cannot make
precise estimates on the number of vacua at this time.

However, to bridge our  results with estimates in string theory,
we make a speculative attempt in \S \ref{HEURISTICS}  to draw
order of magnitude conclusions from Theorem \ref{MAIN}. We will
use the symbol $\simeq$ in an informal sense of `same order of
magnitude' (factorial, exponential and so on).  There we give a
heuristic estimate of $\kcal^\crit(Z) \simeq \frac{1}{b_3!}
(b_3/2)! \mu^{b_3}$ for certain $\mu > 0$. More precisely, we give
heuristic upper and lower bounds with different $\mu$ which are
irrelevant when comparing factorials. To obtain an order of
magnitude for $\frac{f(b_3)}{b_3!}$ one would need to integrate
$\kcal^\crit$ over $\ccal$. At this time, the order of magnitude
of the Weil-Petersson volume $Vol_{WP}(\ccal) $ of $\ccal$ is not
known, even approximately (Z. Lu). We can however make a plausible
estimate for the integral of $\kcal^\crit$ over the region where
the norm of $\Lambda_Z$ is bounded by a uniform constant
(independent of $b_3$).  Since $\Lambda_Z$ is essentially the
Hodge metric, regions where $||\Lambda_Z|| \leq \mu$
 are regions $K_{\mu}$ where the norm of the Ricci curvature of
$\omega_{WP}$)  is bounded above by a uniform constant. It appears
likely that   the volume of such regions is bounded above by the
volume of balls in $\C^{b_3/2}$ of fixed radius (Z. Lu). Since the
volume of balls in $\C^{b_3/2}$  decays like $\frac{1}{(b_3/2)}!$,
we would find that the number of vacua in $K_{\mu}$ would be
approximately $\frac{L^{b_3}}{b_3!} \mu^{b_3}$.

 Now, in
the physical models, $L$ is not a free parameter but is determined
by $X$.  In the case when there exists an involution $g$ of $X$
(an `orientifolding') and  a Calabi-Yau $4$-fold $Z$ which is an
elliptic fibration over $X/g,$ the `tadpole' number is then given
by:
\begin{equation} \label{TADPOLE} \mbox{ tadpole number}: \;\;
 L = \chi(Z)/24. \end{equation}
In  many known examples \cite{KLRY}, one has  $300 < b_3 < 1,000$
and    $L \simeq C b_3$ where $1/3 \leq C \leq 3$. Hence the
number of vacua in $K_{\mu}$ (and possibly in all of $\ccal$) with
the tadpole constraint $L \sim C b_3$ would have exponential
growth $\frac{(C b_3)^{b_3}}{b_3!} \mu^{b_3}$.

Next we turn to the purely  lattice point aspects of the problem.
From a mathematical point of view, this article combines
statistical algebraic geometry in the sense of \cite{BSZ1, DSZ1,
DSZ2} with the study of radial projections of lattice points.  As
far as we know, the radial projection of lattice points  problem
has not been studied systematically before in mathematics (we
thank B. Randol for helping to sort out the historical background
on this problem). The much harder problem of equidistribution of
lattice points of fixed height $R$, i.e. lying on a sphere or
hyperboloid of fixed radius $R$, has been studied by Yu.\ Linnik,
C. Pommerenke \cite{Pom}, W. Duke and others. But the remainders
obtained in this more delicate problem are not as accurate as ours
are for the bulk problem of projecting all lattice points of
height $< R$. Counting projections of lattice points in domains of
a hypersurface is equivalent to counting lattice points in certain
cones, and there are some additional studies of this by methods of
automorphic forms.  In certain right circular cones with a flat
top, Duke and Imamoglu \cite{DO} use Dirichlet series and Shimura
lifts to obtain the leading order asymptotics. Radial projections
of lattice points additionally bear some resemblance to rational
points. Some results and references for this problem are contained
in \cite{DO}. In \cite{Z2}, the general problem of counting radial
projections of lattice points in smooth domains of non-degenerate
hypersurfaces is studied. In \cite{NR}, some further results are
given on radial projections of lattice points, in particular in
the case of hypersurfaces with flat spots or in the case of
polyhedra.

\bigskip\noindent{\it Acknowledgement:\/}
We would like to thank Zhiqin Lu for many helpful comments
regarding the Weil-Petersson and Hodge metrics on the moduli space
of a Calabi-Yau 3-fold. In particular, our discussion of the
Weil-Petersson volume $V_{WP}(\ccal)$ and estimates of the
eigenvalues of $\Lambda_Z$ are based on his remarks.

\section{\label{CY}Background on Calabi-Yau manifolds and complex
geometry}

As mentioned in the introduction, the supersymmetric vacua of type
IIb flux compactifications on a $CY_3$ are critical points of
holomorphic sections of the holomorphic line bundle $\lcal \to
\ccal$ dual to the Hodge bundle $H^{4,0}(X \times T^2)$, where the configuration
space $\ccal$ is the moduli space $\mcal \times \ecal$ of product
complex structures on $X \times T^2$.  In this section, we give the
geometric background necessary for the analysis of critical points
and Hessians of the holomorphic sections $W_G$ of (\ref{WSUBG}).

The most significant aspects of Calabi-Yau geometry in the study
of critical points of flux superpotentials are the following:
\begin{itemize}

\item The  space $\scal_Z$ of flux superpotentials with $\nabla
W_G(Z) = 0$ may be identified with the space $H^3_{Z}(X)$ of
fluxes $G = F + i H$ with the special Hodge decomposition  $F +
\tau H \in H^{2,1}_z (X) \oplus H^{0, 3}_z(X). $ See Proposition
\ref{POS}.

\item The space $H^{2,1}_z (X) \oplus H^{0, 3}_z(X) \subset H^3(X,
\C)$ is a positive complex Lagrangian subspace. See Proposition
\ref{HRPLUS}. Hence, $\scal_Z$ is endowed with an inner product.

\end{itemize}

In addition, we review the relation between holomorphic
derivatives, covariant derivatives and second fundamental forms
for holomorphic frames $\Omega_z$ of the Hodge bundle, and recall
the definition of the prepotential. These results are needed for
the calculations in Lemmas \ref{Lambda} and  \ref{RANGE}. Much of
this material is essentially standard \cite{Can1, St1, DD}, but it
is not always stated precisely in the physics sources. We
therefore present it here for the sake of clarity and
completeness.

\subsection{Geometry of Calabi-Yau manifolds}\label{GCY}

 We recall that a Calabi-Yau
$d$-fold $M$ is a complex $d$-dimensional  manifold with trivial
canonical bundle $K_M$, i.e. $c_1(M) = 0$. By the well-known
theorem of Yau, there exists a unique Ricci flat \kahler metric in
each \kahler class on $M$. In this article, we fix  the \kahler
class, and then the Calabi-Yau metrics correspond to the complex
structures on $M$ modulo diffeomorphisms. We denote the moduli
space of complex structures on $M$ by $\mcal $.

As mentioned in the introduction, the Calabi-Yau manifolds of
concern in this article are the $4$-folds $M = X \times T^2,$
where $T^2 = \R^2/\Z^2$. The $T^2$ factor plays a special role,
and the geometric aspects mainly concern $X$.
 We only consider product Calabi-Yau metrics and complex structures  on $M$.
 Thus, the
configuration space $\ccal = \mcal  \times \ecal$ where $\mcal $
is the  moduli space of complex structures on $X$ and where
$\ecal$ is the moduli space of elliptic curves. We denote a point
of $\ccal $ by $Z = (z, \tau)$ where $z$ denotes a complex
structure on $X$ and where $\tau$ denotes the complex structure on
$T^2$ corresponding to the elliptic curve $ \C/\Z \oplus \Z \tau.$

It is often simplest to view the moduli space of complex
structures on $X$ as the quotient by the mapping class group
$\Gamma$ of the Teichm\"uller space $\tcal \!eich(X)$, where
$$\tcal \!eich(X) =\{\mbox{complex structures on }\; X\}/\Diff_0 $$
where $J \sim J'$ if there exists a diffeomorphism $\phi \in
\Diff_0$ isotopic to the identity satisfying $\phi^*J' = J.$ The
mapping class group is the group of connected components of the
diffeomorphism group,
$$\Gamma_X := \Diff(X)/\Diff_0(X). $$
We shall identify $\mcal $ with a
fundamental domain for $\Gamma_X$ in $\tcal \!eich(X)$, and $\ecal$
with the usual modular domain in $\hcal$.

The mapping class group for a Calabi-Yau
$d$-fold has a representation on
$H^d(M, \R)$ which preserves the intersection form $Q$, which is symplectic in odd dimensions,
and indefinite symmetric in even dimensions. In odd dimensions, this representation gives a
homomorphism $\phi: \Gamma_M \to \mbox{Sp}(b_d(M), \Z)$,
while in even dimensions it gives a homomorphism
to the corresponding orthogonal group.
It was proved by  D. Sullivan \cite{Sul} that if $d
\geq 3$, then $\phi(\Gamma_M)$ is an (arithmetic) subgroup of
finite index (in $\mbox{Sp}(b_d(M), \Z)$ if $d$ is odd), and the kernel of
$\phi$ is a finite subgroup.

On any CY manifold $M$ of dimension $d$, the space $H^{d,0}_z(M)$
of holomorphic  $(d, 0)$ forms for a complex structure $Z$ is
one-dimensional. It depends holomorphically on $Z$ and hence
defines  a complex holomorphic line bundle
$\lcal_\mcal^*=H^{d,0}\to {\mathcal M}$, which we refer to as the
`Hodge bundle.' The Hodge bundle is equipped with the
Weil-Petersson (WP) Hermitian metric of (\ref{HWP}), which we
repeat here:
\begin{equation}\label{hWP}
h_{WP}(\Omega_z, \Omega_z) = i^{d^2}\int_X \Omega \wedge \overline \Omega .
\end{equation}

For a  holomorphic Hermitian line bundle
$(L, h) \to M$ and local holomorphic frame $e_L$ over an open set
$U \subset M$, we write
\begin{equation}\label{Kpot1}|e_L(z)|_h^2 =
e^{-K(z)}. \end{equation} The connection $1$-form in this frame is
the $(1,0)$ form $- \partial K(z)$, and the curvature $(1,1)$
-form is given by
$$\om=\frac{i}{2}\Theta_h= \frac i2 \ddbar K, \qquad K =-\log|e_L|^2_h. $$
The Hermitian line bundle is said to be positive if $\omega$ is a positive
$(1,1)$ form, in which case $K$ is called  the  \kahler potential. The Hermitian
line bundle
$(L,h)$ is negative if
$\omega$ is a negative
$(1,1)$ form.

In particular, the curvature of the  Weil-Petersson metric on $H^{d,0}\to {\mathcal
M}$ is a positive $(1,1)$ form on $\mcal $, and hence it
defines a \kahler form with potential (with respect to the frame $\Om_z$)
\begin{equation} \label{KP} K_{WP} = - \log h_{WP}(\Omega_z, \Omega_z) =-
\log\; i\int_X \Omega \wedge \overline{\Omega}. \end{equation} For
instance, consider the  Hodge bundle $H^{1,0}_{\tau} \to \ecal$.
It has a standard frame $dx + \tau dy$ for which $K = -\log \Im
\tau.$ Here, $\tau$ is the standard coordinate on the upper half
plane. Then $\partial K = - \frac{1}{\tau - \bar{\tau}} d \tau$
and the \kahler form is $- \frac{i}{2 (\tau - \bar{\tau})^2} d
\tau \wedge d \bar{\tau} >> 0.
$

In the product situation of $M = X \times T^2$, $H^{4, 0}_{z,
\tau} (X \times T^2) = H^{3,0}_z(X) \otimes H^{1,0}_{\tau}(T^2)$.
Thus, the line bundle  $H^{4,0}(X \times T^2) \simeq H^{3,0}(X)
\otimes H^{1,0}(T^2) \to \ccal$ is an exterior tensor product  and
the WP metric is a direct product. We  denote an element of
$H^{3,0}_z(X)$ by $\Omega_z$, and an element of
$H^{1,0}_{\tau}(T^2)$ by $\omega_{\tau}$. We often assume that
$\omega_{\tau} = dx + \tau dy$.

\subsection{Variational derivatives and covariant derivatives}\label{derivatives}

The bundle $H^{3, 0}_z(X) \to \mcal $ is a holomorphic line
bundle. Since $H^{3, 0}_z(X) \subset H^3(X, \C)$, one can view a
holomorphically varying family $\Omega_z \in H^{3, 0}_z(X)$ as a
holomorphic map $\mcal \to H^3(X, \C)$ or as a holomorphic section
of $H^{3, 0}_z(X)$. As a holomorphic vector valued function,
$\Omega_z$ may be differentiated in $z$. If $z_1, \dots,
z_{h^{2,1}}$ are local holomorphic coordinates, and if
$\{\frac{\partial}{\partial z_j}\}$ are the coordinate vector
fields, then $\frac{\partial \Omega}{\partial z_j}$ is a
well-defined element of $H^3(X, \C)$.

By the  Griffiths transversality theorem,  see \cite{GHJ},
\cite{Can1}, (5.4)) or \cite{W1, W2},
\begin{equation} \label{FV}
\frac{\partial \Omega_z }{\partial z^{j}} = k_{j}(z) \Omega_z+ \chi_{j}
,\end{equation} where $\chi_{j} \in H^{2,1}_{z}(X)$ and where $k \in
C^{\infty}(\mcal )$. Note that although $\frac{\partial \Omega_z }{\partial
z^{j}}$ is holomorphic, neither term on the right hand side is separately
holomorphic.

We define a Levi-Civita connection on the bundle $H^{3,0} \to
\mcal$ by orthogonally projecting the derivatives $\frac{\partial
\Omega_z }{\partial z^{j}}$ onto $H^{3,0}$. This defines the
Weil-Petersson connection $\nabla_{WP}$ on $H^{3,0} \to \mcal$,
 $$\nabla_{WP} :
C^{\infty}(\mcal, \lcal) \to C^{\infty}(\mcal, \lcal \otimes
T^*).$$  It follows from \eqref{FV} that
\begin{equation} \frac{\partial}{\partial z^{j}} \int_X \Omega_z
\wedge \overline{\Omega}_z  = k_{j} \int_X \Omega_z \wedge
\overline{\Omega}_z, \end{equation} which by (\ref{KP}) implies
\begin{equation} k_{j} = - \frac{\partial K}{\partial
z^{j}}.
\end{equation} Hence,
$$\nabla_{WP} \Omega_z = -\d K \otimes \Omega_z=
\sum k_j dz_j\otimes \Omega_z $$
is the Chern connection of the Weil-Petersson Hermitian metric.

We also define the forms
\begin{equation} \label{DALPHA} \left\{ \begin{array}{l} \dcal_j \Omega_z = \frac{\partial }{\partial
z^{j}}\Omega + \frac{\partial K}{\partial z^{j}}
\Omega\\ \\
\dcal_j \dcal_k  \Omega_z =  (\frac{\partial }{\partial z^{j}} +
\frac{\partial K}{\partial z^{j}})( \frac{\partial }{\partial
z^{k}} + \frac{\partial K}{\partial z^{k}} ) \Omega_z. \end{array}
\right.
\end{equation}
We then have
\begin{equation}\label{Dj} \dcal_j \Omega_{z}=
\frac{\partial \Omega_z }{\partial z^{j}}-k_j\Om_z =  \chi_j \in
H^{2,1}(X_z).
\end{equation}
The operator $\dcal_j \Omega_z$ is analogous to the second
fundamental form $II(X, Y) = (\tilde{\nabla}_X Y)^{\perp}$  of an
embedding, i.e. it is the `normal' component of the ambient
derivative. It is known that the first variational derivatives
span $H^{2,1}$ (see e.g. \cite{W1, W2}. (In the physics
literature, $D_{\alpha}$ is often described as a connection, and
is often identified with  $\nabla_{WP}$, but this is not quite
correct as it is applied to $\Omega_z$).

The Weil-Petersson Hermitian metric $\sum G_{i \bar{j}}dz_i\,d\bar z_{\bar j}$ on
$\mcal$ is the curvature
$(1,1)$-form of the Hodge bundle. From \eqref{KP} and \eqref{Dj}, we have:
\begin{equation}
\label{GIJ} G_{j \bar{k}} = \frac{\partial^2}{\partial z^j\partial \bar z^k}
K(z,
\bar{z}) = -
\frac{\int_{\mcal} \dcal_j  \Omega_z \wedge \overline{\dcal_k \Omega_z}
}{\int_{\mcal} \Omega_z \wedge \overline{\Omega}_z}. \end{equation}

\subsection{Yukawa couplings and special geometry of the moduli space}\label{SGMS}

In formula \eqref{PF},  the density of critical points is
expressed as an integral over a space $\hcal_Z \oplus \C$, where
$\hcal_Z$
 is a subspace  of the complex symmetric matrices
$\sym(h^{2,1} + 1, \C)$ spanned by the special matrices $\xi^j$
given in \eqref{HCALZ}. Their components $\fcal^{\bar j}_{ik}(z)$
are known as  Yukawa couplings and defined as follows: A priori,
$\dcal_k \dcal_j \Omega_z \in H^{2,1} \oplus H^{1,2}$, and
moreover  its  $H^{2,1}$ component vanishes (see e.g.\
\cite[(5.5)]{Can1}). Hence we may define $\fcal_{k j}^{\bar l}$ by
\begin{equation} \label{CANDY} \dcal_k \dcal_j \Omega_z = -\sqrt{-1}\, e^K
\fcal_{k j}^{\bar l} \overline{\dcal_l \Omega}\,\qquad (1\le j,k,l\le h^{2,1}).
\end{equation} See also \cite[(28)]{St1}. It is further shown in
\cite[(37)]{St1}  (see also \cite[(4.8)]{AD},
\cite[Theorem~3.1]{LuS2}) that the Riemann tensor of the
Weil-Petersson metric on the moduli space $\mcal$ of Calabi-Yau
three-folds is related to the Yukawa couplings by
\begin{equation} \label{RIEMANN} R_{i \bar{j} k \bar{\ell}} = G_{i
\bar{j}} G_{k \bar{\ell}} + G_{i \bar{\ell}} G_{k \bar{j}} -
e^{2K} \sum_{p,q} G^{p \bar{q}} \fcal_{i k p} \overline{{\fcal}_{j
\ell q}}\ .
\end{equation}

The Yukawa couplings are related to the  periods of $\Omega_z$ and
to the so-called prepotential of $\mcal$.  We  pause to recall the
basic relations and to direct the reader to the relevant
references.

First, we consider periods.  As a basis of $H_3(X, \R)$ we choose
the symplectic basis consisting of dually paired Lagrangian
subspaces of $A$-cycles $A_a$ and $B$-cycles $B_a$. The
periods of $\Omega_z \in H^{3,0}_z(X)$ over the $A$-cycles
$$\zeta^a= \int_{A_a} \Omega_z\qquad  (1\le a\le h^{2,1}+1=b_3/2)$$
define holomorphic coordinates on $\lcal^*_\mcal=H^{3,0}\to
\mcal$. Alternately, we can view the $\zeta^a$ as `special'
projective coordinates on $\mcal$. The periods of $\Omega_z$  over
the $B$-cycles are then holomorphic functions of the $\zeta^a$.
The principal fact is that the image of $\lcal^*_\mcal$ under the
period map is a complex Lagrangian submanifold of $H^3(M,\C)$, and
thus is determined by a single holomorphic function, the
``prepotential''
 $\fcal=\fcal(\zeta^1,
\dots, \zeta^{b_3/2}):\lcal^*_\mcal\to\C$ such that
\begin{equation} \int_{B_a} \Omega_z = \frac{\partial
\fcal}{\partial \zeta^a}\; .\end{equation} Furthermore, $\fcal$ is
homogeneous of degree $2$ in the periods $\zeta^a$,
$$ \sum_{j = 1}^{b_3/2} \zeta^a\frac{\partial
\fcal}{\partial \zeta^a} = 2 \fcal(z), $$ and hence may be viewed
as a holomorphic section of $\lcal^{\otimes 2}_\mcal$.

The local holomorphic $3$-form $\Omega_z$ may be expressed  in
terms of the Poincar\'e duals of the symplectic basis by:
\begin{equation} \Omega_z = \sum_{a = 1}^{b_3/2} \left(\zeta^{a}
\wh A_a - \frac{\partial
\fcal}{\partial \zeta^a}\, \wh B_a\right)\;. \end{equation}
(See \cite{Can1}, (3.8)). Further, in  these
coordinates, the \kahler potential (\ref{KP}) of the
Weil-Petersson metric may be written as
$$K(z, \bar{z}) = - \log i \left( \sum_{a = 1}^{b_3/2} \zeta^a\overline
{\frac{\partial \fcal}{\partial \zeta^a}} - \bar{\zeta^a} \frac{\partial
\fcal}{\partial \zeta^a} \right). $$

We also have: \begin{equation}\label{fcal}\fcal_{k j}^{\bar l} =
\sum_{r=1}^{h^{2,1}} G^{r\bar l}\frac{\partial^3 \fcal}{\partial
z^r \partial z^j
\partial z^k}\;.\end{equation}
See \cite[(4.5)]{Can1}  and \cite[(64)]{St1}.

\bigskip

In summary, we  reproduce the table from \cite{Can1}:

\begin{equation} \label{CANTABLE} \begin{array}{ll} \mbox{Derivatives of the Basis}
\quad & \mbox{spans}
\\ & \\
\Omega & H^{3,0} \\ & \\
\dcal_j \Omega & H^{2,1} \\ & \\
\dcal_k \dcal_j \Omega = - i e^K \fcal_{k j}^{\bar{\gamma}}
\overline{\dcal_{\gamma} \Omega} & H^{12}  \\ & \\
\dcal_k \dcal_{\bar{j}} \Omega = G_{k \bar{j}} \overline{\Omega} &
H^{03}
\end{array} \end{equation}

\subsubsection{\label{CCALSTUFF}$\ccal$ as the moduli
space of complex structures on $X \times T^2$}

Above, we have reviewed the geometry of the moduli space of
complex structures on the Calabi-Yau three-fold. Our configuration
space $\ccal = \mcal \times \hcal$ may be viewed as (a component
of) the moduli space of complex structures on $X \times T^2$. This
point of view is used in \cite{DD}, but because the $T^2$ factor
plays a distinguished role we do not emphasize this identification
here. Further,  formula (\ref{RIEMANN})
needs to be modified for the moduli space of complex structures on
a Calabi-Yau four-fold. In \cite[Theorem~3.1]{LuS2}), the Riemann
tensor of the Weil-Petersson metric on the moduli space of a
Calabi-Yau manifold of arbitrary dimension is shown to be
\begin{equation} \label{RIEMANN4} R_{i \bar{j} k \bar{\ell}} = G_{i
\bar{j}} G_{k \bar{\ell}} + G_{i \bar{\ell}} G_{k \bar{j}} -
\frac{\langle \dcal_k \dcal_i \Omega, \overline{\dcal_{\ell} \dcal_j \Omega
\rangle }}{\int_\mcal \Omega \wedge \overline{\Omega}} .
\end{equation}
In the case of three-folds, the vectors $\dcal_j \Omega$ form an
orthonormal basis for $H^{2,1}$ and one can write the inner
product in the form (\ref{RIEMANN}).

\subsection{Hodge-Riemann form and inner products}\label{IP}

The Hodge-Riemann bilinear form on $H^3(X, \R)$ is the
intersection form $(\al ,\be)\mapsto \int_X \al\wedge\be$.
We consider the sesquilinear pairing:
\begin{equation} \label{ourQ}(\al,\be)\mapsto Q(\al,\bar\be) = -\sqrt{-1}\,   \int_X
\al\wedge\bar\be\;,\quad \al,\be\in H^3(X,\C)\;.\end{equation}
An important fact is that under the Hodge
decomposition (\ref{HD}) for a given complex structure, the
Hodge-Riemann form is definite in each summand:
\begin{equation} \label{INT} (-1)^p Q(\al,\bar\al)> 0,\quad \al\in H^{p,
3-p}(X,\C),
\end{equation} whose sign depends only on the parity of $p$. (See
\cite[\S7]{GH}.  Note that our definition of $Q$ has the extra sign
$-\sqrt{-1}$.  The inequality \eqref{INT} holds only for
primitive forms, but in our case all harmonic 3-forms are primitive,
since we are assuming that $H^1(M,\C)=0$.)
To restate \eqref{INT}:

\begin{prop}\label{HRPLUS}  Let $\dim X = 3$, and let $b_1(X)=0$. Then for each
$z
\in
\mcal
$, the   Hodge-Riemann form is positive definite on
$H^{2,1}_z \oplus H^{0, 3}_z$ and negative definite on $H^{3,0}_z
\oplus H^{1,2}_z. $ \end{prop}

By Griffiths transversality (see \eqref{FV}),  for any local
holomorphic frame
$\Omega_z$, $\dcal_j \Omega_z \in H^{2,1}_z$ and these elements span
$H^{2,1}_z$. Also, $\overline{\Omega}_z$ spans $H^{0, 3}$.  These forms
provide us with an orthonormal basis for $ H^{2,1}_{z} \oplus H^{0, 3}_{z}$:

\begin{prop} \label{ONB} If $\{z_j\}$ are  coordinates at
$z_0$ such that $\left\{\d/{\d z_j}|_{z_0}\right\}$ are orthonormal, and if
$h_{WP}(\Omega_{z_0},
\Omega_{z_0})= 1$,  then the basis $\{\dcal_j \Omega_{z_0}, \overline{\Omega}(z_0)\}$ is
a complex orthonormal basis of $ H^{2,1}_{z_0} \oplus H^{0, 3}_{z_0}$ with
respect to the Hodge Riemann form $Q.$
\end{prop}

\begin{rem} Here and below, when we say that a basis of a complex vector space  is complex
orthonormal we mean that it is a complex basis and is orthonormal
for the given inner product. By a real orthonormal basis of the
same vector space we mean an orthonormal basis of the underlying
real vector space.

\end{rem}

\begin{proof} It suffices to show that:

$$\begin{array}{rl}
\rm (i) &  Q( \dcal_j \Omega_z, \overline{ \dcal_k  \Omega_z})  =  -i \int_X
\dcal_j \Omega_z
\wedge
\overline{\dcal_k  \Omega_z } = G_{j \bar{k}} e^{- K} \\ & \\
\rm (ii) & Q( \dcal_j  \Omega_z, {\Omega_z})    = -i\int_X
\dcal_{\bar{j}}  \Omega_z
\wedge {\Omega_z} = 0 \\ & \\
\rm  (iii) &  Q(\overline\Omega_z, \Omega_z) = -i\int_X \overline\Omega_z
 \wedge {\Omega_z} = h_{WP}(\Omega_z, \Omega_z)
\end{array}
$$

\smallskip
Equation (i) follows from (\ref{GIJ}), (ii) is by type considerations, and
(iii) follows  from (\ref{HWP}).
\end{proof}

\begin{rem}
In the language of complex symplectic geometry, Proposition \ref{HRPLUS} says
that
$H^{2,1}_z \oplus H^{0, 3}_z$ is a positive complex polarization
of $H^3(X, \C)$.  Let us recall the definitions. The space
$(H^3(X, \R), Q)$ of real $3$-cycles with its intersection form
$Q(\al,\be) = -i\int_M\al\wedge\be$ is a real symplectic
vector space. After complexifying, we obtain the complex
symplectic vector space $(H^3(X, \C), Q). $ In general, if $(V,
\omega)$ is a real symplectic vector space and if
$(V_{\C}, \omega_{\C})$ is its complexification, a complex
Lagrangian subspace $F \subset V_{\C}$ is called a polarization.
The polarization is called real if $F = \overline{F}$ and complex
if $F \cap \overline{F} = \{0\}$. The polarization $F$ is called
{\it positive} if $i \omega(v, \bar{w})$ is positive definite on
$F$.

In our setting,  $(V, \omega)  = (H^3(X, \R), Q) $.  We observe
that for any complex structure $z$ on $X$ (as a complex manifold),
the Hodge decomposition may be written in the form
$$H^3(X, \C) = F \oplus \overline{F}, \;\; F = H^{2, 1} \oplus H^{0, 3}\;\;
\overline F = H^{3,0} \oplus H^{1, 2}, \;\; , $$ where
$F$ is complex Lagrangian. By Proposition \ref{HRPLUS}, this
polarization is positive, i.e.
$$Q(v, \bar v) >0\;, \;\; v \in F\sm\{0\}\;. $$
\end{rem}

\section{\label{BG}Critical points of superpotentials}

In this section, we assemble some basic facts about critical
points and Hessians of flux superpotentials.

\subsection{Flux superpotentials as holomorphic sections}

As discussed in the previous section, $\lcal \to \ccal$ is a
negative line bundle. On a compact complex manifold, a negative
line bundle has no holomorphic sections. However, $(\ccal,
\omega_{WP})$ is a non-compact, incomplete \kahler manifold of
finite Weil-Petersson volume (see \cite{LuS1} for the latter
statement), and the line bundle $\lcal \to \ccal$
 has many holomorphic
sections related to the periods of $X \times T^2$.

The sections relevant to this article are the flux superpotentials
$W_G$ of (\ref{WSUBG})-(\ref{WGG}).  $W_G$  depends on  two real
fluxes $F, H \in H^3(X, \Z)$, which we combine into a complex
integral flux
$$G =  F + i  H \in H^3(X, \Z \oplus \sqrt{-1} \Z).$$
The main reason to form this complex combination is that it
relates the tadpole constraint (\ref{TC}) on the pair $(F, H)$ to
the Hodge-Riemann form (\ref{HRFORM}). However, none of subsequent
identifications preserves this complex structure, and the reader
may prefer to view $G$ as just the pair $G = (F, H) \in H^3(X, \Z)
\oplus H^3(X, \Z)$.  Alternately, we can identifying $G=F+iH\in
H^3(X,\C)$ with the real cohomology class
$$\wt G :=   F \wedge dy - H \wedge dx \in H^4(X
\times T^2, \R)\approx H^3(X,\C)\;.$$

We shall consider the (real-linear) embedding

$$\wcal: H^3(X, \C) \to  H^0(\ccal, \lcal), \qquad G\mapsto W_G\;,$$
where $W_G$ is given by formula (\ref{WGG}); i.e.,
$$\big(W_G(z,\tau),\, \Om_z\otimes\om_\tau\big) = \int _{X\times T^2}
\wt G\wedge
\Om_z\wedge\om_\tau\;.$$  We denote
by
$\scal=$Image$(\wcal)$ the range of this map, and by
$$\scal^{\Z}=\wcal\big(H^3(X, \Z \oplus i\Z)\big)$$ the lattice of
sections satisfying the integrality condition. The map $G
\to W_G$ is not complex linear, so $\scal$ is  not a complex
subspace of
$H^0(\mcal \times \ecal, \lcal)$. Rather, it is a real subspace of
dimension $2 b_3$ (over $\R$) and $\scal^{\Z}$ is a lattice of
rank $2b_3$ in it.     In fact $\scal\approx \R^{2b_3}$ is
totally real in $H^0(\ccal,\lcal)\approx \C^{2b_3}$.

We choose
 local holomorphic frames $\Omega_{z}$ of the Hodge bundle
$H^{3,0} \to \mcal$ and $\omega_{\tau} = dx + \tau dy$ of $H^{1,0}
\to\ecal$  and let $\Omega_{z}^*\otimes\omega_{\tau}^*$ denote
the dual co-frame of $\lcal$. A holomorphic section of $\lcal$
can then be expressed as $W = f(z, \tau)  \Omega_{z}^* \otimes
\omega_{\tau}^*$ where $f \in \ocal(\ccal)$ is a local holomorphic
function.  If   $W = W_G$ is a flux superpotential, then the
corresponding function $f_G$ is given by:
\begin{equation} \label{FGG}  f_{G} (z, \tau)
= \int_{X \times T^2 }  (F \wedge dy -  H \wedge dx) \wedge (
\Omega_{z} \wedge \omega_{\tau}).
\end{equation}
When $\omega_{\tau} = dx + \tau dy$ (on a fundamental domain in
Teichm\"uller space), we obtain the simpler form:
\begin{equation} \label{FSUBG} f_{G}(z, \tau) = \int_X ( F + \tau H)  \wedge
\Omega_{z}. \end{equation}

\subsection{\label{CRITHESS}Critical points and Hessians of
holomorphic sections}

As preparation for critical points of superpotentials, we recall
some basic notations and facts concerning critical points and
Hessians of  holomorphic sections of a general line bundle $L\to M$ (see \cite{DSZ1}).

Let  $(L, h)  \to M$ be a holomorphic Hermitian  line bundle, let
$e_L$ denote a local frame over an open set $U$ and write a
general holomorphic section as $s = f e_L$ with $f \in \ocal(U)$.
Recall that the Chern connection $\nabla_h$ of $h$ is given locally
as   $\nabla (f e_L) = (\partial f - f
\partial K) \otimes e_L$, where $K=-\log \|e_L\|^2_h$, i.e.
\begin{equation}\label{covar}
\nabla s=\sum_{j=1}^m\left( \frac{\d f}{\d z^j} -f \frac{\d K}{\d
z^j}\right)dz^j \otimes e_L=\sum_{j=1}^m e^{K}\frac{\d }{\d
z^j}\left(e^{-K}\, f\right)dz^j \otimes e_L\;.
\end{equation}
The critical point equation thus reads,
$$ \frac{\d f}{\d z^j} -f \frac{\d K}{\d
z^j} = 0. $$

The Hessian of a holomorphic section $s$ of $(L, h) \to M$  at a critical point $Z_0$
is the tensor
$$D \nabla s(Z_0) \in T^* \otimes T^* \otimes L$$
where $D$ is a connection on $T^* \otimes L$. At a critical
point $Z_0$, $D \nabla s(Z_0)$ is independent of the choice of
connection on $T^*$. In a local frame and in local coordinates we
have
\begin{equation}\label{Hjq}
D'\nabla' s(Z_0)=\sum_{j,q}H'_{jq} dz^q\otimes dz^j\otimes
e_L,\qquad D''\nabla' s(Z_0)=\sum_{j,q}H''_{jq} d\bar z^q\otimes
dz^j\otimes e_L\,.\end{equation} The Hessian $D \nabla s(Z_0)$ at a
critical point thus  determines the complex symmetric matrix $H^c$
(which we call the `complex Hessian'):
\begin{equation}\label{HmatriX} H^c:=
\begin{pmatrix} H' &H''\\[6pt] \overline{H''} &\overline{H'}
\end{pmatrix} =\begin{pmatrix} H' &-f(Z_0)\Theta\\[8pt]
-\overline{f(Z_0)\Theta} &\overline{H'}
\end{pmatrix}\;,
\end{equation}  whose components  are given by
\begin{eqnarray}H'_{jq} &=& (\frac{\d}{\d
z^j} - \frac{\d K}{\d z^j}) (\frac{\d}{\d z^q} - \frac{\d K}{\d z^q}) f(Z_0)\;,\label{H'}\\
    H''_{jq} &=& -\left.f \frac{\d^2 K}{\d z^j\d\bar
z^q}\right|_{Z_0}=-f(Z_0)\Theta_{jq}\,,\quad
\Theta_h(Z_0)=\sum_{j,q}\Theta_{jq}dz^j\wedge d\bar
z^q\;.\label{H''}\end{eqnarray}

\subsection{\label{SUSYCRIT}Supersymmetric critical points and the
Hodge decomposition}

We now specialize to the critical point equations for flux
superpotentials $W_G(z, \tau)$. An important observation that is
now standard in the physics literature is
that the complex moduli $(z, \tau)$ at which a flux superpotential
$W_G(z, \tau)$ satisfies $\nabla W_G = 0$ are characterized by the
following special Hodge decomposition of $H^3(X, \C)$ at $z$ (see
\cite{AD}, (3.5)--(3.8)).

 A local holomorphic frame for the
 Hodge bundle $\lcal \to \ccal$ is $e_\lcal= \Om_z^*\otimes
\om_\tau^*$, where $\Omega_z^*$ is dual to  the local frame
$\Omega_z$ of the Hodge line bundle $H^{3,0}\to\mcal$ and
$\om_\tau^*$ is dual to the local frame $\omega_{\tau} = dx + \tau
dy $ of $H^{1,0}\to\ecal$. We let $K(Z)=K_X(z)+K_{T^2}(\tau)$ be
the \kahler potential for the local frame $\Omega_z \otimes
\omega_{\tau}$ of the (positive) Hodge bundle $\lcal^*$. We then
have
\begin{equation}\label{Kpot}|e_{\lcal}(Z)|_h^2 = |\Omega_z \otimes
\omega_{\tau}|_{h_{WP}}^{-2}= e^{K(Z)} = e^{ K_X(z)} e^{
K_{T^2}(\tau)} \;.
\end{equation}

 Hence,  the Weil-Petersson \kahler potential on $\ccal$  is
$$K(Z)=-\log  \int_X \Omega_z \wedge \overline\Omega_z - \log
(\bar{\tau} - \tau). $$  In particular, the  $\tau$-covariant
derivative on $\lcal$ is given in the local frame $e_\lcal$ by
\begin{equation} \label{DELTAU} \nabla_\tau =  \frac{\partial}{\partial
\tau} + \frac{1}{\bar\tau -\tau}. \end{equation}

Hence with $W_G =
f_G \,e_\lcal$, we have
\begin{eqnarray} \label{NABLATAU}   \nabla_{\tau} f_{G}&=&
\int_X \left[H +
\frac{1}{\bar\tau -
\tau} (F + \tau H)\right] \wedge \Omega_z \nonumber  \\
&=&\frac{1}{\bar\tau - {\tau}}  \int_X ( F + \bar{\tau} H) \wedge
\Omega_z  .\end{eqnarray}
 To compute the $z$-derivatives,  we see from \S\ref{derivatives} and
\eqref{FSUBG}--\eqref{covar} that
\begin{eqnarray} \nabla_{z^j} f_G&=&
 \left(\frac{\partial f_G}{\partial
z^{j}} + \frac{\partial K}{\partial z^{j}} f_G\right ) (z,
\tau) = \int_X (F + \tau H) \wedge \left( \frac{\partial
\Omega_z}{\partial z^{j}} + \frac{\partial K}{\partial
z^{j}} \Omega_z\right) \nonumber \\ & = & \int_X (F + \tau H)
\wedge
\chi_{j} =  0, \end{eqnarray}
for $1\le j\le h^{2,1}$.
Thus, the supersymmetric critical point equations for the flux
superpotential $W_G$ read: \begin{equation} \label{CPSYSTEM}
\left\{
\begin{array}{l} \int_X (F + \tau H) \wedge \dcal_j \Omega_z =
0 \qquad (1\le j \le h^{2,1})\\ \\
\int_X (F + {\tau}H) \wedge \overline\Omega_z = 0. \end{array}
\right. \end{equation}

As in (\ref{QZ}), we denote by $\scal_Z$ $(Z = (\tau, z))$ the
space of superpotentials $W_G$ with $\nabla W_G(Z) =0 $. Although
the equation is complex linear on $H^0(\ccal,\lcal)$, $\scal$ is
not a complex subspace of  $H^0(\ccal,\lcal)$, so
$\scal_Z$ is a real but not complex vector space.  Put another way,
for  each $Z= (z, \tau)$, the critical point equation determines a
real  subspace
\begin{equation} \label{H3} H^3_{Z}(X, \C)= \wcal\inv(\scal_Z)
=
\{F + i H,
\;\; F, H \in H^3(X, \R), \;\; (\ref{CPSYSTEM}) \; \mbox{is true}
\}. \end{equation}
 The critical point equations (\ref{CPSYSTEM})  put $b_3=2(h^{2,1} +
1)$ independent real linear conditions on $2 b_3$ real unknowns
$(F,H)$.

\begin{prop} \label{POS} \cite{AD, DD} Let $G = F + i H$ with $F, H \in
H^3(X, \R)$, and let  $\langle W_G(z, \tau), \Omega_z \wedge
\omega_{\tau}\rangle
 = \int_X (F + \tau H) \wedge \Omega_z $ be the associated
 superpotential.
If $\nabla_{z, \tau} W_G(z, \tau) = 0$,  then $(F + \tau H)  \in
H^{2, 1}_z \oplus  H^{0, 3}_z$.  Moreover, the map $$I_{\tau} :
H^3(X, \C) \to H^3(X, \C), \;\;\;\; I_{ \tau}(F + i H) = F +
\tau H$$ restricts to give real linear isomorphisms
$$I_{z, \tau}:  H^3_{z, \tau} \to  H^{2, 1}_z(X) \oplus
H^{0, 3}_z(X), \;\; $$ of  real vector spaces.
\end{prop}

\begin{proof}   We first prove that $(F + i H) \mapsto F + \tau H$ takes
$H^3_Z \mapsto H^{2, 1}_z \oplus H^{0, 3}_z$.  Suppose that $\nabla W_G=0$.
Since the $\chi_j(z)$ span $H^{2,1}_{z}$, we conclude from the first equation of
\eqref{CPSYSTEM} that $(F+\tau H)^{1,2}_z=0$; by the second equation, we also have
$(F+\tau H)^{3,0}_z=0$. Thus $F + \tau H \in H^{2,1}_z \oplus H^{0, 3}_z. $

Since $I_{z, \tau}$ is injective and since $\dim_{\R} H^3_{z,
\tau} = \dim_\R H^{2,1}_z \oplus H^{0, 3}_z = b_3$, it is clearly an
isomorphism.
\end{proof}

\subsection{The map $(z, \tau) \to H^3_{z, \tau}$}

As $(z, \tau)$ varies over $\ccal$, how do the spaces $H^{3}_{z,
\tau}$ move in $H^3(X, \C)$? This question is important in
relating the pure lattice point problem in $H^3(X, \C)$ to the
vacuum distribution problem in $\ccal$. It depends on the
geometry of the diagram
 \begin{equation}\label{DIAGRAM2} \begin{array}{ccccc}
\ical& \hspace{-.6in}\subset \ccal \times H^3(X, \C)  \\
\rho \swarrow \quad \searrow \pi &  \\ \qquad
\ccal \qquad H^3(X, \C), \end{array} \end{equation} where
$\ical = \{(z, \tau, F, H): F
   + i H \in H^3_{(z, \tau)} (X)\},$ which  is a replica of
   (\ref{DIAGRAM}) in which $\scal$ is replaced by $H^3(X, \C)$.

To answer this question, we first note that for
each $(z, \tau) \in
\ccal$,  the real-linear map $$H^3_{z,\tau} \to
H^3(X,\R),\qquad F+iH\mapsto H$$ is bijective. Injectivity follows
by noting that
$$F\in H^3_{z,\tau}  \implies F\in H^{2,1}_z\oplus H^{0,3}_z
\implies F=\bar F\in H^{1,2}_z\oplus H^{3,0}_z \implies F=0.$$
Since both spaces have dimension $b_3$, bijectivity follows. Thus
there  is a real linear isomorphism $\iota_{z, \tau}: H^3(X, \R)
\to H^3_{z, \tau}$ of the form $$\iota_{z, \tau}(H) = F(z, \tau,
H) + i H\;.$$

To describe $F(z, \tau,H)$, we  form the $z$-dependent basis
\begin{equation} \label{basis} \{\Re  D_1 \Omega_z,\dots\,, \Re
D_{h^{2,1}}\Omega_z\,, \Re\Omega_z\,, \Im   D_1 \Om_z\,,\dots,\Im
D_{h^{2,1}} \Omega_z\,,- \Im \Omega_z \} \end{equation} of $H^3(X,
\R).$ We then have \begin{equation}F(z, \tau, H)= J_\tau
H\,,\end{equation}  where $J_\tau$ is given by the block matrix
\begin{equation}\label{J} J_{\tau} = \begin{pmatrix} \Re \tau\, I_m
&  - \Im \tau\, I_m
\\ & \\
\Im \tau\, I_m  &\Re\tau\, I_m \end{pmatrix} \;\;\; (m = h^{2,1} +
1)\,,\end{equation} with respect to the basis \eqref{basis}.

This yields the following proposition:

\begin{prop} \label{H3XR2}
The mapping $(z, \tau, H) \mapsto (z, \tau, \iota_{z,
\tau}(H))$ gives an isomorphism\\ $ \ccal \times
H^3(X, \R) \simeq\ical$.
\end{prop}

An important consequence is:
\begin{prop} For any open subset $U \subset \ccal$, the cone
$\bigcup_{(z, \tau)  \in U} H^3_{(z, \tau)} (X)\sm\{0\}$ is open in
$H^3(X, \C)\sm\{0\}$.

\end{prop}

\begin{proof}  We must show that
$$\pi\big[\ical\cap \{U\times H^3(X,\C)\}\big]\sm\{0\}$$ is open.  By
Proposition
\ref{H3XR2}, it suffices to show that the image of the map
$$\iota:
U
\times [H^3(X,
\R) \sm\{0\}]\to H^3(X, \C), \;\; \iota(z, \tau, H)=\iota_{z,
\tau}(H) = F(z, \tau, H) + i
H\;,$$ is open. We fix $(z_0, \tau_0, H_0)$ and consider the
derivative $D\iota|_{z_0, \tau_0, H_0}$ on $T_{z_0, \tau_0} \ccal
\times H^3(X, \R)$. since the linear map $\iota_{z,\tau}$ is
bijective, if we vary $H$, we get all of
$H^3_{z, \tau}$, so the issue is to prove that we obtain the
complementary space by taking  variations in $\tau, z$.

First, $H^3_{z, \tau} = I_{\tau}^{-1} (H^{2,1}_z \oplus H^{0,
3}_z).$ The  $z$ variations of $H^{2,1}_z \oplus H^{0, 3}_z$ span
this space plus $H^{1,2}_z$. By \eqref{basis}--\eqref{J},
variations in $\Re \tau$, resp. $\Im \tau$, produce $\Re \Omega_z,
\Im \Omega_z$ and hence $H^{3,0}_z=$span$(\Omega_z)$ is also in the
image.

\end{proof}

\begin{rem}
  We could also ask what kind of set is swept out in
$\bigcup_{z \in U} H^{2,1}_z \oplus H^{0, 3}_z $ as $z$ ranges
over an open set $U \subset \mcal$. Since $\dim_{\C} U = h^{2,1}$,
the image of this map is a real codimension two submanifold.

\end{rem}

\subsection{Inner product on $\scal_Z$}

In Theorem \ref{MAIN}, we have expressed $\ncal_{\psi}(L)$ in
terms of a Gaussian type ensemble of holomorphic sections in
$\scal_Z$. We now specify the inner product, Gaussian measure and
\szego kernel on this space.

\begin{prop}\label{CRITPOS}
The Hodge-Riemann Hermitian inner product on $H^3(X, \C)$
restricts for each $Z = (z, \tau)$  to define a complex valued
inner product on $H^3_Z$ which satisfies $Q_Z[G] > 0$ for all $G
\not= 0$. Moreover, the map $I_\tau: H^3_Z \to
H^{2,1}_z \oplus H^{0, 3}_z$ satisfies
$Q[I_\tau G] = \Im\tau\; Q[G]. $
\end{prop}

\begin{proof}  It follows by
Proposition \ref{HRPLUS} that  the symmetric bilinear form
\begin{equation}\label{QQ}Q[F + \tau H] = i^3\int_X (F + \tau H) \wedge
\overline{(F +
\tau H)} = \Im\tau\; Q[F + i H] \end{equation} on
$H^3_{z,
\tau}(X, \C)$ in (\ref{H3}) is positive definite.\end{proof}

Recall that we have the real-linear isomorphisms
\begin{eqnarray} &H^3(X,\C) &\buildrel {\wcal}\over
\longrightarrow \ \scal \subset H^0(\ccal,\lcal)\nonumber\\
&{I_\tau}\!\downarrow &\hspace{1.5in}.\label{realiso}\\
&H^3(X,\C)\nonumber\end{eqnarray} where $I_\tau (F+iH)=F+\tau H$.
Restricting \eqref{realiso} to fluxes with a critical point at
$Z=(z,\tau)$, we have isomorphisms
\begin{eqnarray} &H^3_Z &\hspace{-.3in}\buildrel {\wcal}\over
\longrightarrow \ \scal_Z\nonumber\\
&I_\tau\!\downarrow\quad&\hspace{.5in}.\label{realisoF}\\
&H^{2,1}_z \oplus H^{0,
3}_z\nonumber\end{eqnarray}

We let $\wt Q$ denote the Hermitian inner product on  $H^{2,1}_z
\oplus H^{0, 3}_z$ transported from $(H^3_Z,Q)$ by $I_\tau$; i.e.,
\begin{equation}\label{funnyQ} \wt Q [C] = Q\left[I_\tau\inv
(C)\right]\;, \quad C\in H^{2,1}_z \oplus H^{0,
3}_z\;. \end{equation}  Hence by \eqref{QQ}, we have:
\begin{equation}\label{QfunnyQ}  Q [C] =(\Im\tau)\, \wt Q[C]\;.
\end{equation}

\section{Counting critical points: proof of Proposition
\ref{LNMO}}\label{COUNTING}

We now prove the first result on counting critical points of flux
superpotentials $W_G$ where $G$ satisfies the tadpole constraint
(\ref{TC}). Before starting the proof, we review the geometry of
the lattice point problem and the critical point problem.

We wish  to count vacua in a region of moduli space as $G$ varies
over fluxes satisfying the tadpole constraint. Equivalently, we
count inequivalent vacua in Teichm\"uller space. That is, $\Gamma$
acts on the pairs $(W, Z)$ of superpotentials and moduli by
 $$\gamma \cdot (G, Z) = (\phi(\gamma) \cdot G, \gamma \cdot Z),  $$
Therefore $\Gamma$  acts on the incidence relation (\ref{ICAL}).
We only wish to count critical points modulo the action of
$\Gamma$. To do this, there are two choices: we could break the
symmetry by fixing a fundamental domain $\dcal_{\Gamma} \subset
\ccal$ for $\Gamma$ in $\ccal$, i.e. only count critical points in
a fundamental domain. Or we could fix a fundamental domain for
$\phi(\Gamma)$ in $H^3(X, \C)$ and count all critical points of
these special flux superpotentials. When we do not know the group
$\phi(\Gamma)$ precisely, it seems simpler to take the first
option and that is what we do in Proposition \ref{LNMO} and
Theorem \ref{MAIN}.  We note that the number of critical points of
$W_G$ in Teichm\"uller space equals the number of critical points
of the $\Gamma$-orbit of $W_G$ in $\ccal$.

The level sets $Q[G] = C$ for $C
> 0$ are hyperboloids contained in $\{G: Q[G]
> 0\} $ and thus the tadpole
constraint defines a hyperbolic shell in $\{G : Q[G]
> 0\} .$ The  critical point
equation $\nabla W_G(Z) = 0$ is homogeneous of degree $1$ in $G$.
Hence, summing a homogeneous function  over $G \in \{G : Q[G]
> 0\} $ with $Q[G]
\leq L$ may be viewed as summing a function on the hyperboloid
$Q[G] = 1$ over the radial projections of the lattice points $G$
in the shell $Q[G] \leq L.$ The number which project over a
compact subset of $Q[G] = 1$ is finite.

\subsection{Approximating the sum by an integral}\label{LNMOPROOF}

Our main argument in the proof of Proposition \ref{LNMO} is the following
lemma:

\begin{lem}\label{sumtoint} Let $\psi = \chi_K$ where $K \subset \ical$ is as in
Proposition \ref{LNMO}. Then
$$\ncal_{\psi}(L) = L^{b_3}\left[
 \int_{\scal} \langle C_W,\psi\rangle
  \chi_{Q}(W)\, dW +  O\left( L^{-1/2}\right) \right]\;.
$$
\end{lem}

\begin{proof}  We consider the integer-valued function
$$f(W)=\langle C_W,
\psi
\rangle =
\sum_{\{Z:
\nabla W(Z) = 0\}} \psi(Z,W)= \#\{Z\in\ccal:(Z,W)\in K\}.$$ We  note that the
characteristic function  of the set $\{0 \leq Q[W] \leq L\}$ is
$\chi_Q(W/\sqrt L)$.
Using our symplectic basis to identify $ H^3(X, \Z \oplus
\sqrt{-1} \Z)$ with $\Z^{2b_3}$, we have
$$  \ncal_{\psi}(L)= \sum_{N\in\Z^{2b_3}} f(N) \chi_Q(N/\sqrt L) =
\sum_{N\in\Z^{2b_3}} f(N/\sqrt L) \chi_Q(N/\sqrt L)
=\sum_{N\in\Z^{2b_3}} g(N/\sqrt L)\;,$$ where $$g=f\chi_Q\,.$$

 We note that $f$ is  constant on each connected
component of $\scal\sm[\dcal\cup\pi(\d K)]$. Since the number of
these  components is finite, $f$ is bounded. We let
$S(\scal_Z)=\{N\in\scal_Z:\|N\|=1\}$, where $\|N\|$ denotes the
norm in $\Z^{2b_3}$. Since $Q_{Z}$ is positive definite, the
sphere $S(\scal_Z)$ is contained in the interior of the cone
$\{W\in \scal:Q[W]\ge 0\}$. Let
\begin{equation}\label{K}A_\psi=\sup_{Z\in\rho(\supp\psi)}
\|Q_Z\inv\|<+\infty.\end{equation} Then
\begin{equation}\label{Kinv}\inf \left\{Q[W]: W\in
\bigcup_{Z\in\rho(\supp\psi)} S(\scal_Z)\right\}= 1/A_\psi
>0.\end{equation}  Now let
\begin{equation}\label{Q0}Q_0:= \{W: Q[W]\le 1, |W|\le
A_\psi\}\supset \supp g.\end{equation}

Approximating sums by integrals, we have
$$L^{-b_3} \ncal_{\psi}(L)= L^{-b_3}\sum_{N\in\Z^{2b_3}} g(N/\sqrt L) = \int_{\R^{2b_3}}
g(W)\,dW+\sum_{N\in\Z^{2b_3}} E_{N,L},$$ where \begin{eqnarray*}
E_{N,L}& =& \int_{\rcal_{N,L}} [g(N/\sqrt L)-g(W)]\,dW,\\&& \rcal_{N,L}
=\{W=(W_1,\dots,W_{2b_3})\in\R^{2b_3}: N_j<W_j < N_j+ 1/\sqrt L\}.\end{eqnarray*}

Let $$B=Q_0\cap[\d Q\cup\dcal\cup\pi(\d K)]\;.$$
Since $g$ is locally constant on $\scal\sm B$, the error $E_{N,L}$ vanishes whenever
$\rcal_{N,L}\cap B=\emptyset$.
Hence
$$\sum_{N\in\Z^{2b_3}} E_{N,L} \le (\sup f)
L^{-b_3}\big[ \#\{N: \rcal_{N,L}\cap B\neq \emptyset\}\big]= L^{-b_3}\,O\left
(\sqrt{L}^{2b_3 - 1}\right)=O(L^{-1/2}).$$
\end{proof}

\subsubsection{The index density}

By applying precisely the same argument for $\ical nd_{\psi}(L),$
we obtain

\begin{lem}\label{ICALFIRST} Let $\psi = \chi_K$ where $K \subset \ical$ is as in
Proposition \ref{LNMO}. Then
$$\ical nd_{\psi}(L) =
L^{b_3}\left[
 \int_{\{Q[W]\le 1\}}\langle Ind_W, \psi \rangle \,dW +  O\left( L^{-1/2}\right) \right]\;.
$$
\end{lem}

\medskip
\subsubsection{Non-clustering of critical points} Before concluding
the proof of Proposition \ref{LNMO}, we briefly consider the
question of whether there exist real hypersurfaces $\Gamma \subset
\ccal$ with the property that $\sim \sqrt{L}^{2b_3 - 1}$ critical
points of norm $\leq L$ cluster within a $1/L$ tube around
$\Gamma$. A domain in $\ccal$ whose boundary contained a piece of
$\Gamma$ would attain the remainder estimate in Proposition
\ref{LNMO}.

Since the number of critical points corresponding to $G \in H^3(X,
\Z \oplus \sqrt{-1} \Z)$ is bounded, such clustering of critical
points could only occur if a sublattice of rank $2b_3 - 1$
clustered around the hypersurface  \begin{equation} \label{GAMMA}
\bigcup_{(z, \tau) \in \Gamma} H^3_{z, \tau} \subset H^3(X, \C).
\end{equation}

There do exist real hypersurfaces in $H^3(X, \C)$ for which such
exceptional clustering occurs, namely   hyperplanes containing a
sublattice of rank $2b_3 - 1$. We refer to such a hyperplane as a
rational hyperplane $L$. For instance, any pair of integral cycles
$\gamma_1, \gamma_2$ defines a rational hyperplane
$$L = L_{\gamma_1, \gamma_2} = \{G = F + i H \in H^3(X, \C): \ell(F + i G):=   \int_{\gamma_1} F + \int_{\gamma_2} H  = 0\}.$$
As mentioned in the introduction, projections of the lattice
points $H^3(X, \Z \oplus \sqrt{-1} \Z)$ to $\d Q$ concentrate to
sub-leading order $\sqrt{L}^{2b_3 - 1}$ around the hypersurface of
$\d Q$ obtained by intersecting it with a rational hyperplane.

However, rational hyperplanes never have the form  (\ref{GAMMA}).
Indeed,  under the correspondence $\rho \circ \pi^*$ defined by
the diagram (\ref{DIAGRAM2}), the image of  a hyperplane always
covers a region and not a hypersurface of $\ccal$. That is,
$$ \dim ( L \cap  H^3_{z, \tau}) > 1 \;\; \forall (z, \tau) \in \ccal.  $$
Indeed, under the identification $H^3_{z, \tau} \simeq H^3(X, \R),
$ $L |_{H^3_{z, \tau}}$ becomes the real linear functional  $L(H)
= \int_{\gamma_1} F(z, \tau, H) + \int_{\gamma_2} H$ on $H^3(X,
\R)$. Here, we use that $F(z, \tau, H)$ is linear in $H$. Hence,
$\dim L \cap H^3_{z, \tau} \geq b_3 - 1$ for any $(z, \tau)$.

As will be studied in \cite{Z2}, clustering to order
$\sqrt{L}^{2b_3 - 1}$ can only occur if the second fundamental
form of (\ref{GAMMA}) is completely degenerate. Hence the fact
that rational hyperplanes never have this form is strong evidence
that there are no smooth hypersurfaces $\Gamma \subset \ccal$ for
which lattice points cluster to subleading order around
(\ref{GAMMA}).

\subsection{\label{HDCP}Hessians and density of critical points}

The final step in the proof of Proposition~\ref{LNMO} is to change
the order of integration over $\ccal$ and over $\scal_Z$:

\begin{lem} \label{DSZFORM}  We have:
$$\int_{\{Q[W]\le 1\}}\langle C_W, \psi \rangle \,dW=  \int_{\ccal}
\int_{\scal_Z}\psi(Z,W)\, |\det H^c W(Z)|\,
  \chi_{Q_{Z}}(W) \,dW\,d\vol_{WP}(Z). $$
\end{lem}

Combining the formulas in Lemmas \ref{sumtoint} and \ref{DSZFORM},
we obtain the formula of Proposition~\ref{LNMO}.

The proof of Lemma  \ref{DSZFORM} is in two parts.  The first is an elementary
exercise in changing variables in an integral, which we accomplish below by relating
both sides to pushforwards from the incidence variety in the diagram
(\ref{DIAGRAM}).  The second part involves special geometry, and is given in the
next section.

We may
interpret the integral
$$\int_{\{Q[W]\le 1\}}\langle C_W, \psi \rangle \,dW$$
as an integral over $\ical$ as follows. Implicitly, it defines a
measure
$d\mu_{\ical}$ so that
\begin{equation} \label{CW} \int_{\ical} \psi(Z,W)\, d\mu_{\ical} =
\int_{\{Q[W]\le 1\}}\langle C_W, \psi \rangle \,dW.
\end{equation}

The measure $d\mu_{\ical}$ may be expressed in terms of  the Leray
measure $d \lcal_{\ical}$ defined by  a measure $d\nu$ on $\scal$
and the `evaluation map'
$$\epsilon: (Z,W)  \in  \ccal \times \scal \to \nabla W(Z). $$
The Leray form is the quotient  $d \lcal_{\ical}:= \frac{ dV_{WP}\times d\nu }{d
\epsilon}$, i.e. the unique form satisfying
$$d \lcal_{\ical} \times d \epsilon =  dV_{WP}\times d\nu. $$
This measure is often written $\delta(\nabla W(Z)) dW dV(Z)$ in the physics
literature.

As suggested by the physics formula, $d\mu_{\ical} = \nabla s(Z)^*
\delta_0.$ However, this formula is somewhat ambiguous. If we
regard $s$ as fixed, then it is simply the pullback of $\delta_0$
under $Z \to \nabla s(Z).$  It is then well-known that
\begin{equation} \label{PB} \nabla s^* \delta_0 = \sum_{Z: \nabla s(Z) = 0}
\frac{\delta_Z}{|\det H^c s(Z)|}. \end{equation}  However, when
interchanging the order of integration, we really wish to think of
it as a function of $s$ for fixed $Z$. So we now have a function
$\epsilon_Z(s) = \nabla s(Z)$ which may be viewed as $$
\epsilon_Z: \scal \to \C^m \equiv \R^{b_3}, $$ where $m=h^{2,1}+1=
\half b_3$. So now the zero set $\epsilon_Z^{-1}(0)$ is the
subspace $\scal_Z$ rather than the discrete set $Crit(s)$.

To simplify the notation,
we now consider the general situation where we have a real
$n$-dimensional manifold
$M$ and a space $\scal$ of functions  $F : M_n \to \R^n$. In
our case,
$F=
\nabla s$ and
$M$ is a coordinate neighborhood in $\ccal$ where $M$ has local coordinates
$(x_1,\dots,x_n)$ and
$\lcal$ has a local frame.  Suppose that
$ 0$ is a regular value of $F$,  so that $F$ is
a local diffeomorphism in a neighborhood $U$ of any point $x_0$ of $F\inv(0)$.  Let
$h=F_{|U}\inv$ in a neighborhood of $0$.  Then for
$\phi$ supported in a neighborhood of $x_0$, put
$$\langle F^* \delta_0, \phi \rangle = \langle \delta_0,
\phi(h(y)) |\det d h (y)| \rangle. $$

Let $\dim_\R\scal = d\ge n$.  In our case,   $d =2b_3>n=b_3$, so we introduce a
supplementary linear map: for a point $u\in U\subset M$,
$\scal_u$ is the kernel of $\epsilon_u$,  and we supplement $\epsilon_u$ with
the projection $\Pi_u: \scal \to \scal_u.$ Then,
$$(\epsilon_u, \Pi_u) : \scal \to \R^n  \oplus \scal_u $$
is a linear  isomorphism. Hence it equals its derivative, so
$$\begin{array}{lll} \langle \epsilon_u^{*} \delta_0, \phi \rangle &
=&  \langle \delta_0, \phi((\ep_u, \Pi_u)^{-1}) |\det
(\ep_u, \Pi_u)^{-1} | \rangle.
\end{array}$$
Now, $\scal$ is equipped with an inner product, which induces an inner product
on $\R^n\oplus \scal_u$. We choose an orthonormal basis $\{S_1,
\dots, S_n\}$ of $\scal_u^{\perp}, $ and $\{S_{n + 1}, \dots, S_d \}$ for
$\scal_u$.  Since $\Pi_u : \scal_u \to \scal_u$ is the identity,
$(\ep_u, \Pi_u)$ has a block diagonal matrix relative to the  bases of
$\scal=\scal^\perp_u\oplus \scal_u$ and $\R^n\oplus \scal_u$,   with the identity in
the
$\scal_u $-$\scal_u$ block. Hence,
$\det (\ep_u, \Pi_u) = \det( \ep_u|_{\scal^\perp})$ where the determinant is with
respect to these bases.

 The general case of formula (\ref{CW}) states
that
\begin{equation} d\mu_{\ical} = |\det D W(u)| \times
\frac{\chi_Q du\times dW }{d \epsilon}. \end{equation}
We then compute  the $\ical$ integral as an iterated integral
using the other singular fibration $\pi$, i.e. by first
integrating over the fibers $\scal_u$: \begin{equation}\label{Iint0}\int_{\ical}
\psi(u) d\mu_{\ical} =
\int_U \int_{\scal_u}
\frac{\psi(u)}{|\det( \ep_u|_{\scal_u^\perp})|} \chi_{Q_u}(W) |\det
DW(u)|\,dW\,du\;.\end{equation}

Returning to our case where $F=\nabla s$, \eqref{Iint0} becomes
 \begin{equation}\label{Iint} \int_{\ical} \psi(Z) d\mu_{\ical}=
\int_{\ccal} \int_{\scal_Z}\frac{\psi(Z,W)}{|\det(
\ep_Z|_{\scal_Z^\perp})|}
 |\det H^c W(Z)|
\,  \chi_{Q_{Z}}(W) \,dW\,d\vol_{WP}(Z). \end{equation}

\subsection{Completion of the proof of Lemma \ref{DSZFORM}} To complete
the proof of the lemma, we need to show that
$|\det( \ep_Z|_{\scal_Z^\perp})| = 1$ with respect to normal coordinates
and an adapted frame at $Z_0=(z_0,\tau_0)\in M$.

Recalling \eqref{realiso}--\eqref{realisoF}, we write
$$\wt\scal_Z^\perp=
I_\tau\circ\wcal\inv(\scal_Z^\perp)=H^{3,0}_z\oplus H^{1,2}_z\;.$$
A complex orthonormal basis for $\wt \scal_{Z_0}^\perp$ relative
to $Q$ is $\{\bar\chi_0,\bar\chi_1,\dots,\bar\chi_{h^{2,1}}\}$,
where $\chi_0=
\overline\Om_{z_0}$. A basis (over $\R$) for $\scal_{Z_0}^\perp$
is $$
\overline{U}_j:= \wcal\circ I_\tau\inv(\bar\chi_j),\quad \
\overline{V}_j:= \wcal\circ
I_\tau\inv(\sqrt{-1}\,\bar\chi_j),\qquad 0\le j\le h^{2,1}\;.$$
The basis $\{\overline{U}_j,\ \overline{V}_j\}$ is orthogonal with
respect to $Q_{Z_0}$, but not orthonormal.  By \eqref{QfunnyQ}
\begin{equation}\label{imtau1}Q[\overline{U}_j]=\wt Q[\bar\chi_j]
= \frac 1{\Im\tau }\, Q[\bar\chi_j]=\frac
1{\Im\tau },\qquad Q[\overline{V}_j]=\wt
Q\left[\sqrt{-1}\,\bar\chi_j\right] = \frac
1{\Im\tau }.\end{equation}

To compute $\det( \ep_{Z_0}|{\scal_{Z_0}^\perp})$,  we let
$(z_1,\dots,z_{h^{2,1}})$ be   normal coordinates about
$z_0\in\mcal$, and we let $\nabla_j f$ be given by
$$\nabla_{\d/\d z^j}(fe_\lcal) = (\nabla_j f)\otimes e_\lcal,$$ for
$1\le j\le h^{2,1}$. We find it convenient to use the coordinate
$\tau\in\ecal$, although it is not normal, and we use the
normalized covariant derivative
\begin{equation}\label{nabla0}\nabla_0:=
(\Im\tau)\,\nabla_\tau.\end{equation} Now we write
$$\overline{U}_j=f_j(z)\,\Om_z^*\otimes\om_\tau^*,\quad
\overline{V}_j=g_j(z)\,\Om_z^*\otimes\om_\tau^*,$$  where the local frame
$\Om_z$ is normal at $z_0$, and $\om_\tau = dx+\tau\,dy$.  Note
that the Weil-Petersson norm  $|\om_\tau^*|$ is given by
\begin{equation}\label{imtau2}|\om_\tau^*| = |dx+\tau\,dy|\inv = \frac
1{(\Im\tau)^{1/2}}\ .\end{equation}

Taking into account \eqref{imtau1}--\eqref{imtau2}, the $\Im\tau $
factors cancel out, and we obtain
$$\det(  \ep_{Z_0}|{\scal_{Z_0}^\perp})) =\det\left.
\begin{pmatrix}\Re\nabla_j f_k & \Re\nabla_jg_k\\  \\
\Im\nabla_j f_k& \Im\nabla_jg_k\end{pmatrix}\right|_{{Z_0}},\quad
\mbox{for }\ {0\le j,k\le h^{2,1}} \;.$$

We now evaluate the entries of the matrix. By Proposition \ref{ONB}, we
have
$$\nabla_{k} f_j(Z) =
\int_X
\overline{\dcal_j
\Omega_{z_0}}
\wedge \dcal_k
\Omega_{z}, \quad \nabla_{k} g_j(Z) = \int_X
i\,\overline{\dcal_j
\Omega_{z_0}}
\wedge \dcal_k
\Omega_{z}, $$ and hence
$$\nabla_j f_k(Z_0) =-i\, \delta_{jk}, \quad \nabla_j g_k(Z_0) = \delta_{jk},
\qquad\mbox{for }\ j,k\ge 1.$$
Also $$\nabla_kf_0=\int_X \Om_{z_0}\wedge[\dcal_k\Om_{z_0}-(\d K/\d
z_j)\Om_{z_0}] =0,
\quad \nabla_kg_0=i\nabla_kf_0=0\qquad\mbox{for }\ k\ge 1.$$

By \eqref{NABLATAU}, we have $$\nabla_0(f_j)=(\Im \tau)\,\nabla_\tau(f_j) =\int_X
\dcal_j\Om_{z_0}\wedge \Om_{z_0} =0,
\quad
\nabla_0(g_j) = -i\int_X\Om_{z_0}\wedge \Om_{z_0} =0,\quad j\ge 1,$$
and $$\nabla_0(f_0) = \int_X\overline{\Om_{z_0}}\wedge \Om_{z_0} =i,\quad
\nabla_0(g_0) = \int_X\overline{i\Om_{z_0}}\wedge \Om_{z_0} =1.$$

Therefore,
$$|\det(  \ep_{Z_0}|{\scal_{Z_0}^\perp}))|  =
\left|\det
\begin{pmatrix}0&I\\  \\
D(1,-1,\dots,-1)&0\end{pmatrix}\right|=1\;.$$
\qed

\section{\label{MAINPROOF}Proof of Theorem \ref{MAIN}}

In this section we prove Theorem \ref{MAIN}, which is a combination of an
equidistribution theorem for radial projections of lattice points and an
equidistribution theorem for critical points.

\subsection{\label{LVDC}A local van der Corput Theorem} We first illustrate the
method of proof of Theorem \ref{MAIN} by providing a van der
Corput type asymptotic estimate for the radial distribution of
lattice points (Theorem \ref{localvdC}). The estimate has much in
common with the classical van der Corput estimate of Hlawka,
Randol and others on lattice points in dilates of smooth convex
sets (see for example, \cite{Ran,  H}), and we adapt the proof of
the classical estimate to obtain our asymptotic equidistribution
theorem.

Let $Q\subset \R^n$ ($n\ge 2)$ be a  bounded, smooth, strictly
convex set with $0\in Q^\circ$. Let $|X|_Q$ denote the norm of
$X\in \R^n$ given by \begin{equation}\label{Q}Q=\{X\in\R^n:|X|_Q<1
\}\,.\end{equation} To measure the equidistribution of projections of
lattice points, we consider the sums
$$S_f (t) = \sum_{k \in \Z^n\cap tQ\sm\{0\}} f\left(\frac{k}{|k|_Q}\right),
\quad \mbox{with } \ f \in C^{\infty}(\d Q),\ t>0. $$ We extend
$f$ to $\R^n$ as a homogeneous function of degree $0$, so that
$f(k)=f\left(\frac{k}{|k|_Q}\right)$. Our purpose is to obtain the
following asymptotics of $S_f(t)$:

\begin{theo}\label{localvdC}  $$S_f(t) =
t^n\int_Q f\,dX + O(t^{n - \frac{2n}{n+1}}), \;\; t \to \infty. $$
\end{theo}

From this it is simple to obtain asymptotics of $S_f(t)$ when $f
\in C^{\infty}(\d Q)$ is extended as a homogeneous function of any
degree $\alpha$ to $\R^n$:

\begin{cor}\label{homovdC}  Let $f\in\ccal^\infty(\R^n\sm\{0\})$ be
homogeneous of degree $\al> 0$, and let
$$S_f(t)= \sum_{k\in \Z^n\cap tQ} f(k)\;,\qquad t>0$$
 Then  $$S_f(t) = t^{n + \alpha} \int_Q f\,dX +
O(t^{n - \frac{2n}{n+1} + \alpha}), \;\; t \to \infty.
$$
\end{cor}

\subsubsection{Littlewood-Paley}

To deal with the singularity of $f$ at $x = 0$ we use a dyadic
Littlewood-Paley decomposition in the radial direction. Let $\eta
\in C_0^{\infty}$ with $\eta(r) = 1$ for $r \leq 1$ and with
$\eta(r) = 0$ for $r \geq 2.$ We then  define
$$\rho \in C_0^{\infty}(\R),\;\; \rho (r ) = \eta(r) -
\eta(2 r). $$  Then $\rho(r)$ is supported in the shell $1/2 \leq
r \leq 2$, hence $\rho(2^j r)$ is supported in the shell $2^{- j -
1} \leq r \leq 2^{ -j  + 1}$.  We then have:
$$\eta(r) =  \sum_{j = 0}^{\infty} \rho(2^{j} r),\;\;\; (r \not= 0). $$
Indeed,
$$\sum_{j = 0}^{J} \rho (2^j r) = \eta(r) - \eta(2^J
r) \to \eta(r)$$ by the assumption that $\eta \in C_0^{\infty}$.

We then write
\begin{eqnarray} S_f(t)&  = &
\sum_{k \in \Z^n} f(k) \chi_{[0,1]}(\frac{|k|_Q}{t})\ = \
S'_f(t)+S''_f(t), \nonumber \\&& S'_f(t) =
 \sum_{k \in \Z^n} f(k) \eta(\frac{|k|_Q}{t})
,\\ && S_f''(t)\ = \ \sum_{k \in \Z^n} f(k) (\chi_{[0,1]} -
\eta)(\frac{|k|_Q}{t})).\label{twosums}
\end{eqnarray}

We can assume without loss of generality that $f\ge 0$.  We begin
with the first sum in $ S'_f(t)$:

\begin{lem}\label{first}
$$ S'_f(t) = t^n \int_{\R^n}
f(X)   \eta (|X|_Q) dX + O(\log t).
$$

\end{lem}

\begin{proof}

We write the sum  as
$$  S'_f(t)=\sum_{j = 0}^{\infty} \sum_{k \in \Z^n} f(k)
\rho (\frac{2^j |k|_Q}{t}).$$

We further break up the  dyadic sum into $\sum_{j = 0}^{J(t)} +
\sum_{j = J(t) + 1}^{\infty}$ with $J(t)$ to be determined later.
We first consider
$$S'_1:=\sum_{j = 0}^{J(t)} \sum_{k \in \Z^n} f(k) \rho
(\frac{2^j |k|_Q}{t}).$$ The function $f(X) \rho (2^j |X|_Q) \in
C_0^{\infty}(\R^n)$ when $f$ is homogeneous of degree $0$ and
smooth on $\d Q$. Hence we may apply the Poisson summation formula
to the $k$ sum to obtain
$$ S'_1=\sum_{j = 0}^{J(t)} \sum_{N \in \Z^n} \int_{\R^n} e^{-  i
\langle X, N \rangle} f(X) \rho (\frac{2^j |X|_Q}{t}) dX.$$

The terms with $N = 0$ sum up to \begin{eqnarray*}t^n  \int_{\R^n}
f(X) \left\{ \sum_{j = 0}^{J(t)} \rho(2^j |X|_Q)\right\} dX &=&
t^n \int_{\R^n} f(X) \left\{
\eta(|X|_Q)-\eta(2^{J(t)+1}|X|_Q))\right\} dX\\ &=& t^n
\int_{\R^n} f(X)  \eta(|X|_Q) dX +O(t^n2^{-nJ(t)}),\end{eqnarray*}
where the last estimate is a consequence of the fact that
$\eta(2^{J(t)+1}|X|_Q)$ is supported on $2^{-J(t)}Q$.

 To estimate the remaining terms in the sum $S'_1$, we make the
change of variables  $Y = 2^{j}  X/ t$ in the  integral to obtain
$$ 2^{- n j}  t^n \int_{\R^n}
f(Y) \rho ( |Y|) e^{-    i 2^{-j}  t \langle Y, N \rangle} dY.$$
Since the integrand is smooth, this term has the upper bound $$
c\, 2^{-n j}  t^n (1 + 2^{-j} |N| t)^{-K},\;\;\; \forall K > 0.$$
(Again, we let $c$ denote a constant; $c$ depends on $f$ and $K$,
but is independent of $j,t,N$.) The sum over $N\neq 0$ is then
bounded by
\begin{eqnarray*} c\,t^n \sum_{j \leq J(t)} 2^{- n j} \sum_{N
\not= 0}  (1 + 2^{-j} |N| t)^{-K} & \sim &  t^n  \sum_{j \leq
J(t)} 2^{- n j}  \int_0^{\infty} (1 + 2^{-j} r t)^{-K} r^{n-1} dr
\\ &  = &\sum_{j \leq J(t)}\int_0^{\infty} (1 + s)^{-K} s^{n-1} ds
\ = \ c\,J(t). \end{eqnarray*} Therefore
$$S'_1= t^n \int_{\R^n}
f(X)  \eta(|X|_Q) dX +O(t^n2^{-nJ(t)}) + O(J(t)).$$

Recall that $S'_f(t)=S'_1+S'_2$, where
$$S'_2= \sum_{j = J(t) + 1}^{\infty} \sum_{k \in \Z^n}
f(\frac{k}{|k|_Q}) \rho (\frac{2^j |k|_Q}{t}).$$ Since $$\sum_{j =
J(t) + 1}^{\infty} \rho (\frac{2^j |k|_Q}{t}) =
\eta\left(\frac{2^{J(t)} |k|_Q}{t}\right) \le
\chi_{t2^{-J(t)}Q}\;,$$ the remainder $S'_2$   is bounded by the
total number of lattice points in the shell $|k|_Q \leq 2^{- J(t)}
t$, hence is of order $t^n 2^{- n J(t)}$. It follows that
\begin{equation}\label{firstterm}S'_f(t)= t^n \int_{\R^n}
f(X)  \eta(|X|_Q) dX +O(t^n2^{-nJ(t)}) + O(J(t)).\end{equation} To
balance the terms, we choose $J(t) =  \log_2 t$, and then the last
two terms of \eqref{firstterm} have the form
$$O(t^nt^{-n})+O(\log t)  = O(\log t).$$
\end{proof}

\subsubsection{Stationary phase.}
Theorem \ref{localvdC} is an immediate consequence of Lemma
\ref{first} and the following assymptotics of the second sum
$S''_f(t)$ from \eqref{twosums}:

\begin{lem}\label{remainder}
$$S''_f(t)= t^n \int_{\R^n}
f(X)   (\chi_{[0,1]} - \eta) (|X|_Q) dX + O(t^{n - \frac{2n}{n +
1}}).$$ \end{lem}

\begin{proof}
Let $$g(X)=f(X)(\chi_{[0,1]} -\eta)(|X|_Q)$$ and mollify $g$ by a
radial  approximate identify $\phi_{\epsilon}$ to obtain a smooth
approximation $g_\ep = g*\phi_\ep$.  We claim that
\begin{equation}\label{smoothing}
S_f''(t)=\sum_{k\in\Z^n}g\left(\frac kt\right) =
\sum_{k\in\Z^n}g_\ep\left(\frac kt\right) +O(\ep t^n)\;.
\end{equation}  To see this, we break the sum into two parts.  The
first part is over the lattice points $k$ with $k/t$ in an
$\epsilon$ tube  $T_\ep$ about $\{|X|_Q=1\}$.  The number of such
$k$ is $O(\ep t^n)$, so this part contributes the stated error.
For the remaining sum, the error is
$$\left|\sum_{k\in\Z^n\sm tT_\ep}\left[g\left(\frac kt\right) -
g_\ep\left(\frac kt\right)\right]\right| \le \sum_{k/t\in\supp
g\sm T_\ep}\ep\sup_{|X|_Q>1}|dg(X)| =O(\ep t^n),$$ which verifies
\eqref{smoothing}.

The Poisson summation formula then gives
$$\sum_{k\in\Z^n} g_{\epsilon} (k/t) = t^n \sum_{N\in\Z^n}
\hat{g}_{\epsilon}(2 \pi t N) = t^n \sum \hat{g}(2 \pi t N)
\hat{\phi}(2 \pi t \epsilon N). $$ The term $N = 0$ yields
$$t^N\int_{\R^n}g_\ep(X)dX= t^n \int_{\R^n}
f(X)   (\chi_{[0,1]} - \eta) (|X|_Q) dX +O(\ep t^n),$$ where the
last inequality is by breaking up the integral into two parts as
above.

As for the remainder terms $N\neq 0$, we now show that
\begin{equation}\label{SP} \hat{g}(2 \pi t N) =
O\left( (1+|tN|)^{-\frac{(n+1)}{2}}\right) \;.\end{equation} To
verify \eqref{SP}, we write \begin{eqnarray*} g &=& -f\rho h\ =\
-(f\rho) (h\eta_2)\,, \qquad \mbox{with} \quad \eta_2(X)=\eta
(\textstyle\half|X|_Q)\,,\quad h=\theta\circ\la,\\&& \la(X) =|X|_Q
-1\,,\quad \theta(t)=
\mbox{Heaviside function}=\left\{\begin{array}{ll}0,& \ \mbox{if }\ t<0,\\
1,& \ \mbox{if }\ t\ge 0.\end{array}\right.\end{eqnarray*}  Since
$\wh g = - \wh{f\rho} * \wh {h\eta_2}$ and $\wh {f\rho}$ is
rapidly decaying, it suffices to show that $\wh {h\eta_2}$
satisfies \eqref{SP}.  (Here, we use the elementary estimate
$\|\al*\be\|_{(K)} \le c \|\al\|_{(K+n+1)} \|\be\|_{(K)}$, where
$\|\al\|_{(K)} = \sup_{x\in\R^n}(1+|x|)^K |\al(x)|$.)  Taking partial
derivatives,
$$\dcal_j(h\eta_2) = \dcal_j\eta_2 +(\de_0\circ\la)\,\dcal_j\la.$$
Since the latter term is given by integration over $\d Q$, which
is strictly convex, the standard stationary phase method (see
H\"ormander \cite{Ho}) immediately gives
$(\delta_0\circ\la)\nhat(x) = O(x^{-\frac{(n-1)}{2}})$, and hence
$$x_j\,\wh{h\eta_2}=[\dcal_j(h\eta_2)]\nhat =
O\left(x^{-\frac{(n-1)}{2}}\right),$$ which implies \eqref{SP}.

Hence the  remainder is bounded above by
\begin{equation}\label{sum}c\, t^n \sum_{N \not= 0} (1+|t N|)^{-\frac{(n+1)}{2}}
(1 + | \epsilon tN|)^{-K}.\end{equation} The sum \eqref{sum} can
be replaced  by the integral
\begin{eqnarray*} c\,t^n \int_{\R^n}(1+|t N|)^{-\frac{(n+1)}{2}} (1 + |
\epsilon tN|)^{-K}\,dN &=& c\,t^n \int_0^\infty
(1+t r)^{-\frac{(n+1)}{2}} (1 + \epsilon tr)^{-K}r^{n-1}\,dr\\
&=& c\,\ep^{\frac {1-n}2}\int_0^\infty
(\ep+s )^{-\frac{(n+1)}{2}} (1+ s)^{-K}s^{n-1}\,ds\\
&\le& c\,\ep^{\frac {1-n}2}\int_0^\infty
 (1+ s)^{-K}s^{\frac{n-3}2}\,ds \ =\ c\,\ep^{\frac {1-n}2}.
\end{eqnarray*}

Hence $$S''_f(t)= t^n \int_{\R^n} f(X)   (\chi_{[0,1]} - \eta)
(|X|_Q) dX + O(\ep t^n) +O(\ep^{-(n-1)/2}).$$ To optimize, we
choose $\epsilon$ so that $\ep t^n = \epsilon^{-(n-1)/2}$, i.e. $
\epsilon = t^{-2n/(n+1)}$, which gives the result.   (To be
precise, it is the  sum of the terms in \eqref{sum} with $|N|\ge
\sqrt n$ that is  bounded by the above integral. But there are
only finitely many terms with $|N|<\sqrt n$, and each of these
terms is $<c\,t^{n-\frac{n+1}2}$, which is better than $O(t^{n -
\frac{2n}{n+1}})$ when $n\ge 2$.)\end{proof}

\subsubsection{Van der Corput for homogeneous weights $f$. Proof of
Corollary \ref{homovdC}:} This time, we have
$$S_f (t) = \sum_{k \in \Z^n\cap tQ\sm\{0\}}|k|_Q^{\alpha}\,
f\left(\frac{k}{|k|_Q}\right). $$
The set of norms of lattice points $ \{t_j \in \R^+: \exists k \in
\Z^n \ \ni\ |k|_{Q} = t_j\}$ is  a countable set without accumulation
point. We order the $t_j$ so that $t_j \leq t_{j + 1}$.
We then define the monotone increasing step function on $\R$
$$\mu(T) = \sum_{j: t_j \leq T} \left\{\sum_{k: |k|_Q = t_j}
f\left(\frac{k}{|k|_Q}\right)\right\}. $$

It is clear that
$$\mu(T) = S_{f_0}(T),\;\;\; f_0(x) = \frac{f(x)}{|x|_Q}. $$
Hence, by Theorem \ref{localvdC},\begin{equation}
\label{SF0} S_{f_0}(t) = t^{n } \int_Q f_0\,dX + O(t^{n -
\frac{2n}{n+1}}), \;\; t \to \infty. \end{equation}

We further have
\begin{equation}\label{Sf}S_f(T) = \int_0^T t^{\alpha} d
\mu(t).\end{equation} Indeed,
$$d\mu(t) = \sum_{j} \left\{\sum_{k: |k|_Q = t_j}
f\left(\frac{k}{|k|_Q}\right)\right\} \delta(t_j), $$ and
$$ \int_0^T t^{\alpha} d \mu(t) = \sum_{j: t_j \leq T}\left
\{\sum_{k: |k|_Q = t_j} f\left(\frac{k}{|k|_Q}\right)\right\}
t_j^{\alpha} = S_f(T). $$
Integrating \eqref{Sf} by parts and applying \eqref{SF0}, we get
\begin{eqnarray*}S_f(T)&=& T^{\alpha} \mu(T) - \alpha \int_0^T
t^{\alpha - 1}
\mu(t)dt\\&=&T^{\alpha}\left[ T^{n } \int_Q f_0\,dX + O(T^{n -
\frac{2n}{n+1}})\right]  - \alpha \int_0^T t^{\alpha - 1}
\left[t^{n } \int_Q f_0\,dX + O(t^{n - \frac{2n}{n+1}})\right]
dt\\ &=& T^{n + \alpha} \left[\int_Q f_0\,dX\right]
\frac{n}{\alpha + n} + O(T^{n - \frac{2n}{n+1} +
\alpha})\\
&=&T^{n + \alpha} \int_Q f\,dX +
O(T^{n - \frac{2n}{n+1} + \alpha})\,.
\end{eqnarray*}
\qed

\subsection{Van der Corput for critical points}
We prove Theorem \ref{MAIN} by following the arguments of the
proofs of Theorem
\ref{localvdC} and Corollary \ref{homovdC} with hardly any changes.
We first assume that $\psi$ is  homogeneous of order 0 in
$\scal$.   We let $K_\psi = \rho(\supp\psi)\subset\ccal$, a
compact set.

To begin, we recall that if $W$ has critical points, then $W$ is
in the `light cone' $Q[W]>0$. For $W$ in the light cone, we write
$$|W|_Q = Q[W]^\half,\qquad \mbox{for }\ Q[W]>0.$$
The main difference between this case and our previous one, is
that now the set
$Q$ given by \eqref{Q}, in addition to not being convex, is not compact.
However, since only with those $W$ with critical points in the
support of $\psi$ contribute to the sum, we consider
$$Q_\psi:=Q\cap \scal_\psi,\qquad \scal_\psi=\left(\bigcup_{\tau\in
K_\psi}\scal_\tau\right),$$ which is a compact subset of $\scal$.

We let $f(W)=\langle C_W, \psi \rangle$, which is a smooth
function supported in $\scal_\psi$. Then
$$\ncal_\psi(L)= S_f(L)=  \sum_{k \in \Z^n\cap \sqrt L\,Q\sm\{0\}}
f(k),$$ as before.  Now we follow the previous proof, with
$t=\sqrt L$.  Our first modification is to verify
\eqref{smoothing},  we instead let
$T_\ep$ be the epsilon tube over $\scal_\psi\cap\d Q$.  Finally,
the estimate $(\delta_0\circ\la)\nhat(t) =
O(t^{-\frac{(n-1)}{2}})$, which was based on the convexity of $Q$
in our previous argument, holds in this case, since the phase
$\psi(Y)=L\langle Y,N\rangle$ has (two) non-degenerate critical
points whenever $N$ is in the light cone.  Thus we have
$$\ncal_{\psi}(L) = L^{b_3}\left[
 \int_{\{Q[W]\le 1\}}\langle C_W, \psi \rangle \,dW
   + O\left(L^{-\frac{2b_3}{2b_3+1}}\right)\right].
$$
The case $\al=0$ now follows from Lemma \ref{DSZFORM}, and the
general case then follows exactly as in the proof of Corollary
\ref{homovdC}. \qed

\section{Special geometry and
density of critical points}\label{SG}

The aim of this section is to compute the critical point density
$\kcal^\crit(Z)$ and verify Corollaries \ref{MAIN2}--\ref{MAIN3}. At
the same time, we compute the index density and prove Theorem
\ref{MAININD}. As in \cite{DSZ1}, we do this by pushing forward
the integrand of (\ref{Kcritgauss}) under the Hessian map. The
Hessian map turns out to be an isomorphism, hence the discussion
is more elementary than in \cite{DSZ1}. To make the change of
variables, we first evaluate the image of the Hessian using  the
special geometry of Calabi-Yau moduli spaces and then check how
the Hessian map distorts inner products.  Our discussion gives an
alternate approach to the formulas in the article \cite{DD}, and
connects the special critical point density formula in this
article with the general ones in \cite{DSZ1, DSZ2}.

\subsection{The range of the Hessian map}

We now study the {\it complex Hessian map\/}:
\begin{equation} \label{COMPHES} H^c(Z) : \;\; W \to
\begin{pmatrix}
H' &-x\,\Theta(Z)\\
-\bar x\,\bar\Theta(Z)&\bar H'\end{pmatrix} .  \end{equation} To
describe $H^c(Z)$ in  local coordinates, we fix a point
$Z_0=(z_0,\tau_0)$ and choose normal coordinates
$\{z^1,\dots,z^{h^{2,1}}\}$ at $z_0\in\mcal$. We let $\Om$ be a
local normal frame for $H^{3,0}\to\mcal$ at $z_0$, and we let
$\om=dx+\tau\,dy$. Recall that $\om$ is not a normal frame, since
$|\om_\tau| = (\Im\tau)^{1/2}$. We let $\wt e_\lcal=
(\Im\tau_0)^{1/2}\,\Om^*\otimes \om^*$, so that $|\wt
e_\lcal(Z_0)|=1$.

 As in \S\ref{CRITHESS}, the matrix  $ (H_{jk})$
of the holomorphic Hessian is given by
\begin{equation}\label{HjqCY}
H'(Z_0)=\sum_{j,q}H'_{jq} dz^q\otimes dz^j\otimes \wt
e_\lcal|_{Z_0}\;, \quad 0\le j,q\le h^{2,1}\;,\end{equation} where
$$dz^0|_{Z_0}=\frac 1{\Im\tau_0}\,d\tau|_{Z_0}$$ is the unit
holomorphic cotangent vector (with respect to the Weil-Petersson,
or hyperbolic, metric on $\ecal$) at $\tau_0$.

We wish to express formulas \eqref{HmatriX}--\eqref{H'} for the complex
Hessian in terms of these coordinates and frames. We write
$$(\nabla_j f) \otimes e_\lcal=\nabla_{\d/\d z^j}(fe_\lcal),\quad 1\le
j\le h^{2,1},\qquad (\nabla_0 f)
\otimes e_\lcal=(\Im\tau_0)\nabla_{\d/\d \tau}(fe_\lcal)
\;.$$ ($\nabla_0$ is the normalized covariant
$\tau$-derivative given by \eqref{nabla0}.)
The complex Hessian matrix is given by:
\begin{equation}\label{HmatriXCY} H^c(Z_0)
 =\begin{pmatrix} H'(Z_0) &f(Z_0)\,I\\[8pt]
\overline{f(Z_0)}\,I &\overline{H'(Z_0)}
\end{pmatrix}\;,\qquad H' =\Big(\nabla_j\nabla_q
f\Big)_{0\le j,q\le h^{2,1}}\;.\end{equation}

Identifying the
off-diagonal components with $f(Z_0) \in \C$, we view the image space as a
subspace  of $
\sym(h^{2,1} + 1, \C) \oplus \C$, so we can write the Hessian map in
the form
$$H_{Z_0} : \scal_Z \to\sym(h^{2,1} + 1, \C)\oplus\C,\qquad W\mapsto
\big(H'(Z_0), f(Z_0)\big)\;.$$

\begin{lem} \label{RANGE}  The range of the Hessian map $H_{Z_0}:
\scal_{Z_0}
\to \sym(h^{2,1} + 1, \C)\oplus\C $ is of the form
$\hcal_{Z_0}\oplus\C$, where $\hcal_{Z_0}$ is a real subspace
of $
\sym(h^{2,1} + 1, \C)$ of real dimension $ 2h^{2,1}$
spanned over $\R$ by the  matrices
$$ \xi^k=
\left(
 \begin{array}{cc}
0&    e_k    \\ e_k^t  & \fcal^k(z)

 \end{array}
 \right), \quad\xi^{h^{2,1}+k}= \left(
 \begin{array}{cc}
0&   \sqrt{-1}\,  e_k    \\\sqrt{-1} \, e_k^t  &
-\sqrt{-1}\,\fcal^k(z)
 \end{array}
 \right)\,, \quad 1\le k\le h^{2,1}\,,$$ given by \eqref{HCALZ},
where $e_k$ is the $k$-th standard basis element of $\C^{h^{2,1}}$
and  $\fcal^k(z) \in \sym(h^{2,1}, \C)$ is the matrix $ \left(
\fcal^{\bar k}_{jq} (z) \right)$ of \eqref{CANDY}.
\end{lem}

In other words, $\hcal_{Z_0}$ is the set of matrices of the form
\begin{equation}\label{HZ}\left(\begin{array}{cc}
0&   (\bar v_1, \dots,\bar  v_{h^{2,1}})    \\
 (\bar v_1, \dots,\bar v_{h^{2,1}})^t  &
\sum_{k=1}^{h^{2,1}} \fcal^k(z) v_k
\end{array} \right)\,,\qquad (v_1,\dots,v_{h^{2,1}})\in
\C^{h^{2,1}}\;.\end{equation} We emphasize that $\hcal_{Z} \subset
\sym(h^{2,1} + 1, \C)$ is only a real and not a complex subspace.
We also note that $\dim_\R \hcal_Z = 2h^{2,1}$ and hence $\dim_\R
(\hcal_Z\oplus\C) = b_3=\dim_\R\scal_Z\,$; i.e., $H_Z$ is an
isomorphism.

\medskip\noindent
{\it Proof of Lemma \ref{RANGE}:\/}   We shall use the notation
$1\le j,k,l\le h^{2,1}$, $0\le \al,\be,\ga\le h^{2,1}$. By
\eqref{realisoF}, we have the (real-linear) isomorphism
$$\wt\wcal_{Z_0}=\wcal\circ I_\tau\inv:H^{2,1}_{z_0}\oplus H^{0,3}_{z_0}
\buildrel {\approx}\over \to\scal_{Z_0}.$$  Recall that
$H^{2,1}_{z_0}\oplus H^{0,3}_{z_0}$ has a complex orthonormal
basis $\{\chi_\al\}$ of the form
$$\chi_j= \dcal_j\Omega_{Z_0},\quad 1\le j\le h^{2,1},\qquad \chi_0=
\overline{\Omega}_{Z_0}.$$ By \eqref{QfunnyQ}, a real orthonormal
basis of $\scal_{Z_0}$ is
$$U_\al:=(\Im\tau )^{1/2}\, \wt\wcal_{Z_0}(\chi_\al),\quad V_\al:=
(\Im\tau )^{1/2}\, \wt\wcal_{Z_0}(\sqrt{-1}\,\chi_\al).$$ We write:
$$U_\al=f_\al\,\wt e_\lcal\;,
\quad
V_\al=g_\al\,\wt
e_\lcal\;;$$ equivalently
$$\wt\wcal_{Z_0}(\chi_\al)=f_\al\, e_\lcal\;,
\quad \wt\wcal_{Z_0}(\sqrt{-1}\,\chi_\al)=g_\al\, e_\lcal\;.$$

We must compute the matrices $$H'_{Z_0} (f_\al\wt e_\lcal)=
\big(\nabla_\be\nabla_\ga f_\al\big)|_{Z_0},\quad H'_{Z_0}
(g_\al\wt e_\lcal)= \big(\nabla_\be\nabla_\ga g_\al\big)|_{Z_0},
$$ where $H'_{Z_0}:\scal_{Z_0} \to \sym(h^{2,1} + 1, \C)$ is the
holomorphic Hessian map.

We shall show that: \begin{equation} \label{BIGDISPLAY}
\left\{\begin{array}{rl}
 \rm (i) & \nabla_0^2  f_G(Z_0)  = 0, \;\; \forall G \in H^3_{Z_0}(X, \C) \quad
 (\mbox{where }\ W_G=f_G\,e_\lcal)\\ &\mbox{and thus }\
 \nabla_0^2 f_\al(Z_0)=  \nabla_0^2 g_\al(Z_0)=0,\\ & \\ \rm(ii) &
\nabla_j \nabla_0  f_0 (Z_0) =\nabla_j \nabla_0  g_0 (Z_0)= 0,\\ \\
\rm (iii) &
 \nabla_k \nabla_j f_0(Z_0) = \nabla_k \nabla_j g_0(Z_0) = 0, \\ &
 \\\rm (iv) &
\nabla_k \nabla_0  f_j(Z_0) = -\sqrt{-1}\,\delta_{jk},\quad \nabla_k \nabla_0
g_j(Z_0) = -\delta_{jk},
  \\ & \\\rm (v) &
\nabla_k \nabla_l  f_j (Z_0)  =  \fcal_{kl  }^{\bar j} ,\quad
\nabla_k \nabla_l  g_j (Z_0)  = \sqrt{-1}\, \fcal_{kl }^{\bar
j}.\end{array} \right.
\end{equation}

 First,
\begin{equation} \label{FIRSTTAU} \nabla_0 f_G(z, \tau) =
\frac{|\Im\tau_0|}{\Im\tau } \int_X ( F +
\bar{\tau} H) \wedge \Omega_z.
\end{equation}
It follows that
$$\nabla_0^2 f_G(z_0, \tau_0) =
\frac{|\Im\tau_0|^2}{\Im\tau}\frac
\d{\d\tau} \int_X (F + \bar{\tau} H) \wedge \Omega_z = 0$$ by the
critical point equation $\nabla_0 f_G(z_0, \tau_0)=0$. This proves
(i).

Next, differentiating \eqref{FIRSTTAU} with $f_G=f_\al$, we get
$$\nabla_j \nabla_0  f_\al (Z_0) = \int \overline {\chi_\al} \wedge
\dcal_j\Om_{Z_0}=\int \overline {\chi_\al} \wedge \chi_j =
-i\,\delta_{j\al},$$ and similarly, $$\nabla_j \nabla_0  g_\al
(Z_0) =\int \overline {i\,\chi_\al} \wedge \chi_j = -\delta_{j\al}.$$
This verifies (ii) and (iv).

Finally, we have by \eqref{CANDY},
$$
\nabla_k\nabla_j f_\al = \int \chi_\al\wedge \dcal_k\dcal_j\Om=
-i\sum_l \fcal^{\bar l}_{kj}\int\chi_\al\wedge
\overline{\dcal_l\Om},$$ and hence
$$\nabla_k\nabla_j f_\al(Z_0) =  -i\sum_l
\fcal^{\bar l}_{kj}\int\chi_\al\wedge \overline\chi_l= -i\sum_l
\fcal^{\bar l}_{kj}\de_{l\al} = \left\{ \begin{array}{ll} -
i\fcal^{\bar \al}_{kj} &\quad \mbox{for }\ \al\ge 1\\0&\quad
\mbox{for }\ \al=0\end{array}\right..$$ We also have
$\nabla_k\nabla_j g_\al(Z_0) =i\nabla_k\nabla_j f_\al(Z_0)$,
verifying (iii) and (v).

Thus, the holomorphic Hessian $H'(Z_0)$ maps the orthonormal fluxes
\begin{equation}\label{ortho}iU_1,\ \dots,\ iU_{ h^{2,1}},\ -iV_1,\ \dots,\
-iV_{ h^{2,1}}\end{equation} to the matrices
$\xi^1,\dots,\xi^{2h^{2,1}}$ given by \eqref{HCALZ}. Furthermore,
$$f_0(Z_0)=1,\ H'(U_0)=0,\ g_0(Z_0)=i, \ H'(V_0)=0,$$ while
$$f_j(Z_0)=g_j(Z_0)=0.$$  Thus $H^c(Z_0)$ maps  the orthonormal fluxes
\eqref{ortho} to the elements $\xi^a\oplus 0\in \sym(h^{2,1} + 1,
\C)\oplus\C$, and maps $U_0$ to $0\oplus 1$ and $V_0$ to $0\oplus
i$. \qed

\subsection{Distortion of inner product under the Hessian map}\label{DISTORTION}

 We recall that the space
$\sym(h^{2,1} + 1,
\C)$ of complex symmetric matrices, regarded  as a real vector space, has the
inner product \begin{equation} \label{IPH} (A,B)_\R=\Re\langle A,
B \rangle_{HS} =\Re(\mbox{Trace}\, A B^*)\;. \end{equation}

Recalling that $\scal_Z=\wt\wcal_Z(H^{2,1}_z\oplus H^{0,3}_z)$, we consider
its codimension 1 subspace
$$\scal'_Z=\wt\wcal_Z(H^{2,1}_z).$$ By the proof of Lemma \ref{RANGE},
the  holomorphic Hessian map
\begin{equation} H_Z: \scal'_Z \to \hcal_Z \end{equation}
is bijective, but as a map between inner product spaces, it is not an isometry.
The distortion is given by the positive
 definite operator $\La_Z$.
 We write $$\La_Z\xi^a=\sum_{b = 1}^{2
h^{2,1}}\La_{ab}\xi^b,$$ so that
$$(\xi^a,\xi^b)_\R = (\La_Z\inv \La_Z\xi^a,\xi^b)_\R =
\sum_c\La_{ac}(\La_Z\inv \xi^c,\xi^b)_\R=\sum_c\La_{ac}\de_{cb} =
\La_{ab}.$$

Tracing through the definitions, we obtain that $(\Lambda_{ab})$ is the
matrix
\begin{equation} \label{Lambda}  \begin{pmatrix} \La' & \La'' \\\La'' &
\La'\end{pmatrix}, \qquad \La'_{jk}= 2\de_{jk} + \Re \;
\mbox{Tr}\; \fcal^j \fcal^{k*},\ \  \La''_{jk}= \Im \; \mbox{Tr}\;
\fcal^j \fcal^{k*}\;.\end{equation} of Hilbert-Schmidt  inner
products of the matrices in Lemma \ref{RANGE}. Hence,
\begin{equation} \Lambda_{jk}' +\sqrt{-1}\, \Lambda_{jk}''=
2\de_{jk} +  \mbox{Tr}\; \fcal^j \fcal^{k*},\end{equation}

\medskip To tie this discussion
together with that in \cite{AD} and \cite[\S 2.1]{DSZ2}, we note
that we can consider $\hcal_Z$ as a complex vector space by
redefining complex multiplication in $\hcal_Z$:
$$c\odot \begin{pmatrix} 0 & u\\u^t & A\end{pmatrix}=
\begin{pmatrix} 0 & \bar c u\\\bar cu^t & cA\end{pmatrix}.$$  We
then define a Hermitian inner product on $\hcal_Z$:
$$\left( \begin{pmatrix} 0 & u\\u^t & A\end{pmatrix},
\overline{\begin{pmatrix} 0 & v\\v^t & B\end{pmatrix}}\right) =
2\bar u \cdot v + \mbox{Tr}(AB^*).$$

We recall from \eqref{defC} that \begin{equation}
\label{LAMBDADEF2} \Lambda_Z  = \sum_{j = 1}^{ h^{2,1}} \xi^j
\otimes \xi^{j*},
 \end{equation} where the
 $\xi^j$ are $(h^{2,1}+1) \times
(h^{2,1}+1)$ matrices.  Each term $\xi^j \otimes \xi^{j*}$ in
$\Lambda_Z$  may be expressed in matrix form as $\left( \xi^j_{ab}
\,\bar\xi^{j}_{cd} \right)$; i.e.,
\begin{equation}\label{LZ}(\Lambda_ZH)_{kl}=
\sum_{p,q}[\Lambda_Z]_{kl}^{pq}H_{pq}, \quad [\Lambda_Z]_{kl}^{pq}=
\sum_{j=1}^{h^{2,1}}\xi^j_{kl}\,\bar\xi^j_{pq}, \quad 0\le k,l,p,q
\le h^{2,1} .\end{equation} As in \cite[\S 2.1]{DSZ2},    the
result may  be expressed  in terms of the \szego kernel $\Pi_Z$,
i.e.\ the  kernel of the
 orthogonal projection onto $\scal_Z.$ By (\ref{BIGDISPLAY}) and
 (\ref{LAMBDADEF2}),we have
\begin{equation}\label{lambdaszego}\left[\Lambda_Z\right]^{p q}_{kl} =
\nabla_{\zeta_k}\nabla_{\zeta_l}\nabla_{\bar\eta_{
p}}\nabla_{\bar\eta_{ q}}
F_Z(\zeta,\eta)|_{\zeta=\eta=Z},\end{equation} where $F_Z$ is the
local representative of $\Pi_Z$ in a frame (cf. \cite{DSZ2}).

In addition,  $\Lambda_Z$ determines an  operator
$\tilde{\Lambda}_Z$ on the space $\hcal^c$ of  complex matrices of
the form
\begin{equation}\label{HmatriX2} H^c:=
\begin{pmatrix} H & x I\\ \\
\overline{x} I &\overline{H}
\end{pmatrix}\;, \;\; H \in \sym(h^{21}, \C),
\end{equation}
defined by
\begin{equation}\label{HmatriX3} \tilde{\Lambda}_Z
\begin{pmatrix} H & x I\\ \\
\overline{x} I &\overline{H}
\end{pmatrix} = \begin{pmatrix} \Lambda_Z H & x I\\ \\
\overline{x} I &\overline{ \Lambda_Z H}
\end{pmatrix}
\end{equation}

We now relate the  $(1,1)$-form  $\om_\Lambda$ of
(\ref{LAMBDAFORM}) and the operator $\Lambda$  to the curvature of
the Weil-Petersson metric on $\ccal$.

\begin{prop}\label{LAMBDARICCI} We have:
\begin{enumerate}

\item[i)] $[\Lambda_Z]^{j q}_{j' q'} = - G^{q\bar p}R^{j}_{j'  q'\bar p}
 +\delta^j_{j'} \delta_{q'}^q
+ \delta_{ q'}^j \delta_{j'}^q $, where $R$ is the curvature
tensor of the Weil-Petersson metric on $\ccal$;

\item[ii)]  $\om_\Lambda = (m + 3) \omega_{WP} +   Ric(\omega_{WP})$
where $Ric$ is the Ricci curvature $(1,1)$ form of the
Weil-Petersson metric of $\mcal$, i.e.
$$Ric(\omega_{WP}) = \frac i2\sum_{i \bar{j}} Ric_{i \bar{j}} dz^i \wedge
d\bar{z}^j, \;\;\; Ric_{i \bar{j}} := - G^{k \bar{\ell}} R_{i
\bar{j} k \bar{\ell}}. $$ Thus, $\om_\Lambda$ is the Hodge metric
\cite{Lu, W2}.
\end{enumerate}
\end{prop}

\begin{proof}
To prove (i), it suffices to combine  (\ref{LZ}) and
(\ref{RIEMANN4}), raising and lowering indices as appropriate. (In
\eqref{LZ}, a normal frame at $Z$ is assumed.)

For (ii) we note that the $(1,1)$-form \begin{equation}
\om_\Lambda= \frac i2 \sum \left[2\de_{i j} + \mbox{Tr}\;
\fcal^i(Z) \fcal^{j*}(Z)\right] dz^i \wedge d\bar z^j
\end{equation} On the other hand, by (\ref{RIEMANN}),
\begin{equation} \label{RICCI}\begin{array}{lll}  Ric_{i \bar{j}} & = & - G^{k
\bar{\ell}} \left[ G_{i \bar{j}} G_{k \bar{\ell}} + G_{i
\bar{\ell}} G_{k \bar{j}} - \frac{1}{\int_\mcal\Omega\wedge \bar \Omega} \;
\sum_{p,q} G^{p \bar{q}} \fcal_{i k p} \overline{{\fcal}_{j \ell
q}} \right]  \\ &&\\
& = &   -  (m + 1) G_{i \bar{j}} + Tr \fcal^i \fcal^{j *}
\end{array}\end{equation}\end{proof}

\begin{rem} To facilitate comparison with \cite{AD, DSZ1}, we
note that our notational conventions are the same as in
\cite{DSZ1}. In \cite{AD}, the \szego kernel $\Pi_Z$ is denoted
$G_Z$. The formulas in  \cite{AD} (4.8) are the same as
(\ref{LZ}), resp.\   Proposition \ref{LAMBDARICCI}(1). Also
$F_{ab|\bar{c} \bar{d}} = \Lambda_{ab}^{pq}G_{p\bar c}G_{q\bar
d}.$ The coefficients $F_{a\bar{b}| c \bar{d}}$ in \cite{AD}
correspond to the off-diagonal blocks of $\tilde{\Lambda}$.
\end{rem}

\subsection{Proof of Theorem \ref{MAININD}}

All but one of the ingredients of the proof are precisely the same
as in Theorem \ref{MAIN}.  We first define the analogue of
\eqref{Kcritgauss} and \eqref{PF} for the signed sum:
\begin{eqnarray}   \ical nd(Z) &:=& \int_{\scal_{Z}}
\det H^c W(Z)\,
  \chi_{Q_{Z}} dW  \nonumber\\& =
& \frac 1 {b_3!\,\sqrt{\det \La_Z}} \int_{\hcal_Z \oplus \C} \det
\left( H^*H - |x|^2 I\right)\, e^{-(\La\inv_Z H, H)_\R -
|x|^2}\,dH\,dx\;.\label{INDDEN}
\end{eqnarray}
By Lemma \ref{ICALFIRST} and the proof of Lemma \ref{DSZFORM}, we
conclude that
\begin{equation} \ical nd_{\chi_K}(L)= L^{b_3}\left[\int_K \ical
nd(Z)\,d\vol_{WP} + O(L^{-1/2})\right]\;.\end{equation}

To complete the proof of Theorem \ref{MAININD}, we evaluate the
integral in \eqref{INDDEN}:

\begin{lem} \label{INDCURV} We have
$$  b_3!\, \ical nd(Z)\,d\vol_{WP}
=\frac {\pi^{2m}}{2^m}\, c_m(T^{* (1,0)}(\ccal)\otimes
\lcal,\om_{WP}\otimes h^*_{WP})= \left(\frac \pi 2\right)^m
\det\left(-R - \omega \otimes I \right)\;.$$ \end{lem}

\begin{proof}

This follows by a supersymmetric formula for the determinant, used
in this context  in \cite{AD} and also in \cite{BSZ2}. We briefly
review the fermionic formalism referring  to \cite{BGV, BSZ2} for
further details in a similar setting.

Let
 $M=\left(M^{j}_{j'}\right)$ be an $n \times n$ complex matrix.
 Then,
\begin{equation} \det M= \int^{B^{2n}}
 e^{-\langle
M\eta,\bar\eta\rangle} d\eta\,,\qquad  \langle
M\eta,\bar\eta\rangle = \sum_{j,j'}\eta_j M^{j}_{j'} \bar\eta_{j'}
\,,\end{equation} where $\eta_j,\bar\eta_j$ ($1\le j\le n$) are
anti-commuting (or ``fermionic") variables.   The integral
$\int^B = \int^{B^{2n}}$  is the  Berezin integral, a notation for the   linear
functional $\int^B:\bigwedge^\bullet \C^{2n}\to \C$ defined by
$$\int^B|_{\bigwedge^t \C^{2n}}=0\quad \mbox{for \ } t<2n\,,\quad
\textstyle \int^B \left(\prod_{j}\bar\eta_j\eta_j\right)=1\,.$$

We now apply this formalism to $\det \left( H^*H - |x|^2 I\right)
= \det H^c$ where $H^c$ is defined as in (\ref{HmatriX2}) and
refer to the discussion in \S \ref{DISTORTION}. The matrix $H^c$ is
of rank $b_3$, and  we write
\begin{equation}\label{susy-det} \det H^c = \int^{B^{2b_3}}
  e^{-\langle H^c
(\eta, \bar\eta ),(\theta,  \bar\theta )\rangle} d\eta d\theta
\,,\end{equation} where $\eta=(\eta_1,\dots,\eta_{b_3/2}),\
\theta=(\theta_1,\dots,\theta_{b_3/2})$, and
$$\langle H^c (\eta, \bar\eta ),(\theta,  \bar\theta )\rangle= \sum
\left(H_{jk}\eta_j\theta_k+ x\de_{jk}\eta_j\bar\theta_k+ \bar
x\de_{jk}\bar\eta_j\theta_k + \bar
H_{jk}\bar\eta_j\bar\theta_k\right).$$ The quadratic form
$(\La\inv_Z H, H)_{\R} + |x|^2$ in the exponent of the Gaussian
integral may be expressed in the form $\frac{1}{2}
(\tilde{\Lambda}_Z^{-1} H^c, H^c)$, where $\tilde{\Lambda}_Z$ is
the restriction of the operator defined in (\ref{HmatriX3}) to
$\hcal_Z^c$. Indeed, both quadratic forms are equivalent to
$Q_Z(W, W)$ under a linear change of variables ($W \to H_Z(W) $ in
the case of $\Lambda_Z$ and $W \to H^c(W)$ in the case of
$\tilde{\La}_Z$).

Then
\begin{equation} \label{fourier} b_3! \;\ical nd(Z) =
 \frac{1}{\sqrt{\det\tilde \La_Z}}  \int_{\hcal^c_Z}
\int^{B^{2b_3}} e^{-  \langle H^c (\eta, \bar\eta ),(\theta,
\bar\theta )\rangle- \langle \tilde{\Lambda}_Z^{-1} H^c, H^c
\rangle } dH^c d\eta d\theta. \end{equation} We let $$\Omega=
(\eta, \bar{\eta})\otimes (\theta, \bar{\theta})^t =
\begin{pmatrix} (\eta_j\theta_k) &  (\eta_j\bar\theta_k)\\
(\bar\eta_j\theta_k) &  (\bar\eta_j\bar\theta_k)\end{pmatrix},$$
so that $\langle H^c (\eta, \bar\eta ),(\theta, \bar\theta
)\rangle = \left(H^c, \Omega \right)= \mbox{Tr}\,H^c \Omega^t$.
Then the $dH^c$ integral in \eqref{fourier} becomes the Fourier
transform of the Gaussian function $e^{-\langle
\tilde{\Lambda}^{-1} H^c, H^c \rangle }$ evaluated at $i\Omega$.
Recalling that the Fourier transform of $e^{- \langle A x, x
\rangle/2}$ equals $(2 \pi)^{n/2} (\det A)^{-1/2} e^{- \langle
A^{-1} \xi, \xi \rangle/2}$, we have that the  $dH^c$ integral
equals $(\det \tilde{\Lambda})^{\half} e^{-\frac{1}{4} \langle
\tilde{\Lambda} \Omega, \Omega \rangle}$. After cancelling $(\det
\tilde{\Lambda})^{\half}$, we obtain \begin{equation} b_3!\ical
nd(Z) = \pi^m \int^{B^{2b_3}} e^{-  \frac{1}{4} ( \tilde{\Lambda}
\Omega, \Omega )_\R} d\eta d\theta,
\end{equation}
 where in normal
coordinates, we have (by (\ref{HmatriX3}) and Proposition
\ref{LAMBDARICCI}) \begin{eqnarray*} ( \tilde{\Lambda}_Z \Omega,
\Omega )_\R &= &
 \mbox{Trace} \left[\begin{pmatrix} \Lambda_Z \eta \otimes \theta &
   \eta \otimes \bar{\theta}\\
 \bar{\eta} \otimes \theta &\bar\Lambda_Z \bar\eta \otimes
\bar\theta
\end{pmatrix}  \begin{pmatrix} \eta \otimes \theta  &
\eta \otimes \bar{\theta}\\  \bar{\eta} \otimes \theta &\bar\eta
\otimes \bar{\theta}
\end{pmatrix}^*\right] \\ &= &
 \sum_{jq j'q'}\left (\Lambda^{jq}_{j'q'} \eta_j  \theta_q
\bar{\eta}_{j'} \bar{\theta_{q'}} + \bar\Lambda^{jq}_{j'q'}
\bar\eta_j \bar\theta_q {\eta}_{j'} {\theta_{q'}}\right)
+\sum_{jq} \left(\eta_j\bar\theta_q \bar\eta_j\theta_q
+\bar\eta_j\theta_q \eta_j\bar\theta_q \right)
\\ &= &
 2\sum_{jq j'q'}\left (\Lambda^{jq}_{j'q'}  - \delta_{j j'}
\delta_{q q'} \right)\eta_j  \theta_q
\bar{\eta}_{j'} \bar{\theta_{q'}}\\
&= &2 \sum_{jq j'q'} \left(  R_{j \bar{j'}q \bar{q'} }  +
\delta_{jq} \delta_{j'q'} \right) \eta_j \theta_q\bar{\eta}_{j'}
\bar{\theta_{q'}}.
\end{eqnarray*}  (Here we used the fact that $\bar
\Lambda^{jq}_{j'q'}=\Lambda^{j'q'}_{jq}$; see \eqref{LZ}.) Thus
\begin{eqnarray*} b_3!\ical nd(Z) &= &
\pi^m \int^{B^{2b_3}} e^{  -   \half\left(R_{j \bar{j'}q \bar{q'}
} + \delta_{jq} \delta_{j'q'} \right) \eta_j \bar{\eta}_{j'}
\theta_q \bar{\theta}_{q'} } d\eta d\theta \\
 & = & \left(\frac{\pi}{2}\right)^m\; \frac{\det \left(- R -
\omega \otimes I \right) }{d\vol_{WP}}\;.
\end{eqnarray*}\end{proof}

\begin{rem}  The index density computation in special geometry is
closely   related to the asymptotics in \cite[\S 5]{DSZ2} for
critical point densities for powers of a positive line bundle $L$
on a compact \kahler manifold $M$.  The expansions in \S 5.1 of
\cite{DSZ2} can be used to show that the (first few) terms in the
asymptotic expansion of the index density equal those of the Chern
form corresponding to $ c_m(T^{*1,0}\otimes L^N)$.
\end{rem}

\subsection{\label{EXAMPLES}Examples}
We describe in this section the critical point distribution for
the cases where the dimension $h^{2,1}(X)$ of the moduli space is
0 and 1, i.e.\ when $\dim\ccal$ is 1 and 2, respectively.

 \subsubsection{\label{EXZERO}$h^{2,1}(X) = 0$}

The simplest example is the case where the Calabi-Yau manifold $X$
is rigid, i.e. $\mcal  = \{pt\} $. (See \cite{AD, DD} for further
details and computer graphics of critical points in this case.)
Then only the parameter $\tau \in \hcal$ varies. Let $G=F+iH$, and
consider the flux superpotential $W_G$. Its critical point
equation is $$F+\tau H\in H^{0,3}$$ (since in this case
$H^{2,1}(X,\C)=0$). So we write $$F=A\Om+\overline {A\Om}\;,\quad
H=B\Om+\overline {B\Om}\;,\qquad A=a_1+ia_2,\
B=b_1+ib_2\in\Z+\sqrt{-1}\,\Z\;.$$ Then writing $W_G=W_{A,B}$, we
have
$$\nabla W_{A,B}=0 \iff F+\tau H\in H^{0,3} \iff A+\tau B = 0 \iff \tau=-\frac
AB.$$

Each flux superpotential  $W_{A,B}\in\scal$ (with $A, B \in \C$)
has a unique critical point in $\hcal$, which may or may not lie
in the fundamental domain $\ccal$. In the notation of
(\ref{DIAGRAM}),
$$\pi(\scal) = \{W_{A, B}: - { \frac{A}{B}} \in \ccal \}$$
is a domain with boundary in $\C^2$.  Each $SL(2, \Z)$-orbit of
fluxes (or superpotentials) contains a unique element whose
critical point lies in $\ccal$, so $\pi(S)$ is a fundamental
domain for the action of $\Gamma$ on $\scal$.

Thus, counting critical points is equivalent to counting $SL(2,
\Z)$ orbits of superpotentials satisfying the tadpole constraint.
 The pair $(A, B)$ corresponds to the element $\left(
\begin{array}{ll} a_1 & b_1  \\ a_2 & b_2 \end{array} \right)
\in GL(2, \Z)$ and the Hodge-Riemann form  quadratic form
 may be identified with the indefinite
quadratic form
$$Q[(A, B)] = a_1 b_2 - b_2 a_1$$
on $\R^4 $.
 The
modular group $SL(2, \Z)$ acts by the standard diagonal action on
$(A, B) \in \R^2 \times \R^2$ preserving $Q[(A, B)]$ or
equivalently by left multiplication preserving $\det$. Thus, the
set of superpotentials satisfying the tadpole constraint is
parametrized by:
$$\left\{\left( \begin{array}{ll} a_1 & b_1  \\ a_2 & b_2 \end{array}
\right) \in GL(2, \Z): 0<\det \left( \begin{array}{ll} a_1 & b_1 \\
a_2 & b_2
\end{array} \right) \leq L \right\}, $$
and we want to count the number of
 $SL(2, \Z)$-orbits in this set.
Counting the  number of $SL(2, \Z)$ orbits   in $\dcal_L$ is
equivalent to determining the average order of  the classical
divisor function $\sigma(m)$, see for instance Hardy-Wright
\cite[Theorem 324]{HW}:
\begin{equation} \label{ONEP} \ncal^\crit(L) = \sum_{m = 1}^L \sum_{k | m} k = \sum_{m = 1}^L \sigma(m) \sim \frac{\pi^2}{12} L^2
+ O(L \log L). \end{equation} As verified in \cite{DD} (and as
follows very simply from Theorem \ref{MAIN}), the critical points
are uniformly distributed relative to the hyperbolic area form.

\subsubsection{$h^{2,1}(X)=1$}
 We now illustrate our notation and results with the case
where the moduli space of complex structures on $X$ is
one-dimensional over $\C$. (This case is also studied in \cite{DD}
from a slightly different point of view.) In this case, there is a
single Yukawa coupling $\fcal_{11}^{\bar 1} (z)$ defined by $D_z^2
\Omega_z = \fcal^{\bar 1}_{11} (z) \overline{D_z \Omega_z}. $

The space $\scal_{z, \tau} \simeq
H^{2,1} \oplus H^{0, 3} \simeq \C^2$. The space is spanned as a
real vector space  by four superpotentials $U_0, U_1, V_0,V_1$
corresponding to $\{\overline{\Omega_z}, \dcal_z \Omega_z,i\overline{\Omega_z},
i\dcal_z \Omega_z\}$. By the proof of Lemma \ref{RANGE}, the holomorphic
Hessians of $U_0$ and $V_0$ at a critical point equal zero, so we only need to
consider the holomorphic Hessian map on $U_1$ and $V_1$. The corresponding
space of Hessians is the real $2$-dimensional subspace $\hcal_Z$ of $\sym(2,
\C)$ spanned by
$$ \xi^1 =  \begin{pmatrix} 0 & 1  \\
 1& F(z) \end{pmatrix}, \;\;\;\;\xi^2 =  \sqrt{-1}
\begin{pmatrix} 0 & 1
 \\
 1 &  -F(z) \end{pmatrix} , $$
where we write $F=\fcal_{11}^{\bar 1}$.
  Hence, we may parameterize the space $\hcal_Z$ of holomorphic Hessians by
 $$w = y_1 + i y_2  \mapsto H(w) =
\left( \begin{array}{ll} 0 & w \\ & \\
 w & F(z) \bar{w}\end{array} \right). $$

By \eqref{Kcritgauss}, we have:
$$\kcal^\crit (Z) = \frac 1{2!} \int_{\C\oplus  \C} |\det(H(w)^*H(w)  - |x|^2 I)
|\;\;  e^{- |w|^2+|x|^2} dw dx . \;\;$$ We note that
$$\det(H(w)^*H(w)  - |x|^2 I) =  |w|^4 + |x|^4 -  (2+|F(z)|^2)  |x|^2
|w|^2. $$ Hence
$$\kcal^\crit (Z) = \frac 1{2!} \int_{\C\oplus  \C} \left| |w|^4 + |x|^4 -  (2+|F(z)|^2)  |x|^2
|w|^2\right|\,e^{- |w|^2+|x|^2}\, dw\, dx , $$ agreeing with
(3.19) of \cite{DD}. There, the integral is evaluated as
$$\kcal^\crit(Z) = \frac{\pi^2}{2}\left(2 - |{F}|^2 + \frac{2
|{F}|^3}{\sqrt{4 + |\tilde{F}|^2}}\right). $$

\begin{rem} In this example, the discriminant variety is given by
$$ \wt\dcal = \{(Z,x\,W_0(Z)+w\,W_1(Z))\in\ical:
|w|^2-|x|^2= \pm |wxF(z)^2|\},$$ where $W_\al = U_\al +iV_\al$.
The matrix $\La$ is given by
$$\La=\begin{pmatrix} 2+|F|^2 &0\\0&  2+|F|^2\end{pmatrix}.$$
\end{rem}

\section{\label{FCFP}Problems and heuristics on the string theory landscape}

In this section, we continue the discussion begun  in \S
\ref{RELATIONS} on  the bearing of our methods and results on the
physicists' picture of the string theory landscape. We briefly
review some of the heuristic estimates in the physics discussions,
and then discuss a number of mathematical pitfalls in the
heuristics. In \S \ref{PROBLEMS}, we state some mathematical
problems suggested by the heuristics and by rigorous vacuum
statistics.  In \S \ref{HEURISTICS}, we give our own (tentative)
heuristic estimate of the dependence of the critical point density
$\kcal^\crit(Z)$ on the dimension $b_3/2$ of $\ccal$.

\subsection{Complexity of the  string theory landscape}

As mentioned in \S \ref{RELATIONS}, the possible vacua in string/M
theory are often represented as valleys in a complex string theory
landscape, and the number of valleys is often estimated at
$10^{500}$.

  L. Susskind and others have argued that such  a large number of
possible vacua should  essentially be a consequence of the large
number of variables in the potential. A common and general
argument to arrive at this number of vacua without specifying any
particular string theory model is to reason that the potential
energy is a function of roughly $1000$ variables.
 A generic polynomial $f$ of degree $d$ on $\C^m$  has
  $(d - 1)^m$ critical points since critical
points are solutions of the $m$ equations $\frac{\partial
f}{\partial z_j} (w) = 0$ of degree $d - 1$. Thus, the number of
critical points would seem to grow at an exponential rate in the
number of variables. Such an exponential growth rate of critical
points also appears in the physics of spin glasses, where the
growth in the number of metastable states (local minima of the
Hamiltonian) in terms of the number of variables is often used to
measure  the complexity of the energy landscape. In special model
of random Hamiltonians on domains in $\R^N$, exponential growth of
the number of local minima in $N$
 has recently been proved rigorously \cite{Fy}.

 In the specific models of  type IIb flux compactifications on
a CY $3$-fold $X$, the number of variables is $b_3(X)$. As
mentioned above, for  a typical $CY$ $3$-fold,  $b_3$ is often
around $300$ and sometimes as high as $1000$ (cf.\ \cite{ GHJ,
Can1}), and therefore the scalar potential $V_W$ in (\ref{V}) is a
function of this number of variables. By naive counting of
variables one would thus arrive at a figure like $10^{500}$ for
such models. The  more sophisticated estimate $N_{vac} \simeq
\frac{L^{b_3}}{b_3!} f(b_3)$ in flux compactifications (see \S
\ref{RELATIONS} for the notation) does not supplant the naive
counting argument since the order of magnitude of $f(b_3) $
 is unknown. We recall that it is the integral over $\ccal$ of the Gaussian
 integral in  \eqref{PF} (see (\ref{FB3}). The Gaussian integral for
 $\kcal^\crit$ in that line  resembles to some extent
 the integral formula for the expected number of
critical points in spin glass theory, which has exponential growth
(see e.g. \cite{Fy}).

Although the naive counting of variables or the analogy to
complexity of energy landscapes bring some insight into vacuum
counting, we now  point out some pitfalls in estimating numbers of
vacua or the coefficient $f(b_3)$ in flux compactifications on
this basis.

\begin{enumerate}

\item The critical point equation (\ref{CRITSET}) is $C^{\infty}$
but not holomorphic, so vacua are critical points of a real system
of equations, and it is not obvious how many connection critical
points to expect even a polynomial of a given degree to have. This
number depends on the connection, and  is studied in detail in
\cite{DSZ1, DSZ2} and in the present paper.

\item A flux superpotential $W$
 is not a polynomial and it is not clear how to assign it a `degree' which
 reflects its number  of critical points on all of Teichm\"uller
 space, or equivalently, the number of critical points in $\ccal$
 corresponding to the $\Gamma$-orbit of $W$.
 Examples (e.g. in \S \ref{EXZERO}) show that this number can be relatively small.

 \item It seems reasonable to say that the number of fluxes rather than the number of
 critical points per flux that dominates the number of vacua.
In  flux compactifications,  the landscape should  therefore be
viewed as the graph of the scalar potential $V_W(Z)$ on $\ccal
\times \scal$, i.e. as a function of both variables $W, Z$, and
the local minima should be viewed as   pairs $(W_G, Z)$ with $G
\in H^3(X, \Z \oplus \sqrt{-1} \Z)$ and with $Z \in Crit(W_G). $

\item However (see the problems below) it is not straightforward

 to define `per vacua',
since  the tadpole constraint is
 hyperbolic,  and the total number of lattice points in the shell $0 < Q[G] <
L$ is  infinite.

\item In estimating $\kcal^\crit(Z)$ we are fixing $Z$ in the
interior of $\ccal$. But there could exist singular points of
$\ccal$ at which $\kcal^\crit(Z)$ blows up (see \cite{DD} for
discussion of conifold points). It would also be interesting to
study $\kcal^\crit(Z)$ as $Z \to \d \ccal$.

 \item As mentioned in \S \ref{RELATIONS} (see also \S \ref{HEURISTICS}), there  may be a significant difference between the order of
 magnitude of the density of critical points and of the number of
 critical points, since
$\ccal$ is an incomplete \kahler manifold of possibly quite small
volume. See \cite{LuS1} for the current state of the art on the
volume.  There is no analogue of the small volume of the
configuration space in spin glass complexity.

\item The tadpole constraint (\ref{TC}) becomes much more highly
constraining as the number $b_3$ of variables increases for fixed
$L$ and is responsible for the factor $\frac{1}{(b_3)!}$ in
Theorem \ref{MAIN}. Again, no such feature exists in complexity
estimates in spin glasses.

\end{enumerate}

\subsection{\label{PROBLEMS}Problems}

The issues mentioned above (and the detailed heuristics in \S
\ref{HEURISTICS}) suggest a number of problems. The ultimate goal
is:

\begin{prob} \label{BIGMAINPROB}
Does string theory contain a
vacuum consistent with the standard model, and if so, how many?
Find examples of Calabi-Yau manifolds, and any other postulated structures,
for which it is certain that such a vacuum exists.
\end{prob}

Now testing consistency with the standard model requires elucidating
far more structure of a candidate vacuum -- the gauge group, the matter
content, and so forth -- than we are considering here.  To address this
ultimate problem, one would need many more statistical results, along
the lines set out in \cite{Doug}.  However one can make arguments
(admittedly quite speculative at this point) that the dominant
multiplicity in vacuum counting arises from the multiplicity of flux vacua
we are discussing here.
An important problem in this context is

\begin{prob} \label{MAINPROB}
How large does $L$ need to be to ensure that there exists a
vacuum with
\begin{equation} \label{cosbound}
|W_G(Z)|^2 \le \lambda_*
\end{equation}
for a specified $\lambda_*$ ? In that case, how many such
vacua are there?  Find examples of Calabi-Yau manifolds where it is
certain that such a vacuum exists.
\end{prob}

To solve this problem for type IIb flux compactifications, we
would need to sharpen Theorem \ref{MAIN} in many ways which lead
to the subsequent problems stated below.

The constraint (\ref{cosbound}) on $|W_G(Z)|^2$ is a simple
example of `consistency with the standard model.'  If the real
world were (counter-factually) exactly supersymmetric, this would
be the constraint that the vacuum should have a cosmological
constant $V_W(Z) = -3|W_G(Z)|^2$ (as in (\ref{V})) consistent with
the known value. While the physical discussion requires taking
supersymmetry breaking into account, as discussed in \cite{DD2},
vacua can exist in which supersymmetry is broken by effects not
taken into account here, making additional contributions to the
vacuum energy which lift the exact vacuum energy to be consistent
with the known value (essentially, zero). For such a vacuum, the
quantity $3|W_G(Z)|^2$ would be the mass squared of the gravitino,
a quantity which could be constrained by physical observations.

An independent motivation for (\ref{cosbound}) is that some proposals
for stabilizing the moduli we did not discuss, such as that of \cite{KKLT}, are
believed only to work under such a constraint.

In any case, as discussed in \cite{DD} (\S 3.3), one can count such vacua by
choosing the test function to be $\theta ( \lambda_* - |W_G(Z)|^2)$
where $\theta (x) = 1$ for $x> 0$ and $= 0$ for $x
\leq 0.$ This test function is not homogeneous but can be handled
by the methods of this paper (loc. cit.).

Theorem \ref{MAIN}  is asymptotic in $L$ and we have also analyzed
to some degree the $b_3$ dependence. But as mentioned in \S
\ref{RELATIONS},  $L$ depends on the topology of $X$. There, we
stated that in many examples $L \simeq C b_3$ with $1/3 \leq C
\leq 3$. To bridge one gap between Theorem \ref{MAIN} and Problem
\ref{MAINPROB}, we state:

\begin{prob}  How are the order of magnitudes of $b_3(X)$ and $L$ of
(\ref{TADPOLE}) related as $X$ varies over topologically distinct
Calabi-Yau manifolds?
\end{prob}

We have already mentioned the importance of obtaining effective
estimates in $b_3$ of the coefficient (\ref{LEADDEN}) in Theorem
\ref{MAIN}:

\begin{prob}   Obtain an effective estimate of $\kcal^\crit(Z)$ and of
its integral over $\ccal$ in $b_3$.  Also, obtain such an estimate
of the remainder.
\end{prob}

Among the difficulties with this problem is that $\kcal^\crit(Z)$
depends on special features of the moduli space $\ccal$ which
depend on more than just the dimension $b_3$ and which may change
in an irregular way as the dimension increases. We consider this
problem below in \S \ref{HEURISTICS}.

To gain insight into the size of the leading coefficient
(\ref{LEADDEN}), one could write the principal term in Theorem
\ref{MAIN} in the form $\frac{L^{b_3}}{b_3!} \times f(b_3)$ that
is often used in string theory (cf. \S \ref{RELATIONS}), with
$f(b_3)$ the Gaussian integral in \eqref{PF}. As mentioned above,
it is natural to try to separate out the effects of the number of
fluxes and the number of vacua per flux, or more precisely:

\begin{enumerate}

  \item the number of fluxes $G$  satisfying the tadpole constraint
   with a critical point in a compact subset  $\kcal \subset \ccal$;

  \item the number of critical points `per flux', or more precisely per $\Gamma$-orbit of fluxes,  in
  $\kcal$ (see \S \ref{EXZERO} to clarify this distinction);

  \item the total number of critical points in $\kcal$ of all fluxes satisfying the tadpole constraint.

  \end{enumerate}

  We can define the first quantity precisely as
 the sum
$$\Theta_K(L) = \sum_{G  \in H^3(X,
\Z \oplus i \Z): Q[G] \leq L} \theta\left( \sum_{Z \in \ccal:
\nabla W_G(Z) = 0} \chi_K(Z)\right). $$  Thus, the problem we pose
is:
\begin{prob}  Determine the asymptotics of $\Theta_K(L)$ as $L \to
\infty$.
\end{prob}

 The second quantity is the ratio $\ncal_{K}(L)/\Theta_K(L).$ A
possibly more tractable way to restate this problem is in terms of
the  `average number of critical points' of a superpotential $W_G$
in $\kcal$. To define `average' we need to introduce a probability
measure on $\fcal$ which is compatible with $\chi_Q dW$.  The most
natural probability measures seem to be the normalized Gaussian
measures $\gamma_{Z_0}$ on the spaces $\scal_{Z_0}$ defined by the
inner product $Q_{Z_0}$.Thus, we ask for the average number of
critical points of $W \in \scal_{Z_0}$ with respect to
$\gamma_{Z_0}$. It would be interesting to study the number of
critical points in a fixed $\kcal \subset \ccal$ or in all of
$\ccal$ or indeed in all of Teichm\"uller space (which corresponds
to counting critical points in $\ccal$ for a $\Gamma$-orbit of
fluxes).

We observe that  $W \in \scal_{Z_0}$ has a critical point at $Z$
if and only if  $W \in \scal_{Z_0} \cap \scal_Z$. In the case of
flux superpotentials, $\dim \scal_{Z_0} = \frac{1}{2} \dim \fcal$
so for generic pairs $Z, Z_0$, $\scal_{Z_0} \cap \scal_Z = \{0\}$.
Thus, $\E_{Z_0} (\# Crits(W))$ will be an integral over the
special variety $\Sigma_{Z_0} = \{Z \in \C: \dim \scal_{Z_0} \cap
\scal_Z
> 0\}$.  This variety is obviously stratified by
$h^{2,1}$ strata $\Sigma_d$ on which the dimension $d$ takes the
values $d = 1, 2, \dots, h^{2,1}$, and $\E_{Z_0} (\# Crits(W))$ is
a sum of integrals over each strata.

\begin{prob}  Determine the asymptotics of
 $\E_{Z_0} (\chi_{Q_{Z_0} (G/ L)} \# Crits(W_G))$
\end{prob}

We also recall that in Theorem \ref{MAIN} we ignored the effect of
the discriminant variety and the boundary of the region of
$\ccal$.

\begin{prob}
Estimate the remainder if $\psi$ does not vanish near the
discriminant variety $\dcal$, or if $\psi$ is a characteristic
function of a smooth region $K \subset \ccal.$ Investigate the
boundary behavior as $\kcal$ fills out to $\ccal$.
\end{prob}

An analogue problem about studying accumulation of lattice points
around boundaries of domains on non-degenerate surfaces is studied
in \cite{ZZ}.

\subsection{Heuristic estimate of the critical point
density}\label{HEURISTICS}

We now present a heuristic estimate on the  $b_3$-dependence of
the critical point density (relative to the Weil-Petersson volume
form)
\begin{eqnarray}\label{Kcritgauss2}
\kcal^\crit(Z)  &=& \frac 1 {b_3!\sqrt{\det \La_Z}} \int_{\hcal_Z
\oplus \C} \left|\det H^*H - |x|^2 I\right|\;\; e^{-(\La\inv_Z H,
H)_\R - |x|^2}\,dH\,dx  \end{eqnarray} for $Z$ in regions of
moduli space where the norm of $\La_Z$ satisfies bounds
independent of $b_3$.
 We recall (cf. Proposition \ref{LAMBDARICCI}) that $\Lambda_Z$ is the
Hodge metric,  hence we are studying the density of critical
points in regions $K \subset \ccal$ where the absolute values of
the eigenvalues of the Ricci curvature of the Weil-Petersson
metric $\omega_{WP}$ are bounded by a uniform constant. In the
notation $N_{vac}(L) \sim \frac{L^{b_3}}{b_3!} f(b_3)$, we have
\begin{equation} \label{FB3} f(b_3) = \int_{\ccal} \chi_K(Z)
\frac{1}{\sqrt{\det \La_Z}} \int_{\hcal_Z \oplus \C} \left|\det
H^*H - |x|^2 I\right|\;\; e^{-(\La\inv_Z H, H)_\R - |x|^2}\,dH\,dx
,\end{equation}
 where $\kcal$ is the region in which we are  counting the critical points.

Our heuristic estimate  is that the Gaussian integral (i.e. $b_3!
\kcal^\crit(Z)$)  has growth rate $(b_3/2)! N_{\mu}^{b_3}$ for $Z$
in a region $K = K_{\mu}$ of moduli space where $||\Lambda_Z||
\leq \mu$. Here, $N_{\mu}$ is a constant depending only on $\mu$.
It follows that $\kcal^\crit(Z)$ would have  the decay rate
$b_3^{-b_3/2}$ for $Z$ in $K_{\mu}$. We note that this heuristic
estimate is consistent with the heuristic estimate given by
Ashok-Douglas \cite{AD}  that $\kcal^\crit(Z)$ should have the
same order of magnitude as $\ical nd (Z)$ (\ref{INDDEN}).  By
Proposition \ref{INDCURV}, $b_3 ! \ical nd(Z)$  is a differential
form depending polynomially on the curvature. The density of $b_3
! \ical nd(Z)$  relative to $dVol_{WP} =
\frac{\omega_{WP}^{b_3/2}}{(b_3/2)!}$ thus has the growth
 $(b_3/2)! N_{\mu}^{b_3}$  we predict. We present the new
 heuristic to give evidence that the absolute value only changes
 the coefficient and not the order of magnitude in vacuum
 counting.

Before going into the heuristic estimate, we first discuss the
consequences for vacuum counting.
 As mentioned in the
introduction, it  has been tentatively conjectured at this  time
of writing (Z. Lu)  that the Weil-Petersson volume of $K_{\mu}$ is
bounded above by the volume of  a ball of radius $r(\mu)$ in
$\C^{b_3/2}$ depending only on $\mu$, and the latter volume decays
like $\frac{1}{(b_3/2)!}$. Thus it would appear that
 $N_{vac, K_{\mu}}(L) \sim \frac{(C_1 L
N_{\mu})^{b_3}}{b_3!}$. We include a constant $C_1$ to take into
account the dependence on various parameters including $r(\mu)$,
factors of $\pi$ and so on. If we then take the (often) observed
value $L \sim C b_3$ with $C \in [\frac{1}{3}, 3]$, then the
number  of vacua in $K_{\mu}$ satisfying the tadpole constraint
would grow at an exponential rate in $b_3$.

We now explain the heuristic estimate regarding the  order of
magnitude of $\kcal^\crit(Z)$ (\ref{LEADDEN}): the latter depends
on two inputs, the subspace $\hcal_Z$ (or equivalently the
orthogonal projection $P_Z$ onto $\hcal_Z$) and the eigenvalues of
$\La_Z$. To obtain upper and lower bounds on $\kcal^\crit(Z)$ we
note that
\begin{equation} \label{MUMIN} 2 P_Z \leq \La_Z \leq \mu_{\max}(Z)
P_Z, \end{equation} where $\mu_{\max}(Z)$ is the maximum
eigenvalue of $\La_Z$.
 We recall here that $\Lambda_Z$ is the matrix of the
Hodge metric (see (\ref{LAZY})), and its eigenvalues can be
estimated in terms of the Weil-Petersson metric and its curvature
(cf. \cite{Lu}).  In particular, its minimum eigenvalue satisfies
$\mu_{\min}(Z) \ge 2$, and that explains the lower bound $2 P_Z$
 in (\ref{MUMIN}).  For most CY
$3$-folds $X$, the Weil-Petersson metric on $\ccal$ is incomplete,
and  $\mu_{\max} (Z)\to \infty$ as $Z$ tends to the boundary (Z.
Lu).

By (\ref{MUMIN}), we have
 \begin{equation} \label{BOUNDS} J_{-} (\mu, P_Z) \leq (b_3!) \kcal^\crit(Z)
 \leq J_+(\mu, P_Z), \;\;\;(\forall \mu \geq \mu_{\max}(Z))\end{equation}  where  \begin{eqnarray}  J_+(\mu, P_Z) :
& = & \frac 1 {^{b_3/2 - 1} } \int_{\hcal_Z \oplus \C} \left|\det
H^*H - |x|^2 I\right|\;\; e^{- \left( \mu^{-1} \mbox{Tr} H^*H -
|x|^2 \right)}\,dH\,dx,
\end{eqnarray}
and where
 \begin{eqnarray}  J_-(\mu, P_Z) : & = &
\frac 1 { \mu^{(b_3/2 - 1)} } \int_{\hcal_Z \oplus \C} \left|\det
H^*H - |x|^2 I\right|\;\; e^{- \left( 2^{-1} \mbox{Tr} H^*H -
|x|^2 \right)}\,dH\,dx,
\end{eqnarray}
Thus we obtain upper and lower bounds for the density in regions
$K_{\mu} \subset \ccal$ for which the absolute values of the
eigenvalues of the Hodge metric relative to the Weil-Petersson
metric  satisfy $\mu_{\max}(Z) \leq \mu$. We have bounded the
determinant of $\Lambda$ by a power of an extremal eigenvalue, but
it could also be identified with the  volume density of the Hodge
metric.  We note that the lower bound tends to zero and the upper
bound tends to infinity in $\sim \pm b_3$ powers of
$\mu_{\max}(Z)$  as $Z \to \d \ccal$ when the Weil-Petersson
metric is incomplete and the norm of the  Ricci curvature of
$\omega_{WP}$ tends to infinity.

We now estimate   $J_{\pm} (\mu, P_Z)$ under the assumption that
$\hcal_Z$ is  a `sufficiently random' subspace.
 The subspace
$\hcal_Z$ is a real subspace of dimension $b_3 - 2$ of $\sym(b_3/2
- 1, \C) $, but by modifying the definition of the complex
structure it becomes a complex $b_3/2$-dimensional one. Hence, we
may view $Z \to \hcal_Z$  as a map $\ccal \to Gr(b_3/2 - 1,
\sym(b_3/2 - 1, \C))$ to the complex Grassmannian of $b_3/2 - 1$
dimensional complex subspaces. Lacking knowledge of the
distribution of the image of $Z \to \hcal_Z$, we make the
assumption that it is random, or more precisely we  approximate
$J_{\pm}(\mu, P_Z)$  by the expected  value of $J_{\pm}(\mu, P)$,
where $P$ is the projection corresponding to a random element
$\hcal \in Gr(b_3/2 - 1, \sym(b_3/2 - 1, \C))$.

This approximation by the expected value seems to be reasonable
because
 Grassmannians $Gr(k, N)$  are examples of Gromov-Milman 'Levy
families' of Riemannian manifolds for which concentration of
measure phenomena hold as $N \to \infty$  \cite{GM, T}.
Concentration of measure refers to a metric space $(X, d)$ with a
probability measure $P$ and a concentration function $\alpha(P,
t)$, which is the smallest number such that the measure of a set
$A$ and the metric tube $A_t = \{x: d(x, A) < t\}$ around $A$ are
related by $P(A) \geq 1/2 \implies P(A_t) \geq 1 - \alpha(P, t).$
If $f$ is a Lipschitz function and if $M_f$ is a median for $f$,
we  put $A = \{x: f(x) \leq M_f\}$, and then $P(|f - M_f| > t)
\leq 2 \alpha(P, \frac{t}{||f||_{Lip}}). $ Concentration of
measure occurs if $\alpha(P, t)$ decays rapidly in $t$, and thus
$f$ is highly concentrated around its median. In a  L\'evy family
$(X_N, d_N)$, the functions $\alpha_N(P, t)$ decay at ever faster
rates depending on $N$. For instance on the unit $N$-sphere $S^N$,
the rate is (a universal constant times) $e^{- \frac{(N - 1)}{2}
t^2}$.

  In our setting, the family consists of Grassmannians $Gr(b_3/2 - 1,
\sym(b_3/2 - 1, \C))$
  equipped with the invariant probability measure $d\nu$ and with  the
standard bi-invariant
  metric. It is pointed out in \cite{GM} that $Gr(k, N)$ is a
  L\'evy family for fixed $k$ (see section (3.3) of \cite{GM}), and the same argument should apply to
  $k_N \sim N/2$. Moreover, $\{U(N)\}$ with its Haar probability
  measure and bi-invariant metric is L\'evy, and by section (2.1)
  of \cite{GM}
  its quotients should be.
 The  function $f$ is $J_{\pm}(\mu, P)$ for fixed $\mu$. Since we are mainly interested in factorial
 dependencies, we set $\mu = 1$ and change the exponent $2^{-1}$ to $1$
  to make the Gaussian measure a probability measure. In general, the result would
 be modified by a $\pm  b_3$ power of $\mu$.  In this heuristic
discussion, we will not attempt to determine $\alpha_N(P, t)$ or
$M_f$  but will assume that $ \alpha(P, \frac{t}{||f||_{Lip}})$
has rapid decrease in $t$ which improves with the dimension. We
also note that  when $\alpha(P, t)$ is small, we can replace the
median of $J_{\pm}(\mu, P)$ (with $\mu = 1$) by its mean
$$ \int_{Gr(b_3/2 - 1, \sym(b_3/2 - 1,
\C)} \left\{\int_{\hcal \oplus \C} |\det (H^*H - |x|^2 I)| e^{- Tr
H^* H - |x|^2} dH dx\right\} d\nu(\hcal)$$ with a small error
(cf.\ \cite{T}). This mean equals \begin{equation} \label{MEAN}
\int_{\sym(b_3/2 - 1, \C) \oplus \C} |\det (H^*H - |x|^2 I)| e^{-
Tr H^* H - |x|^2} dH dx \end{equation} since both measures are
invariant probability measures and are therefore equal. Here we
ignore factors of $(2\pi)$ (etc.) for the sake of simplicity,
since we are primarily interested in the factorially growing
quantities. Due to the concentration of measure, the spaces
$\hcal_Z$ would have to be very `rare events' if $J_{\pm}(\mu,
P_Z)$ differed appreciably from its mean. We note that since
$H^3_Z$ is a complex polarization, $P_Z$ has special features that
do not hold for random subspaces, but we have no reason to believe
that these special features
 bias $J(\mu, P_Z)$ away from its mean.

 We now observe that (\ref{MEAN}) (with any choice of $\mu$)
  is
similar to the integral for the density of critical points for
holomorphic sections of $\ocal(N) \to \CP^m$ with $m = b_3/2 - 1$
with respect to the Fubini-Study connection  for a fixed degree
$N$ \cite{DSZ2} (\S 4). There, the $\Lambda_Z$ matrix was (for
every $Z$) a two-block diagonal matrix with a large scalar block
and a $1 \times 1$ scalar block. When $\mu = 1$ (\ref{MEAN})
agrees with that $\ocal(N) \to \CP^m$ density in the case
 $N = 1$.  As noted in \cite{DSZ2}, the total number of critical points
of a given Morse index appears to grow at a rate $N^m$ times a
rational quantity in $m$ as $m \to \infty$. This growth rate may
also be easily verified for the Euler characteristic $\chi(T^{*
1,0} \otimes \ocal(N))$, i.e. the alternating sum over the Morse
indices, which is given by
\begin{eqnarray*}\chi(T^{* 1,0} \otimes \ocal(N))&=& \left(
\frac{c(\ocal(N-1))^{m+1}}{c(\ocal(N))}, [\CP^m]\right)\ =\ \frac
{(N-1)^{m+1}+(-1)^m}{N}\;.\end{eqnarray*} Since the  volume of
$\CP^m$ is $\frac{1}{m!}$, this would imply that the density of
critical points grows like $m!$ with the dimension. On this basis,
we would expect that $J_{\pm}(\mu, P_Z)$ for $\mu \simeq 1$ grows
with the dimension at the rate $(b_3/2)! N_{\mu}^{b_3}$ for some
$N_{\mu} >0$.

We note that the Ashok-Douglas heuristic that  the density of
critical points should have the same order of magnitude as  the
 index density is indeed correct in the setting of  $\ocal(N)
\to \CP^m$.  Further, the origin of the factorials $(b_3/2)!$ is
essentially in both the $\ccal$ and $\CP^m$ settings.

Thus our heuristics give  $\kcal^\crit(Z) \sim \frac{(b_3/2)!
N_{\mu}^{b_3}}{b_3!}$. If we integrate over $K_{\mu}$ and apply
the conjectural volume bound $\frac{1}{(b_3/2)!)}$ for $K_{\mu}$,
we would get roughly $\frac{L^{b_3}  N_{\mu}^{b_3}}{b_3!}$.
Further applying the observed relation $L \sim C b_3$ with $C \in
[1/3, 3]$ gives an exponential growth rate for numbers of vacua in
$K_{\mu}$.


\begin{thebibliography}{HHHH}



\bibitem[AD]{AD} S. Ashok and M. R. Douglas, Counting Flux Vacua, JHEP 0401
(2004) 060 (hep-th/0307049).

\bibitem[BGV]{BGV} N. Berline, E. Getzler, and M.  Vergne,
{\it Heat kernels and Dirac operators}. Grundlehren der
Mathematischen Wissenschaften  298. Springer-Verlag, Berlin, 1992.

\bibitem[BSZ1]{BSZ1}  P. Bleher, B. Shiffman and S. Zelditch,  Universality
and scaling of correlations between zeros on complex manifolds,
{ Invent.\ Math.}  142 (2000) 2, 351--395.


\bibitem[BSZ2]{BSZ2} P. Bleher, B. Shiffman and S. Zelditch,
Correlations between zeros and supersymmetry, Commun.\ Math.\
Phys. 224 (2001) 1, 255-269.


\bibitem[BP]{BP} R. Bousso and J.  Polchinski,
Quantization of four-form fluxes and dynamical neutralization of
the cosmological constant,  J. High Energy Phys. 2000, no.\ 6,
Paper 6.

\bibitem[CHSW]{CHSW} P. Candelas, G.  Horowitz, A. Strominger, and E.  Witten,
Vacuum configurations for superstrings. Nuclear Phys. B 258
(1985), no. 1, 46--74.

\bibitem[CO]{Can1} P. Candelas, X. C. de la Ossa,
Moduli space of Calabi-Yau manifolds. Nuclear Phys. B 355 (1991),
no. 2, 455--481; also appeared in {\it Strings '90}
(College Station, TX, 1990), 401--429, World Sci.\ Publishing, River
Edge, NJ, 1991.

\bibitem[DD1]{DD} F. Denef and
M. R. Douglas, Distributions of flux vacua,  J. High Energy Phys.
2004, no.\ 5, 072 (hep-th/0404116).

\bibitem[DD2]{DD2}  F. Denef and
M. R. Douglas, Distributions of nonsupersymmetric flux vacua,
hep-th/0411183.



\bibitem[DGKT]{DGKT}  O.  DeWolfe, A.
Giryavets, S.  Kachru and W. Taylor,  Enumerating Flux Vacua with
Enhanced Symmetries (hep-th/0411061).

\bibitem[Do]{Doug} M. R. Douglas,  The statistics of string/M theory vacua.
J. High Energy Phys. 2003, no.\ 5, 046 (hep-th/0303194).

\bibitem[DSZ1]{DSZ1} M. R. Douglas, B. Shiffman and S. Zelditch,
Critical Points and supersymmetric vacua I,  Comm.\ Math.\ Phys. 252
(2004), no.\ 1-3, 325--358.

\bibitem[DSZ2]{DSZ2} M. R. Douglas, B. Shiffman and S. Zelditch,
Critical Points and supersymmetric vacua II: Asymptotics.


\bibitem[DO]{DO} W. Duke and  O. Imamoglu, Lattice points in cones and
Dirichlet series, IMRN 53 (2004).


 \bibitem[Fy]{Fy}  Y. V. Fyodorov,  Complexity of Random Energy Landscapes,
Glass
 Transition and Absolute Value of Spectral Determinant of Random Matrices
 Physical Review Letters v. 92 (2004), 240601; Erratum: ibid.\ v. 93 (2004), 149901
(cond-mat/0401287).

\bibitem[GKP]{GKP} S.B. Giddings, S.  Kachru, and J. Polchinski,
Hierarchies from fluxes in string compactifications.  Phys.\ Rev.\ D
(3) 66 (2002), no. 10, 106006.


\bibitem[GKT]{GKT}  A. Giryavets, S.
Kachru and P. K. Tripathy,  On the taxonomy of flux vacua, JHEP
0408:002, 2004 (hep-th/0404243).


\bibitem[GH]{GH} P. Griffiths and J. Harris, {\it Principles of Algebraic
Geometry\/}, Wiley-Interscience, New York (1978).

\bibitem[GM]{GM} M.  Gromov and V. D. Milman,
A topological application of the isoperimetric inequality. Amer.\
J. Math. 105 (1983), no.\ 4, 843--854.

\bibitem[GHJ]{GHJ} M. Gross, D. Huybrechts, and D. Joyce, {\it
Calabi-Yau Manifolds and Related Geometries}, Springer
Universitext, Springer, New York (2003).

\bibitem[GVW]{GVW} S. Gukov, C.  Vafa, and E. Witten,
 CFT's from Calabi-Yau four-folds. Nuclear Phys. B 584 (2000), no.\ 1--2, 69--108.

\bibitem[HW]{HW}  G. H. Hardy and E. M. Wright, {\it An introduction to the
theory of numbers.} Fifth edition. The Clarendon Press, Oxford University
    Press, New York, 1979.

\bibitem[Hl]{H} E. Hlawka, \"Uber Integrale auf konvexen K\"orpern I, II,
Monatsh.\ Math. 54 (1950), 1--36, 81--99.

\bibitem[Ho]{Ho} L. H\"ormander, {\it  The analysis of linear partial differential
operators. I. Distribution theory and Fourier analysis.}  Classics in Mathematics.
Springer-Verlag, Berlin, 2003.

\bibitem[KKLT]{KKLT}
S.~Kachru, R.~Kallosh, A.~Linde and S.~P.~Trivedi, De Sitter vacua
in string theory, {\it Phys.\ Rev.\ D}  68, 046005 (2003),
(hep-th/0301240).

 \bibitem[KL]{KL} R. Kallosh and A.  Linde,  Landscape, the scale of SUSY breaking,
 and inflation, hep-th/0411011.

\bibitem[KLRY]{KLRY}
A.~Klemm, B.~Lian, S.~S.~Roan and S.~T.~Yau, Calabi-Yau fourfolds
for M- and F-theory compactifications, Nucl.\ Phys.\ B { 518}, 515
(1998) (arXiv:hep-th/9701023).


\bibitem[Lu]{Lu} Z. Lu, On the Hodge metric of the universal
deformation space of Calabi-Yau threefolds.  J. Geom.\ Anal. 11
(2001), no.\ 1, 103--118.



\bibitem[LS1]{LuS1} Z. Lu and X. Sun,
  On the Weil-Petersson volume and the first Chern
Class of the moduli space of Calabi-Yau manifolds,
math.DG/0510021.

\bibitem[LS2]{LuS2}  Z. Lu and X. Sun,
 Weil-Petersson geometry on moduli space of polarized
Calabi-Yau manifolds, J.  Inst.  Math. Jussieu (2004) 3(2),
185-229
 (math.DG/0510020).


\bibitem[NR]{NR} M. Nechayeva and B. Randol, Approximation of
measures on $S^n$ by discrete measures (preprint, 2005).



\bibitem[Po]{Pom} Ch. Pommerenke, \"Uber die
Gleichverteilung von Gitterpunkten auf $m$-dimensionalen Ellipsoiden.  Acta Arith. 5
(1959), 227--257.




\bibitem[Ra]{Ran} B. Randol, A lattice-point problem. Trans.\ Amer.\
Math.\ Soc.\ 121 (1966), 257--268.

\bibitem[Si]{Sil}  E.
Silverstein,  AdS and dS Entropy from String Junctions, in {\it
From Fields to Strings: Circumnavigating Theoretical Physics}, Ian
Kogan Memorial collection,  vol. 3* Shifman, M. (ed.), World
Scientific,  pp.\ 1848--1863  (hep-th/0308175).



\bibitem[St1]{St1} A. Strominger,
Special geometry. Comm.\ Math.\ Phys. 133 (1990), no.\ 1,
163--180.


\bibitem[St2]{St2} A. Strominger, Kaluza-Klein compactifications,
supersymmetry and Calabi-Yau manifolds, in {\it Quantum fields and
strings: a course for mathematicians}, P. Deligne et.\ al., eds.,
American Mathematical Society, Providence, RI; Institute for
Advanced Study (IAS), Princeton, NJ, 1999, Vol.~2, pp.\
1091--1115.

\bibitem[Sul]{Sul} D. Sullivan,
Infinitesimal computations in topology. Inst. Hautes \'Etudes Sci.,
Publ.\ Math.\ No.\ 47, (1977), 269--331 (1978).

\bibitem[Sus]{S} L. Susskind, The anthropic landscape of string theory,
hep-th/0302219.

\bibitem[Ta]{T} Michel Talagrand,
A new look at independence,
 Ann.\ Probab.  24 (1996), no.\ 1, 1--34.


\bibitem[Wa1]{W1} C.-L. Wang, On the incompleteness of the Weil--Petersson
metrics along degenerations of Calabi--Yau manifolds, Math.\ Res.\ Lett. 1
(1997), 157--171.

\bibitem[Wa2]{W2}
C.-L. Wang, Curvature properties of the Calabi--Yau moduli, Doc.\
Math. 8 (2003), 577--590.



\bibitem[WB]{WB} J. Wess and J.  Bagger, {\it Supersymmetry and supergravity}.
Second edition. Princeton Series in Physics. Princeton University Press, Princeton, NJ,
1992.


\bibitem[Ze1]{ZZ} S. Zelditch, Angular distribution of lattice
points (in preparation).

\bibitem[Ze2]{Z2} S. Zelditch, Random complex geometry and vacua, or: How to count
universes in string/M theory (preprint, 2005).


\end{thebibliography}
\end{document}